\documentclass{aa}

\usepackage{graphicx}
\usepackage{txfonts}
\usepackage{natbib}
\usepackage{multirow}
\usepackage{longtable,lscape}
\usepackage{hyperref}

\usepackage[normalem]{ulem}

\bibpunct{(}{)}{;}{a}{}{,} 

\begin{document}

  \title{The Great Observatories Origins Deep Survey\thanks{Based on
  observations using the Very Large Telescope at the Paranal
  Observatory of the European Organisation for Astronomical Research
  in the Southern Hemisphere, Chile, under ESO programmes 168.A-0485,
  64.O-0643, 66.A-0572, 68.A-0544, and 73.A-0764.}}

  \subtitle{VLT/ISAAC Near-Infrared Imaging of the GOODS-South Field}

  \author{J. Retzlaff\inst{1} \and P. Rosati\inst{1} \and
  M. Dickinson\inst{2} \and B. Vandame\inst{3} \and
  C. Rit\'{e}\inst{4} \and M. Nonino\inst{5} \and C. Cesarsky\inst{6} \and the GOODS Team}

  \offprints{J. Retzlaff, \email{jretzlaf@eso.org}}

  \institute{European Organisation for Astronomical Research in the
  Southern Hemisphere (ESO), Karl-Schwarzschild-Str.\ 2, D-85748
  Garching bei M\"unchen, Germany
  \and
  NOAO, 950 N. Cherry Avenue, Tucson, AZ 85719, USA
  \and
  Canon Research Centre, rue Touche Lambert, F-35510 Cesson S\'{e}vign\'{e}, France
  \and
  Minist\'{e}rio da Ci\^{e}ncia e Tecnologia -- Observat\'{o}rio Nacional, 
  Rua Gal. Jose Cristino 77, Sao Cristovao, Rio de Janeiro -- RJ, Brasil
  \and
  INAF -- Osservatorio Astronomico di Trieste, Via Tiepolo 11, 
  I-34131 Trieste, Italy
  \and
  CEA Saclay, Haute-commissaire a l'Energie Atomique, F-91191 Gif-sur-Yvette, France
  }


\abstract
   {}  
   { We present the final public data release of the VLT/ISAAC
   near-infrared imaging survey in the GOODS-South field.  The survey
   covers an area of 172.5, 159.6 and 173.1 arcmin${}^2$ in the $J$,
   $H$, and $K_\mathrm{s}$ bands, respectively.  For point sources total
   limiting magnitudes of $J=25.0$, $H=24.5$, and $K_\mathrm{s}=24.4$
   ($5\sigma$, AB) are reached within 75\% of the survey area. Thus
   these observations are significantly deeper than the previous EIS
   Deep Public Survey which covers the same region.  The image
   quality is characterized by a point spread function ranging between
   0.34\arcsec{} and 0.65\arcsec{} FWHM.  The images are registered to a
   common astrometric grid defined by the GSC\,2 with an accuracy of
   $\sim\! 0.06\arcsec$ RMS over the whole field.  The overall
   photometric accuracy, including all systematic effects, adds up to 0.05~mag. 
   The data are publicly available from the ESO science
   archive facility. }
   { We describe the data reduction, the calibration, and the quality
   control process.  The final data set is characterized in terms of
   astrometric and photometric properties, including the PSF and the
   curve of growth.  We establish an empirical model for the sky
   background noise in order to quantify the variation of limiting
   depth and statistical photometric errors over the survey area.  We
   define a catalog of $K_\mathrm{s}$-selected sources which contains
   $JHK_\mathrm{s}$ photometry for 7079 objects. Differential aperture
   corrections were applied to the color measurements in order to
   avoid possible biases as a result of the variation of the PSF.  We
   briefly discuss the resulting color distributions in the context of
   available redshift data.  Furthermore, we estimate the completeness
   fraction and relative contamination due to spurious detections for
   source catalogs extracted from the survey data. For this purpose,
   an empirical study based on a deep $K_\mathrm{s}$ image of the Hubble Ultra
   Deep Field is combined with extensive image simulations. }
   { With respect to previous deep near-infrared surveys, the surface density of
   faint galaxies has been established with unprecedented accuracy by
   virtue of the unique combination of depth and area of this survey.
   We derived galaxy number counts over eight magnitudes in flux
   up to $J=25.25$, $H=25.0$, $K_\mathrm{s}=25.25$ (in the AB system). Very
   similar faint-end logarithmic slopes between 0.24 and 0.27~mag${}^{-1}$
   were measured in the three bands.  We
   found no evidence for a significant change in the slope of the
   logarithmic galaxy number counts at the faint end. }
   {}

\keywords{Cosmology: observations -- Cosmology: large scale structure of the universe 
-- Galaxies: evolution -- Infrared: galaxies -- Surveys}

\maketitle

\section{Introduction}

The Great Observatories Origins Deep Survey (GOODS) aims at combining
the best and deepest data from X-ray through radio wavelengths and
making them publicly available to the community. The combined data
enable studies of the distant universe in terms of the formation and
evolution of galaxies and active galactic nuclei, the distribution of
dark and luminous matter at high redshift and the origin of
extragalactic background radiation. Observations are centered on the
two target fields: the Hubble Deep Field North and the Chandra Deep
Field South (CDF-S), each of them covering an area of
160~arcmin${}^2$.  The data are obtained using the most powerful
facilities in space and on the ground, i.~e.{} the NASA Great
Observatories, the Spitzer Space Telescope, Hubble (HST), and Chandra,
ESA's XMM-Newton, the 8.2\,m-aperture Very Large Telescopes at the ESO
Paranal Observatory, the twin 10-meter Keck Telescopes, the 8.2-m
Subaru Telescope, and the 8-m telescopes of the Gemini Observatory.
For an overview of the GOODS project, see
\cite{2003mglh.conf..324D}, \cite{2004ApJ...600L..93G}, and
\cite{2005NewAR..49..440G}.

To study the evolution of galaxies, selection at near-infrared (NIR)
wavelengths has a number of advantages over optical selection, the
most significant one being the small dependence on type out to
redshifts of about three \citep[e.~g.][]{1994ApJ...434..114C}.  The
discovery of a population of distant galaxies based on NIR color
selection \citep[see,
e.~g.,][]{2003ApJ...587L..79F,2004ApJ...617..746D} which would have
been missed by the Lyman break technique demonstrates the importance
of deep NIR imaging for the understanding of the history of mass
assembly in the universe.

While deep ground-based optical imaging has become largely obsolete
due to the GOODS ACS data set, high-quality NIR observations had to be
executed from a ground-based facility because the small field of view
of NICMOS on HST would have made NIR imaging of the GOODS region
overly costly.  However, NIR imaging remains a key requirement for the
scientific goals of the GOODS program because sampling of the NIR
spectral range is crucial if spectral energy distributions have to be
fitted in order to derive accurate photometric redshifts and stellar
population parameters of galaxies.  ESO has dedicated a substantial
amount of observing time in order to support the community with
imaging and also VLT spectroscopy for the target field CDF-S
\citep{2003mglh.conf..332R}.

There were four public incremental releases of the GOODS ISAAC data:
version 0.5 (April 2002), version 1.0 (April 2004), version 1.5
(September 2005), and version 2.0 (September 2007).  In this
publication we present the final version (2.0) of the fully reduced
and calibrated images which have been obtained with the Infrared
Spectrometer And Array Camera (ISAAC) in the $J$, $H$, and
$K_\mathrm{s}$ bands. This data set was released via the ESO science
archive facility as part of the ESO/GOODS project.\footnote{To be
downloaded from\\
\texttt{http://archive.eso.org/cms/eso-data/data-packages/\\
goods-isaac-final-data-release-version-2-0}\label{url}}

The paper is organized as follows. Sect.~\ref{obs} outlines the
observations, Sect.~\ref{datareduction-calibration} describes the
image data reduction and calibration process.  The final survey
images, their properties and an assessment of the photometric
calibration are given in Sect.~\ref{finalimages}. In
Sect.~\ref{catalogs}, we describe how photometric source catalogs were
defined and discuss their characteristics in term of color-magnitude
and color-color diagrams.  In Sect.~\ref{analysis}, first, catalog
completeness and contamination is discussed, and thereafter deep
galaxy number counts in the three survey bands are presented.
Finally, we conclude in Sect.~\ref{summary}.

Note that throughout this work magnitudes are expressed in the AB photometric system
\citep{1983ApJ...266..713O} unless otherwise noted.

\section{Observations}
\label{obs}

\begin{figure}
  \includegraphics[width=\hsize]{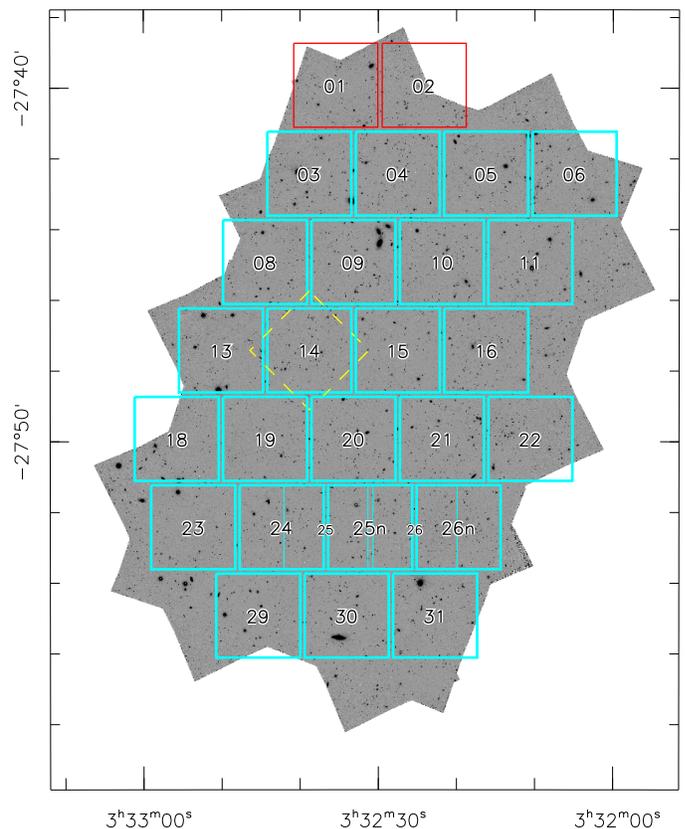}
  \caption{Tiling of the CDF-S region with
  $2.5\arcmin\times 2.5\arcmin$ ISAAC fields displayed over the
  HST/ACS GOODS $z$-band mosaic \citep{2004ApJ...600L..93G}. The
  fields on the very top, F01 and F02, are covered in $J$ and
  $K_\mathrm{s}$, the other 24 fields (F03 through F31) in all
  three bands $J$, $H$, and $K_\mathrm{s}$.  The dashed diamond
  centered on field F14 is the footprint of the deep UDF/ISAAC
  $K_\mathrm{s}$ image.\label{fig:acsz-layout}}
\end{figure}

\begin{table}
\caption{Nominal center positions of the GOODS/ISAAC survey fields.\label{fieldcoords}}
\centering
\begin{tabular}{ccc} \hline \hline
Field &   R.A.  &  Decl.  \\
      & (J2000) & (J2000) \\ \hline
F01  & $3^\mathrm{h}32^\mathrm{m}35\fs{}4$ & $-27\degr 39\arcmin 55\arcsec$ \\
F02  & $3^\mathrm{h}32^\mathrm{m}24\fs{}1$ & $-27\degr 39\arcmin 55\arcsec$ \\
F03  & $3^\mathrm{h}32^\mathrm{m}38\fs{}8$ & $-27\degr 42\arcmin 25\arcsec$ \\
F04  & $3^\mathrm{h}32^\mathrm{m}27\fs{}5$ & $-27\degr 42\arcmin 25\arcsec$ \\
F05  & $3^\mathrm{h}32^\mathrm{m}16\fs{}2$ & $-27\degr 42\arcmin 25\arcsec$ \\
F06  & $3^\mathrm{h}32^\mathrm{m} 4\fs{}9$ & $-27\degr 42\arcmin 25\arcsec$ \\
F08  & $3^\mathrm{h}32^\mathrm{m}44\fs{}4$ & $-27\degr 44\arcmin 55\arcsec$ \\
F09  & $3^\mathrm{h}32^\mathrm{m}33\fs{}1$ & $-27\degr 44\arcmin 55\arcsec$ \\
F10  & $3^\mathrm{h}32^\mathrm{m}21\fs{}8$ & $-27\degr 44\arcmin 55\arcsec$ \\
F11  & $3^\mathrm{h}32^\mathrm{m}10\fs{}5$ & $-27\degr 44\arcmin 55\arcsec$ \\
F13  & $3^\mathrm{h}32^\mathrm{m}50\fs{}1$ & $-27\degr 47\arcmin 25\arcsec$ \\
F14  & $3^\mathrm{h}32^\mathrm{m}38\fs{}8$ & $-27\degr 47\arcmin 25\arcsec$ \\
F15  & $3^\mathrm{h}32^\mathrm{m}27\fs{}5$ & $-27\degr 47\arcmin 25\arcsec$ \\
F16  & $3^\mathrm{h}32^\mathrm{m}16\fs{}2$ & $-27\degr 47\arcmin 25\arcsec$ \\
F18  & $3^\mathrm{h}32^\mathrm{m}55\fs{}7$ & $-27\degr 49\arcmin 55\arcsec$ \\
F19  & $3^\mathrm{h}32^\mathrm{m}44\fs{}4$ & $-27\degr 49\arcmin 55\arcsec$ \\
F20  & $3^\mathrm{h}32^\mathrm{m}33\fs{}1$ & $-27\degr 49\arcmin 55\arcsec$ \\
F21  & $3^\mathrm{h}32^\mathrm{m}21\fs{}8$ & $-27\degr 49\arcmin 55\arcsec$ \\
F22  & $3^\mathrm{h}32^\mathrm{m}10\fs{}5$ & $-27\degr 49\arcmin 55\arcsec$ \\
F23  & $3^\mathrm{h}32^\mathrm{m}53\fs{}6$ & $-27\degr 52\arcmin 25\arcsec$ \\
F24  & $3^\mathrm{h}32^\mathrm{m}42\fs{}3$ & $-27\degr 52\arcmin 25\arcsec$ \\
F25  & $3^\mathrm{h}32^\mathrm{m}36\fs{}7$ & $-27\degr 52\arcmin 25\arcsec$ \\
F25n & $3^\mathrm{h}32^\mathrm{m}31\fs{}0$ & $-27\degr 52\arcmin 25\arcsec$ \\
F26  & $3^\mathrm{h}32^\mathrm{m}25\fs{}3$ & $-27\degr 52\arcmin 25\arcsec$ \\
F26n & $3^\mathrm{h}32^\mathrm{m}19\fs{}7$ & $-27\degr 52\arcmin 25\arcsec$ \\
F29  & $3^\mathrm{h}32^\mathrm{m}45\fs{}3$ & $-27\degr 54\arcmin 55\arcsec$ \\
F30  & $3^\mathrm{h}32^\mathrm{m}34\fs{}0$ & $-27\degr 54\arcmin 55\arcsec$ \\
F31  & $3^\mathrm{h}32^\mathrm{m}22\fs{}7$ & $-27\degr 54\arcmin 55\arcsec$ \\ \hline
\end{tabular}
\end{table}

The Infrared Spectrometer and Array Camera (ISAAC) mounted on the
first of the four 8.2-m VLT unit telescopes at Cerro Paranal is
equipped with a $1024\times 1024$ pixels HgCdTe Rockwell Hawaii array
having a pixel scale of 0.148\arcsec, corresponding to a
$2.5\arcmin\times 2.5\arcmin$ field of view
\citep{1999Msngr..95....1M}.  Fig.~\ref{fig:acsz-layout} shows how the
ISAAC pointings were laid out to assemble a mosaic of the CDF-S
region.  A contiguous area of 24 fields, 150~arcmin${}^2$
respectively, are covered in the $J$, $H$, and $K_\mathrm{s}$ bands,
plus 2 additional fields (F01 and F02) at the top of the survey area
which have $J$ and $K_\mathrm{s}$ but no $H$ data.  In the early
stages of the project, in an attempt to maximize the overall survey
efficiency, fields F25 and 26 were re-arranged and replaced with F25n
and 26n.  The $K_\mathrm{s}$ band data which was obtained for fields
F25 and 26 has been included in the survey data processing in the same
way as the other fields despite their shallowness. Later on, this
auxiliary data proved to be very valuable for the purpose of
validation of the internal photometric consistency
(Sect.~\ref{sect:photval}).  See Table~\ref{fieldcoords} for the field
coordinates.

The GOODS ESO ISAAC program was originally planned to reach $5\sigma$
limiting magnitudes of 25.2, 24.7, and 24.4 in $J$, $H$, and
$K_\mathrm{s}$, respectively, in order to roughly match pre-launch
expectations for Spitzer IRAC sensitivity at 3.6 to 8~$\mu$m.  The
intent was to provide NIR flux measurements for the large majority
of IRAC-detected galaxies, as well as images with higher angular
resolution in order to facilitate the deblending of sources in crowded
regions of the lower-resolution IRAC data.  In practice, Spitzer IRAC
in-flight performance at 3.6 and 4.5~$\mu$m significantly exceeded the
pre-launch expectations, making the GOODS data at those wavelengths
substantial deeper than any ground-based NIR program could
reasonably match.  Given the instrumental sensitivity, a total
integration time per tile of 3.5, 5, and 6 hours was estimated to
reach the intended depth.  To achieve accurate sky background
subtraction which is most crucial for deep NIR imaging, the commonly
practiced jitter imaging technique was used.  A jitter box size of
$25\arcsec$ was chosen within which the control system automatically
offsets the telescope between subsequent integrations.  Detector
integration times (DITs) between 10 and 30~s were used, according to
the typical sky brightness (darker at $J$, brighter at $H$ and
$K_\mathrm{s}$). The number of detector integrations (NDIT) was chosen
so that total integration times (DIT $\times$ NDIT) between 60 and
180~s were used.  The survey was executed in observation blocks (OBs)
each of which comprising the integration of a single field in one
filter for a total integration time of normally 3600~s or 1800~s, and,
thus, resulting in between 15 and 40 images.  The total actual
integration time of the data that were combined in the final survey
images amounts to 360 hours.  Given that some data were discarded and
taking into account observing overheads the whole program was
allocated about 500 hours of observing time.  The observations have
been executed in service mode in 216 nights between October 1999 and
January 2007.

The data were obtained under the ESO large programme 168.A-0485, led
by C.~Cesarsky, in direct support to the GOODS project. Data covering
four ISAAC fields in $J$ and $K_\mathrm{s}$ bands were also drawn from
the ESO programmes 64.O-0643, 66.A-0572 and 68.A-0544, led by
E.~Giallongo \citep[see][]{2001A&A...375....1S}.

\section{Image data reduction and calibration}
\label{datareduction-calibration}

\subsection{Overview}

\begin{figure}
  \includegraphics[width=\hsize]{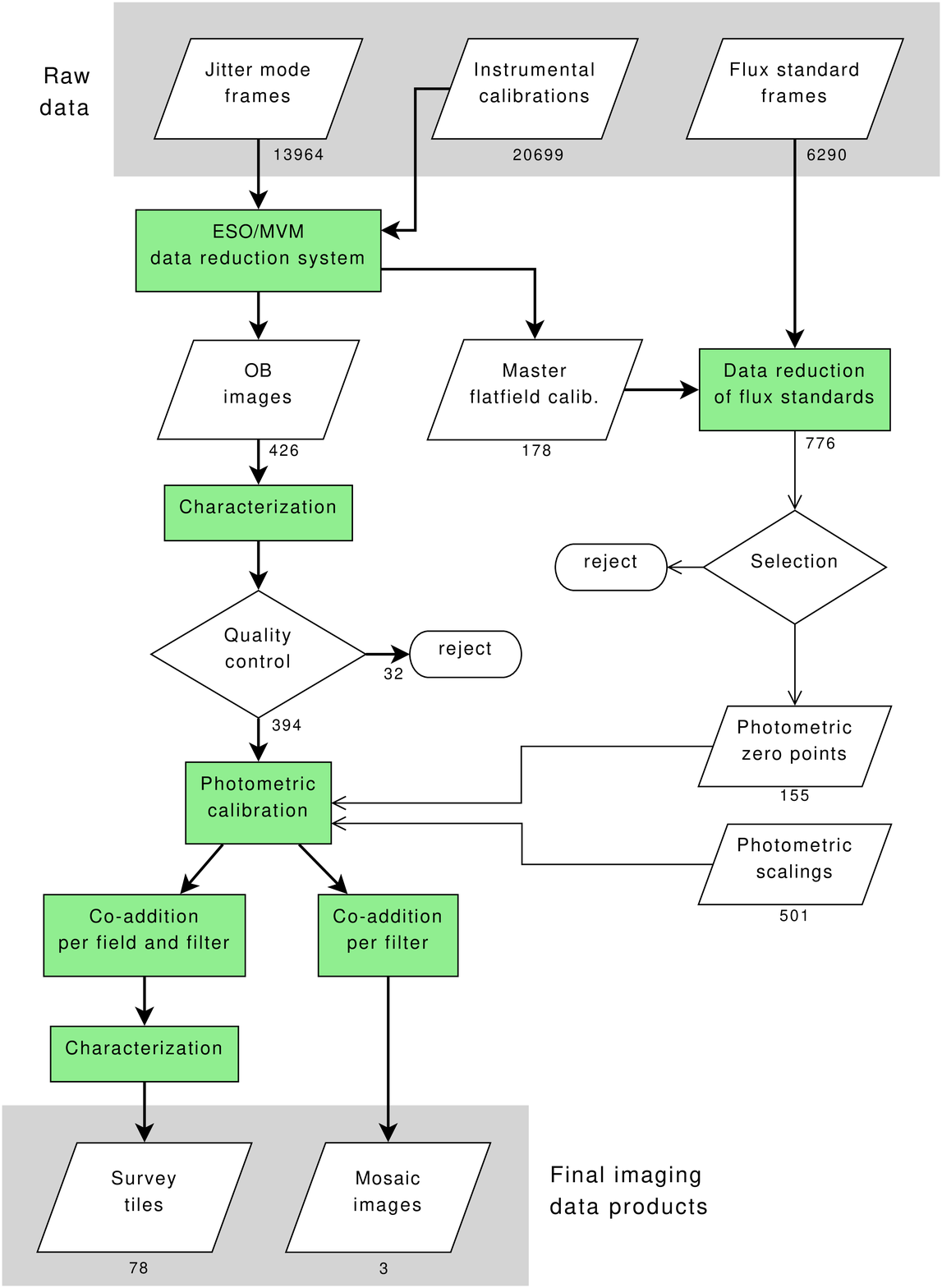}
  \caption{Schematic flow of the data reduction and calibration
  process from raw data to final, fully calibrated survey data
  products (from top to bottom).  Processing steps are represented by
  rectangles, input/output data by parallelograms, and thick arrows
  indicate the image data flow.  Numbers represent the cardinality of
  each data set.  Refer to the text for a detailed
  description.\label{fig:procflow}}
\end{figure}

The overall survey data reduction and calibration process is
schematically illustrated in Fig.~\ref{fig:procflow}.  To begin with,
the ESO/MVM data reduction system was used to process the raw jitter
mode observations using the required calibration files to correct for
instrumental signatures and to produce \emph{OB images}, that is one
co-added image with associated weight map per OB being astrometrically
registered to the predefined astrometric grid of the survey.  The
fully automated processing of the entire data set of 13964 raw jitter
mode frames and 20699 calibration files resulted in 426
astrometrically calibrated OB images (Sect.~\ref{sect:dataproc},
\ref{sect:astrom}).
Then, the OB images were characterized in terms of PSF width,
photometric uniformity, and depth in order to identify and remove
images of low-quality which would have degraded the
quality of the final survey products (Sect.~\ref{sect:qc}).

For the photometric calibration of the OB images, at first, the raw
standard star observations were reduced to instrumental photometric
zero points using the master flatfield calibrations generated by
ESO/MVM.  From this data set, a subset of ``best'' zero points was
constructed.  Then, photometric scalings of OB images belonging to the
same tile and between adjacent tiles were measured and combined with
the zero points to determine -- in a robust way -- an accurate
photometric calibration for all OB images equalized over the whole
survey field (Sect.~\ref{sect:photcal}).  That followed, the
photometrically calibrated OB images were co-added to produce the
\emph{survey tiles} which subsequently underwent a final step of image
quality characterization.

\subsection{Data processing}
\label{sect:dataproc}

The ESO/MVM software, version 1.3.4, has been used for image data
reduction.  ESO/MVM follows well established procedures for the
reduction of NIR imaging observations taken in jitter mode by
featuring a two-pass scheme to mask out sources prior to the final
background estimation.  In the following, we will outline the main
reduction steps executed by ESO/MVM. Comprehensive documentation of
ESO/MVM including all the reduction steps and algorithmic details
being implemented is beyond the scope of this publication and may be
found in \cite{2004ThesisVandame} which serves as reference
documentation and describes in length the implemented algorithms,
parameter settings, instrument specific configurations, and sample
applications\footnote{The full software package may be downloaded
from\\ \texttt{http://archive.eso.org/cms/eso-data/data-packages/\\
eso-mvm-software-package}}.

First, the raw calibration data, namely dark frames and twilight sky
flats, were combined into master calibration files and used to correct
the raw scientific images for the basic instrumental signatures.  No
attempt was made to correct for detector nonlinearity. However, the
overall effect is expected to be rather small because the scientific
OBs for the survey program and, by the same token, the OBs for
photometric standards had been generally designed with the objective
to operate the detector in the linear regime under nominal
conditions. The magnitude of possible nonlinearity effects on the
final results is discussed in Sect.~\ref{sect:photerr}.

In a few cases, overly long OBs were split in two blocks, or,
particularly short, partially completed, OBs were merged prior to data
reduction in order to optimize the quality of the reduction.
Sky background images were computed from groups of between 13 and 15
consecutive jitter images (depending on filter) using sigma-clipped
pixel-by-pixel image combination.  Each science image was
sky-subtracted using the linear interpolation in time of the two
corresponding successive sky background images.  Possible transparency
variations from image to image were monitored by means of the
signal-to-noise ratio (SNR) based on which outliers with exceptionally
low SNR were automatically discarded. An individual rescaling of the
images has not been done.

Then, for each OB, the individual background-subtracted images were
astrometrically registered to each other with sub-pixel accuracy and
co-added (see below) to form a preliminary version of the OB images.
The purpose of these first-pass images is the creation of masks that
mask the astronomical sources in order to improve the sky background
computation in the second pass. To this end all sources exceeding
$\sim\! 3$ times the local RMS noise were detected on the
PSF-convolved image and -- to also exclude the wings of the source
profile -- each source's region was artificially enlarged by a factor
of 2 (linearly) before creating its mask.  Thus, in the second pass,
the sky background images were re-computed exclusively from image
pixels that belong to background regions whereas the rest of the
procedure was repeated unaltered.

\subsection{Astrometric calibration}
\label{sect:astrom}

The astrometric calibration is based on a dense reference catalog
which was generated by the GOODS team from a deep R-band image of the
CDF-S and was also used for the production of the GOODS/ACS image
mosaics \citep{2004ApJ...600L..93G}.  The image was obtained with the
Wide Field Imager mounted at the 2.2-m MPG/ESO telescope at La Silla,
and astrometrically calibrated using the Guide Star Catalog (GSC\,2).
Each OB image of the GOODS/ISAAC survey was astrometrically registered
using the reference catalog whereas image distortions were modeled
with a $3^\mathrm{rd}$ order polynomial. This resulted in an internal
astrometric accuracy between 0.05\arcsec{} and 0.06\arcsec{} RMS as
measured for sources brighter than 20~mag in the final mosaic images
in all three filters.  The relative registration between bands is
accurate to $\sim\! 0.03\arcsec$.

For the astrometric grid of the OB images the same projection as for
the GOODS/ACS images has been adopted and a pixel size of
0.15\arcsec{} has been chosen so that one GOODS/ISAAC pixel subtends
exactly a block of $5\times 5$ GOODS/ACS pixels.  The individual
jitter images were resampled to this grid using the Lanczos-3
interpolation kernel.  Pixel-to-pixel noise correlation, the
interpolation process has introduced, is quantified in
Sect.~\ref{how-to-compute-skynoise}.  During the process of image
co-addition, bad-pixel masks were taken into account and each
contribution was recorded pixel-by-pixel to build up respective weight
maps.

The comparison of the final mosaics with the calibrated data from the
Hubble Space Telescope (HST) Advanced Camera for Surveys (ACS) in its
version 1.1 incarnation yields astrometric offsets of typically
$\sim\! 0.1\arcsec$ RMS across the entire area which can be attributed
primarily to zonal residuals that are known to be present in the
astrometric calibration of the HST/ACS mosaics.

\subsection{Quality control of OB images}
\label{sect:qc}

For each OB image the quality was quantified in terms of the FWHM of
stellar sources (``seeing''), the sky noise as a function of aperture
size, and the photometric scaling of images belonging to the same tile
and filter.  Then, based on the inspection of these quality parameters
with respect to the whole sample, images of significantly low quality
were identified and sorted out, followed by a visual inspection, to
ensure that no corrupted image slipped through.

In order to determine the average FWHM of stellar sources, we made use
of a list of ca.\ $400$ sources that appear point-like in the
GOODS/ACS $z$-band image. Based on this list we pre\-selected
appropriate candidate sources for the PSF characterization.

\label{how-to-compute-skynoise}
The sky noise was measured for each OB image in a similar fashion as
described by \cite{2003AJ....125.1107L}, that is, first, a
noise-equalized image was created by multiplication with the square
root of the weight map. Then, the image was segmented into background
and sources using a $\sim\! 2.6\sigma$ significance threshold plus a 5
pixel wide safety margin around each source segment. Then, the
background flux was sampled with non-overlapping circular apertures
with diameters between 0.5\arcsec{} and 4.0\arcsec{} and a linear
relation of the form $(a+bs)s$ was fitted to the dispersion to obtain
the two parameters $a$ and $b$. Here, the linear aperture size, $s$,
is defined as the square root of the aperture area in pixel units.  It
has been verified that the adopted parameterization describes the
actual data adequately.  Therewith, the sky noise, that is the
statistical fluctuation of the flux measured within an aperture of
size $s$ at any point of the image can be expressed by the Gaussian
dispersion
\begin{equation}
\sigma_\mathrm{sky}(s) = \frac{\sigma_0}{\sqrt{w_j}}( a+b s)s ,
\label{sigmasky}
\end{equation}
where $\sigma_0$ is the RMS of all background image pixels, and $w_j$
is the relative weight\footnote{For the \emph{relative weight}, $w_j$,
a normalization to unity at the \emph{median pixel} has been adopted,
i.~e., $w_j\equiv W_j/\mathrm{median}(W_j)$ with $W_j$ denoting the
inverse variance weight map and the median being taken of all pixels
of the given image having $W_j>0$.  Thus, $\sigma_0$ refers to the
median pixel. This notion, using the index zero, has been adopted
throughout the text for quantities scaling with the weight map, namely
the limiting depth and the gain conversion factor.}  of pixel $j$.
This empirical model of the sky noise accounts for the pixel-to-pixel
correlation which results from the preceding image interpolation step
and which breaks the simple scaling with aperture scale
$\sigma_\mathrm{sky}\propto s$.  According to the interpolation
kernel, the correlation length is not more than a few pixels.  To
quantify the amount of this small-scale contribution, the total $a+b$
is typically about 1.15, that is, at the scale of a single pixel,
i.~e.{} $s=1$, the direct estimation $\sigma_0$ underestimates the
flux dispersion by about 15\%.  The other contribution which affects
slightly larger scales is due to imperfect background subtraction,
that is residual spatial modulations of the image background.  If the
night sky background is highly variable with a time scale being
shorter than the one implicit to the averaging process that is part of
the image data reduction, then such residuals may remain -- mostly
apparent in the form of ripples or speckles.  Not surprisingly, the
parameters $a$ and $b$ are sensitive indicators for the presence of
such residuals. Therefore, we have employed them to identify and
reject OB images which were manifestly affected above average.

In the course of the quality control process, 15 OB images were rejected
because of the seeing being worse than 0.85\arcsec{} FWHM, nine for being
very noisy and showing particularly strong residuals in the sky
background ($b>0.15$), five for being exceptionally ``shallow'' with the photometric
zero point of more than 0.3~mag below average, and three for other
reasons such as corrupted data.
In total, 394 OB images, which account for 93.7\% of the total
integration time of reduced OBs, passed the quality control criteria
and underwent the following steps of photometric calibration, image
stacking, and final characterization which we will describe in the
next section.

\begin{figure}
  \resizebox{\hsize}{!}{\includegraphics{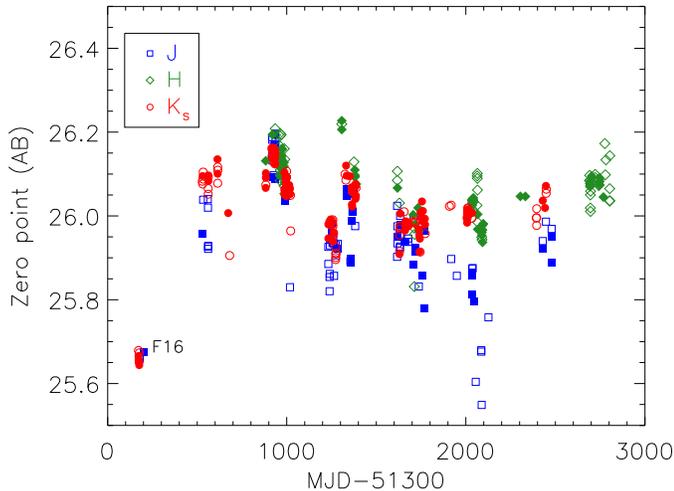}}
  \caption{Variation of the photometric zero point of the GOODS/ISAAC
  OB images from October 1999 until January 2007.  Filled symbols
  correspond to OBs with flux standard star calibrations, open symbols
  refer to OBs whose zero points have been inferred by photometric
  scaling. Zero points have been converted to unit airmass to produce
  this plot. However, this correction is negligibly small compared to
  the gross variation. 
\label{sci-zp-plot}}
\end{figure}

\subsection{Photometric calibration}
\label{sect:photcal}

\subsubsection{Photometric zero points}

As the GOODS/ISAAC program was spread over a large time frame, the
substantial variation of the instrumental photometric zero point on
various time scales is not surprising (Fig.~\ref{sci-zp-plot}).  These
variations are caused by instrumental interventions, the steady
degradation of the mirrors' reflectivity, mirror re-coating events,
and, finally, the variation of atmospheric transparency.  The
significantly low instrumental zero point at the beginning of VLT
operations, for instance, when the $J$ and $K_\mathrm{s}$ data for tile F16 were
being obtained, is likely due to accumulated dust on the mirror of UT1
caused by the ongoing construction of the other UTs at that time.  In
order to establish an accurate and consistent photometric calibration
across the whole survey area, we have combined photometric zero points
(ZP) obtained from standard stars with relative photometric scalings
between survey tile images.

Instrumental ZPs were obtained using faint NIR standard stars which
are being observed on a nightly basis as part of the observatory's
instrument calibration plan. Most of the flux standards were drawn
from \cite{1998AJ....116.2475P} and
\cite{1998AJ....115.2594H}, complemented by a small number of UKIRT standards 
\citep{2001MNRAS.325..563H}.

The available data of photometric flux standard observations was
carefully selected to minimize the potential bias due to detector
nonlinearity.  For each image of a photometric standard star the peak
signal including the background was used as quality indicator and
outliers exceeding 15000 ADU which corresponds to a formal
nonlinearity of 1\% \citep{2002IsaacDataReductionGuide} were
eliminated.  The remaining set of photometric calibrations from which
the ZPs were finally determined is dominated to 85\% by data with peak
fluxes below 10000 ADU which corresponds to a formal nonlinearity of
ca.\ 0.5\% while the remaining 15\% of the data have peak fluxes above
10000 ADU.  Additionally, we have examined more than 5000 individual
photometric calibrations in $J$, $H$, and $K_\mathrm{s}$ band as a
function of the maximum signal level to empirically check for a
possible systematic trend in the derived ZPs.  Up to a signal level of
15000 ADU, we found no evidence that ZPs are systematically biased
low. Merely beyond 20000 ADU, i.~e.{} for data that was finally not
used for calibration, a decline by $\ga$0.03~mag becomes apparent in
all three bands.  Hence, we can conclude that the ZPs which were
eventually used to anchor the survey's photometry are affected by
detector nonlinearity by less than 0.01~mag even in the worst case.

We started out by selecting all the standard star observations within
an interval of plus or minus 2 nights around each science observation
and derived a total of 776 zero points -- by far more than what went
into the final photometric solution (see below).  ZPs were computed
from the instrumental flux measured within an aperture of 10\arcsec{}
diameter in compliance with \cite{1998AJ....116.2475P}.

Since flux standards had not always been observed immediately before
or after each science integration, we adopted the following scheme for
the association of ZPs aiming at a practical compromise between
conservatively minimizing the effect of possible transparency
variations and ending up with a sufficient number of ZPs.  For each OB
the ZP whose flux standard had been acquired closest in time to the
science observation within a time interval of less than 2 hours and an
airmass difference below 0.5 was selected. If these conditions were
not met by any ZP, the respective OB was not assigned a ZP but the
photometric calibration was purely inferred from the scaling relative
to other OBs belonging to the same or to neighboring tiles (see
below).  Using these criteria, 210 out of the 394 OB images were
associated with 165 ZPs, 50\% of which were obtained in a time
interval of plus or minus 45 minutes with respect to the corresponding
science observation.

Then, we have inspected each associated ZP in the context of all the
other ZPs obtained within an interval of 5 nights so as to identify
non-photometric conditions and, hence, we have dismissed 10 ZPs for
being low by $\ga 0.05$~mag.

The resulting 155 instrumental ZPs were converted into ZPs for 199 OB
stacks using the atmospheric extinction coefficients for the ISAAC
instrument, 0.09, 0.04, and 0.06~mag per unit airmass for $J$, $H$,
and $K_\mathrm{s}$, respectively, as determined by \cite{Mason2008}.

The match between the ISAAC filters ($J$, $H$, and $K_\mathrm{s}$) and
those used to establish the faint IR standard star system of
\cite{1998AJ....116.2475P} is quite good and one expects color terms
which differ from 0 by less than 0.01
\citep{2002IsaacDataReductionGuide}.  As the color transformation
between ISAAC magnitudes and those of LCO have never been
experimentally verified, and our data does not allow for it either, we
have chosen not to apply any color correction.

\begin{figure}
  \includegraphics[width=\hsize]{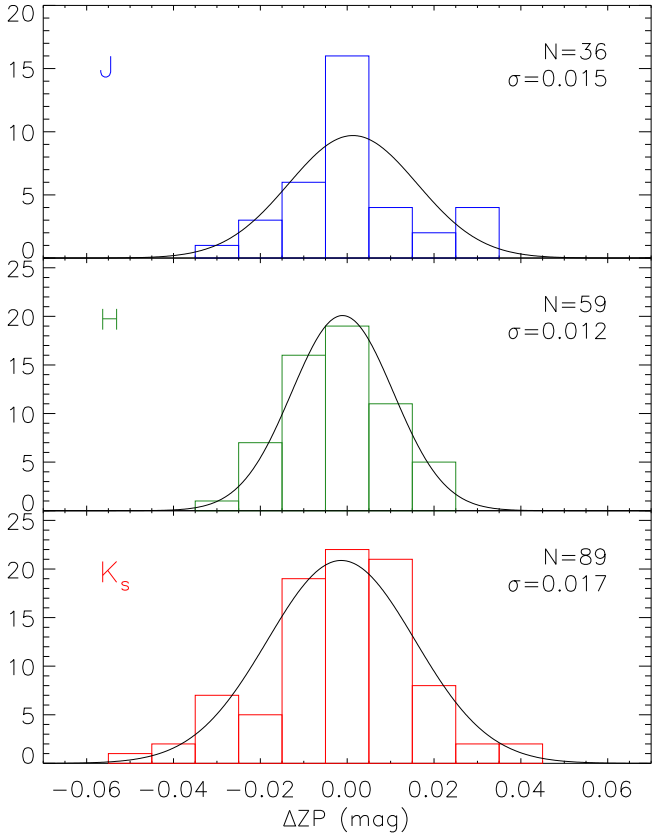}
  \caption{\label{fig:zpsol-diff}Residuals of photometric zero points
  of OB images as derived from flux standards with respect to the
  final global photometric solution for $J$, $H$, and $K_\mathrm{s}$
  (from top to bottom).  For each passband, the total number of OB
  images having photometric zero points based on flux standards, $N$,
  and the RMS of the residuals about the mean, $\sigma$, is reported.
  Gaussian distributions with the same mean and dispersion as the
  actual data are superposed.}
\end{figure}

\subsubsection{Global photometric calibration}
\label{sect:globphotcal}

To determine a survey-wide photometric solution per filter,
photometric scalings between individual OB images were computed.  As
it was straightforward to determine the relative flux scaling for
images belonging to the same tile and filter with sub per-cent
accuracy once PSF matching had been done, the scaling for neighboring
tiles turned out to be difficult to measure, because, usually, the
overlapping area was too small to find a sufficient number of high
signal-to-noise sources for accurate photometry.  Therefore, we have
employed public data from the ESO science archive which covers the
survey area in the same bands in order to establish accurate relative
photometric calibrations between adjacent survey fields.  The data had
been obtained using the SOFI instrument
\citep[][]{1998Msngr..91....9M} mounted on the New Technology
Telescopy at La Silla.  SOFI's field of view of about
$4.9\arcmin\times 4.9\arcmin$ guarantees sufficient overlap between
ISAAC and SOFI images in general, while other instrument
characteristics are similar by construction, in particular, the
instrumental response.  The $J$ and $K_\mathrm{s}$ observations that
were used belong to the infrared part of the ESO Imaging Survey
\citep[e.~g.][]{2006A&A...452..119O}, the $H$ band data were
originally obtained within a program led by D.~Rigopoulou
\citep[see][]{2003A&A...403..493M}.\footnote{The observations have been obtained under
ESO programmes 60.A-9005, 164.O-0561, 66.A-0451, and 68.A-0375.}  The
layout of the $JHK_\mathrm{s}$ SOFI observations in the CDF-S
generally includes some overlap of adjacent images, and, additionally,
a series of low-exposure ``calibration images'' arranged in a pattern
being offset to the deep integrations had been acquired in $J$ and
$K_\mathrm{s}$, so that accurate photometric scalings could be
measured not only between ISAAC and SOFI but also between adjacent
SOFI images.  To this end, we have processed the raw SOFI data in the
same way as we did with the ISAAC data, employing ESO/MVM for the data
reduction followed by careful examination of the image quality
(Sect.~\ref{sect:qc}).  After all, 39 best-quality images in terms of
seeing, noise, and background homogeneity were used in $J$, $H$, and
$K_\mathrm{s}$, plus 19 shallower ``calibration images'' in $J$ and
$K_\mathrm{s}$.  To measure the photometric scaling between two
images, at first, the image with the better seeing was smoothed with a
Gaussian kernel to match the PSF of the other image with poorer
seeing.  Because typically the ISAAC seeing is better than the SOFI
seeing (0.5\arcsec{} vs.{} 0.7\arcsec{} median), for scalings between
ISAAC and SOFI images in the majority of cases the ISAAC image is
smoothed to match the PSF of the SOFI image.  Then, instrumental
magnitudes of all isolated, high signal-to-noise (SNR$>$20) sources
were measured using SExtractor's auto-scaling aperture magnitudes in
double-frame mode \citep{1996A&AS..117..393B} and the magnitude
differences were averaged to obtain the relative photometric scaling.
In this way, photometric scalings with respect to SOFI images could be
measured with a typical uncertainty between 1 and 2\%, slightly
depending on the filter.  We have minimized a possible bias due to the
relative nonlinearity of the two instruments as far as possible.
Objects being relatively bright for ISAAC were discarded based on
their peak flux corresponding to an effective flux cut at $\sim\!
16.5$~mag (AB).  Furthermore, the photometric scaling was computed by
averaging over all suitable sources, i.~e.{} typically more than 10
sources, stars and galaxies, between ca.\ 18 and 20~mag (AB)
contribute to the resulting scaling instead of being based on a few
bright stars only.  For objects fainter than 17~mag, ISAAC's detector
nonlinearity is not an issue (Sect.~\ref{sect:photerr}). As another
precaution, we visually inspected the differential ISAAC-SOFI
photometry but did not see any case of flux dependent bias. Therefore,
we can exclude that the measured photometric scalings between ISAAC
and SOFI data are biased significantly, that is by more than 0.01~mag.
That followed, we have applied a global $\chi^2$-minimization
technique \citep[][]{1998PASP..110.1464K}, to find, for each passband,
the photometric solution adjusted across the whole survey.

The residual photometric differences of the photometric ZPs as given
by the flux standards with respect to the global photometric solution
were examined to identify implausible ZPs. Consequently, 5 ZPs were
removed, and the final photometric solution was calculated based on
155 photometric ZPs and 501 photometric scalings.  This means that on
average, more than one photometric ZP contributed to the final
photometric solution per OB image, namely 1.4, 2.5, and 3.2 for $J$,
$H$, and $K_\mathrm{s}$, respectively, thereby reducing the impact of
possible errors of individual photometric ZP measurements by virtue of
averaging.  As an additional check, we have verified that the
residuals are not correlated with observational or instrumental
parameters (seeing, sky background level, DIT).  The photometric
residuals appear to be normally distributed, and indicate an internal
photometric accuracy of better than 0.02~mag RMS from tile to tile in
all three bands (Fig.~\ref{fig:zpsol-diff}).  The individual
inspection of a set of non-saturated, isolated stars has revealed in a
few cases photometric discrepancies of 3--5\% between SOFI and ISAAC
images which were not attributable to photon noise but are presumably
due to systematic errors left over after flat field correction.

Finally, in order to convert from the Vega to the AB system, AB
corrections of 0.9603, 1.426, and 1.895~mag for $J$, $H$, and
$K_\mathrm{s}$, respectively, were applied.  These numbers were
obtained from the ESO Magnitude-to-flux-converter\footnote{Accessible
at \texttt{http://archive.eso.org/apps/mag2flux}} (version 1.07) which
makes use of the instrumental transmission curves of the ESO Exposure
Time Calculators.

\begin{figure}
  \includegraphics[width=\hsize]{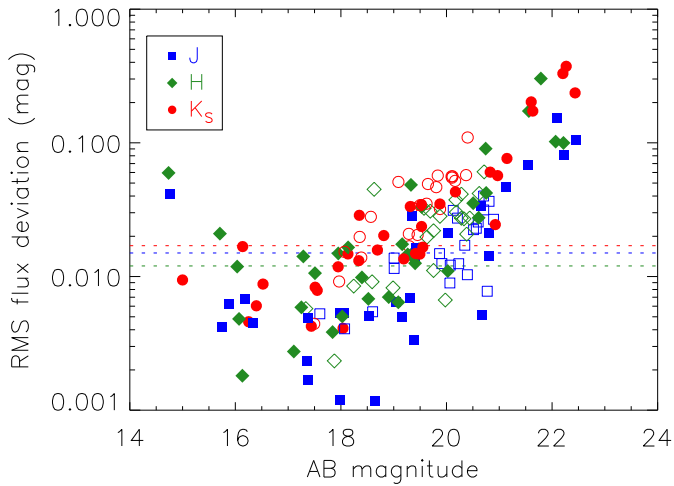}
  \caption{Reproducibility of flux measurements on OB images as a
  function of AB magnitude.  Each data point corresponds to one of 56
  individually selected test sources measured in one of the three
  filters, $J$, $H$, and $K_\mathrm{s}$, discriminated by different
  plot symbols, square, diamond and circle, respectively.  The
  root-mean-square (RMS) flux deviation from the mean was computed
  from 3 to 30 measurements on the photometrically calibrated OB
  images.  Solid symbols denote unresolved sources, while open symbols
  correspond to extended sources.  For comparison the $1\sigma$ errors
  of the photometric solution of the OB images (Sect.~\ref{sect:globphotcal})
  are shown as horizontal dotted
  lines.\label{fig:magrms}}
\end{figure}

\subsubsection{Photometric uncertainties}
\label{sect:photerr}

In order characterize the effect of detector nonlinearity on the final
survey data we have performed photometric checks of individual objects
on the calibrated OB images as a function of source properties (flux,
extent), observational conditions (seeing, background brightness), and
instrumental parameters (DIT).  To this end five survey tiles were
selected within which aperture fluxes (4\arcsec{} diameter) of 56
isolated objects were measured in $J$, $H$, and $K_\mathrm{s}$ from
between 3 and 30 OB images.  Fig.~\ref{fig:magrms} summarizes the
results of this study as a function of source flux.  The increasing
RMS flux deviation towards the bright end indicates that detector
nonlinearity starts to take effect for unresolved sources at total
magnitudes between 16 and 16.5 AB.  The comparison with the calibrated
SOFI images, which are expected to be significantly less affected by
nonlinearity than the ISAAC data, allows to directly estimate the net
flux bias for these objects.  On average the ISAAC photometry
underestimates the flux by between 0.02~mag at $\sim\! 16.5$~mag and
ca.{} 0.05~mag at 14.7~mag AB irrespective of the pass band.  The
maximum flux bias of 0.07~mag was found for the $H$ flux of the
brightest star in the test sample ($H=14.7$).  Note that there are
actually very few objects that bright in the whole survey, namely 5
having $K_\mathrm{s} \le 15$ and 23 having $K_\mathrm{s}
\le 16.5$, almost all of them being stars with the exception of two
bright galaxies of $K_\mathrm{s}=16.1$ and 16.5.  For the bulk of objects having
total fluxes of 17~mag AB and fainter, no indication for nonlinearity
effects has been found.  In fact, the overall trend of the flux
uncertainty, which resolved and unresolved objects 
seem to follow in the same way, can be explained purely in
terms of statistical errors (see also Sect.~\ref{sect:limdepth}).

Fig.~\ref{fig:magrms} also indicates the errors of the photometric
solution to illustrate the level of systematic photometric errors in
proportion to the statistical errors.  For instance, between ca.\ 16
and 18~mag, $H$ and $K_\mathrm{s}$ photometry is typically limited by
the systematic uncertainties intrinsic to the photometric solution,
or, photometry fainter than $J \sim 20$ becomes dominated by
statistical errors.  Remember that the data points in
Fig.~\ref{fig:magrms} refer to the $1\sigma$ fluctuations of
photometric measurements on OB images, whereas the final co-added
survey images are expected to have reduced photometric errors,
typically by a factor of between 2 and 2.5.

Exceptionally high sky background brightness in combination with a
comparatively large detector integration time (DIT) has resulted in a
few OBs in $H$ and $K_\mathrm{s}$ for which the detector was not
operated in the linear regime.  In fact, the most extreme cases are 8
out of 294 OBs in $H$ and $K_\mathrm{s}$ for which the sky background
signal induces a nominal nonlinearity of between 2 and 3\%.  To
compensate for this effect which is effectively resulting in a loss of
sensitivity, or a lowered photometric ZP, the respective OBs were not
directly anchored to ZPs but their photometric calibration was
inferred from relative photometric scalings with respect to the other
OBs of the same tile and filter (see Sect.~\ref{sect:globphotcal}).
Finally, we have verified that the photometric measurements on the
calibrated ``nonlinear'' OBs are indistinguishable from the other
data.

To estimate the overall error, we linearly add up the individual
contributions, that is the internal accuracy of the photometric
solution ($\le$\,0.017~mag), the residual flat field error (0.025~mag,
$1\sigma$), and the possible nonlinearity bias ($\la$\,0.01~mag) for
objects not brighter than 17~mag (AB), resulting in the total
photometric error of up to 0.05~mag ($1\sigma$).

\begin{figure*}
  \includegraphics[width=\textwidth]{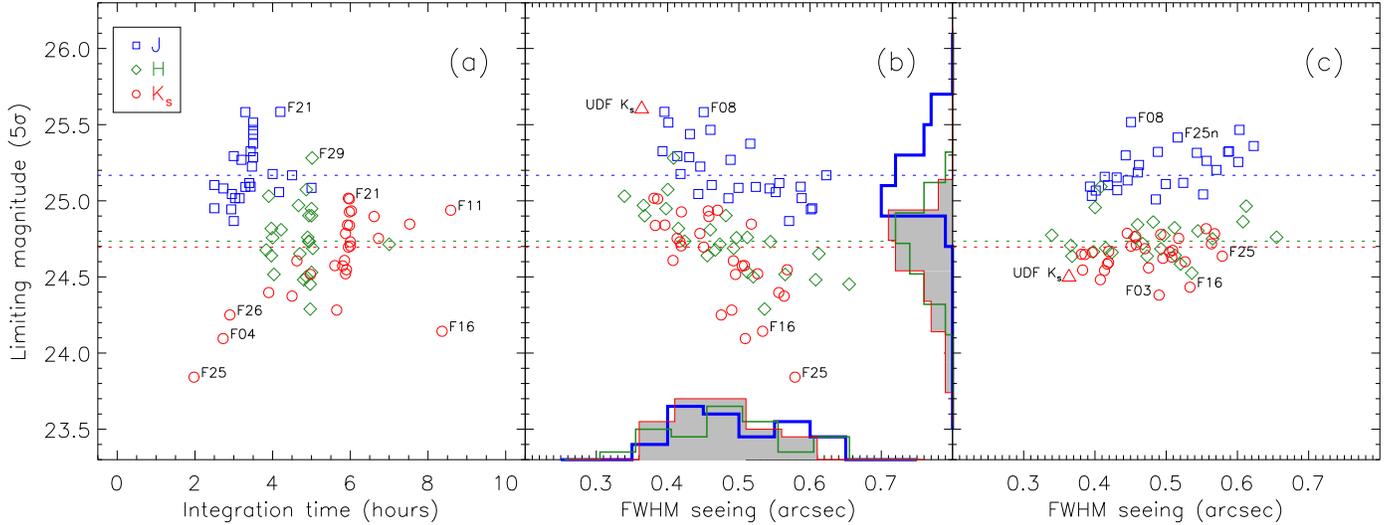}
  \caption{Limiting magnitude ($5\sigma$, total for point sources) of
  the final survey tiles as a function of \emph{(a)} total integration
  time, and, \emph{(b)} PSF size (``seeing'').  Panel \emph{(c)} shows
  the limiting magnitude corrected for the variation of the
  instrumental zero point and scaled by exposure time and seeing so
  that the dependencies formally cancel out (see text).  The survey
  tiles are marked with square, diamond and circle symbols as
  indicated in the legend.  The triangle symbol represents the
  UDF/ISAAC $K_\mathrm{s}$ image. Dotted lines denote the median depth of the
  GOODS/ISAAC tiles. Histograms of seeing and limiting magnitude are marked
  by thick line (blue), moderately thick line (green), and 
  normal line (red) plus grey shade for $J$, $H$, and $K_\mathrm{s}$, respectively.
\label{fig:stackpar3}}
\end{figure*}

\subsection{Final image co-addition}
\label{sect:coadd}

The co-addition of the 378 OB images into 78 survey tile images marks
the final step of the image data reduction process (cf.{}
Fig.~\ref{fig:procflow}).  There is no resampling involved in this
step because the OB images are already defined on the final
astrometric grid.

At first, OB images were rescaled to the fiducial ZP of 26.0~mag (AB),
which in fact is quite close to the typical value for all three bands
(see Fig~\ref{sci-zp-plot}).  Relative weights between OB images
contributing to the same tile were chosen to be proportional to the
inverse variance of the flux integrated within the circular aperture
that maximizes the total limiting magnitude for point sources (cf.{}
Sect.~\ref{sect:limdepth}).  The weight map associated with each OB
image was employed to consistently accounted for the positional
dependence of the variance.  By comparison with different weighting
schemes, we have experimentally verified that the adopted scheme
indeed optimizes the resulting image depth for point sources as
intended.  The associated final weight maps are inverse variance maps
and were rescaled to the actual, empirically determined pixel-to-pixel
variance of the sky background.  For the final survey images the gain
factor, $G_0$, that is the number of detector electrons per data unit,
is reported in Table~\ref{tileparams}, last column.  As $G_0$ refers
to the median pixel, the effective gain at pixel $j$ can be obtained
by scaling with the relative (i.~e.{} median-normalized) weight,
$w_j$, according to $G_j = G_0 w_j$.

In addition to the individual image tiles, the data release also
includes mosaics of the co-adjoined tiles as single FITS files in $J$,
$H$, and $K_\mathrm{s}$ bands, as well as the corresponding weight
maps.  The net area is 172.5, 159.6 and 173.1 $\mathrm{arcmin}^2$ in
$J$, $H$, and $K_\mathrm{s}$, respectively.  The WCS information and
accuracy of the individual tiles is preserved in these mosaics. A
uniform ZP of 26.0~mag can be used (e.~g.{} with SExtractor) across
the entire field, however, it is important to note that the PSF varies
from tile to tile within each mosaic. In the absence of proper
aperture corrections or PSF matching procedures, this would lead to
biases when creating multi-color catalogs.  A possible procedure for
coping with the PSF variations without resorting to image convolution
is outlined in Sect.~\ref{colorcat}.

\section{Final survey images}
\label{finalimages}

\subsection{General properties}
\label{genprops}

In the following, we summarize and discuss the properties of the final
survey images as presented in Table~\ref{tileparams}.  The recorded
parameters are: total integration time, total number of raw images
which were combined to form the final product, the period during which
observations where conducted, the FWHM of the PSF (``seeing'') in
arcseconds, the $5\sigma$ limiting magnitude correct to the total flux
assuming a point source profile (``image depth''), the aperture
diameter to which the depth refers, the 3 parameters of the noise
model $\sigma_0$, $a$, $b$, and the effective gain.

Summing the actual exposure time from the data that were combined in
the final stacks yields a total amount of integration time of 359.9
hours.  Regarding the distribution of integration time that went into
the final images, one notices many tiles standing out with respect to
the nominal values (Fig.~\ref{fig:stackpar3}a).  This is because OBs
which were rejected in the course of the final quality control process
could not be re-scheduled for observations leading to reduced
integration time for those tiles with respect to the nominal values
for the survey.  For instance, in the most extreme case
(F04$K_\mathrm{s}$), three out of six OBs were rejected -- two OBs for
high noise, one for bad seeing.  The two ``supplementary'' tiles, F25
and F26, are an exception in the sense that they have just a
fractional amount of observation time by design.  On the other side,
there are tiles with integration times in excess of the nominal
values, most notably F11 and F16 in $K_\mathrm{s}$, which is primarily
due to the fact that data from several observing programs have been
combined.

The PSF of the final images, quantified by its FWHM using the same
methodology as detailed in Sect.~\ref{sect:qc}, varies between
0.34\arcsec{} (F23$H$) and 0.65\arcsec{} (F15$H$) with a median value
of 0.48\arcsec{} and appears to be independent of the pass band
(Fig.~\ref{fig:stackpar3}b).  We have checked that the final FWHM does
not show any correlation with the number of raw images or the total
integration time, which indicates that the image reduction process
does not degrade the PSF quality.  Generally, we have found a low
level of PSF anisotropy.  SExtractor-measured ellipticities are
typically smaller then 0.05 -- sometimes even clearly below 0.03.

Based on the analysis of the curve of growth of unsaturated isolated
stars using circular apertures between 0.7\arcsec{} and 10\arcsec{}
diameter, we have quantified the aperture corrections for point source
photometry, $\delta{}m_\mathrm{ap}$, for each tile
(Table~\ref{tileapcors}).  10\arcsec{} has been adopted as the
reference aperture for the practical reason, that this turns out to be
the upper limit out to which the correction can be traced in case of
most favorable conditions, i.~e., if sufficiently bright stars are
present in the image.  In other cases -- when appropriate stars are
lacking -- a power-law extrapolation of the curve of growth had to be
applied to obtain the data for the largest apertures.  For all images
the curve of growth falls significantly below the 1\% level at the end
which justifies to consider the 10\arcsec{} flux as the total flux,
and to infer the total magnitude, $m_\mathrm{tot}$, from the aperture
magnitude, $m_\mathrm{ap}$ by means of
$m_\mathrm{tot}=m_\mathrm{ap}-\delta{}m_\mathrm{ap}$ as is done
throughout the rest of the work.

\begin{figure}
  \resizebox{\hsize}{!}{\includegraphics{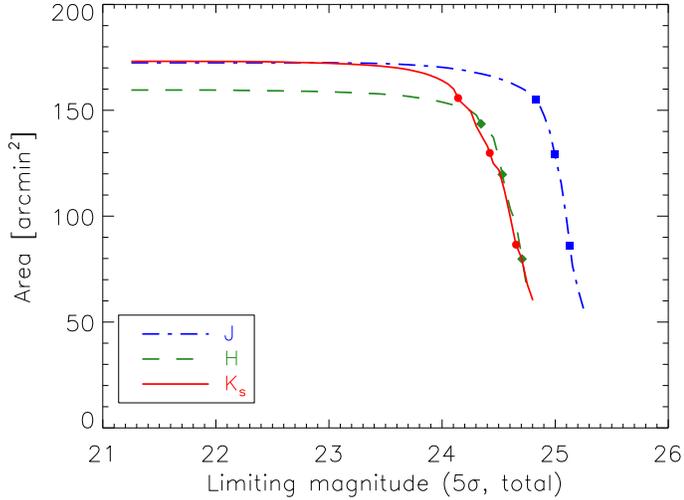}}
  \caption{Effective survey area vs. limiting magnitude for $J$,
  $H$, and $K_\mathrm{s}$ bands.  Plot symbols mark 90\%, 75\%, and 50\% of
  the total survey area for which the limiting magnitudes are given in
  the text.\label{area-depth-plot}}
\end{figure}

\subsection{Sky background and limiting magnitude}
\label{sect:limdepth}

First, we have evaluated the sky background noise using the same
methodology as detailed in Sect.~\ref{sect:qc}. That is, for each
image, we have obtained a model for the sky noise according to
Eq.~\ref{sigmasky} in terms of the 3 parameters $\sigma_0$, $a$, and
$b$.  Below, we will also employ these numbers to estimate flux errors
(Sect.~\ref{detect-photom}).  Then, the total magnitude for point
sources being associated with the sky background fluctuation was
computed taking into account the aperture correction, $\delta
m_\mathrm{ap}$, and the photometric ZP,
\begin{equation}
m_\mathrm{lim} = -2.5 \log \sigma_\mathrm{sky} - \delta m_\mathrm{ap} + \mathrm{ZP} ,
\end{equation}
and its maximum with respect to the aperture diameter was identified
as the point source limiting magnitude of the given image at $1\sigma$
significance level.  As customary, we refer to the 1.747~mag brighter
$5\sigma$ limiting magnitude in the following.  The variation across
the image is given by the weight map, $w_j$, according to
$m_\mathrm{lim} = m_\mathrm{lim}^{(0)} +1.25 \log w_j$, whereas
$m_\mathrm{lim}^{(0)}$ refers to the median pixel by convention.  The
median limiting magnitude and the diameter of the corresponding
circular aperture are listed in Table~\ref{tileparams}.

The primary dependencies of the limiting magnitude of the final images
on total integration time and atmospheric seeing are shown in
Fig.~\ref{fig:stackpar3}\emph{a}, \emph{b}.  However, there are
considerable deviations for individual tiles.  F16$K_\mathrm{s}$, for
instance, turns out to be much shallower than expected based on its
substantial total integration time which is a consequence of the
joined effect of worse-than-average seeing and the lower instrumental
sensitivity in this period (see also Fig.~\ref{sci-zp-plot}).  The
median of the depth of the final images is 25.2 for $J$, and 24.7 both
for $H$ and $K_\mathrm{s}$, which is in good agreement with the
original survey goals (cf.{} Sect.~\ref{obs}).  Scaling the limiting
magnitude to correct for the primary determining factors, namely
integration time ($t_\mathrm{intg}$), FWHM, and the effective
variation of the instrumental zero point for this image,
$\Delta\mathrm{ZP}$, according to
\begin{equation}
m_\mathrm{lim} -1.25\log t_\mathrm{intg} +2.5\log \mathrm{FWHM} -\Delta \mathrm{ZP} ,
\end{equation}
leaves a variation about the mean of between 0.11 and 0.14~mag RMS
(Fig.~\ref{fig:stackpar3}\emph{c}).  This is caused by sky brightness
fluctuations. There is the simple effect that the sky noise is higher
when the sky is brighter -- i.~e., shot noise variations due to the
fact that the sky is brighter on some nights than on others.
Moreover, short-term fluctuations in the atmospheric sky brightness on
a time scale similar to the integration time of an OB ultimately limit
the accuracy with which the sky background can be subtracted during
image data reduction.

To quantify the depth of the survey as a whole, the effective area,
that is the cumulative area distribution as a function of limiting
magnitude, has been computed (Fig.~\ref{area-depth-plot}).  As the
overlap of tiles is taken into account, the effective area corresponds
with the final survey mosaics. One reads off that within 90\% of the
nominal survey area the limiting magnitude is 24.8, 24.3, and 24.1,
within 75\% it is 25.0, 24.5, and 24.4, and within 50\% it is 25.1,
24.7, and 24.7 for $J$, $H$, and $K_\mathrm{s}$, respectively.

\begin{figure*}
  \includegraphics[width=\textwidth]{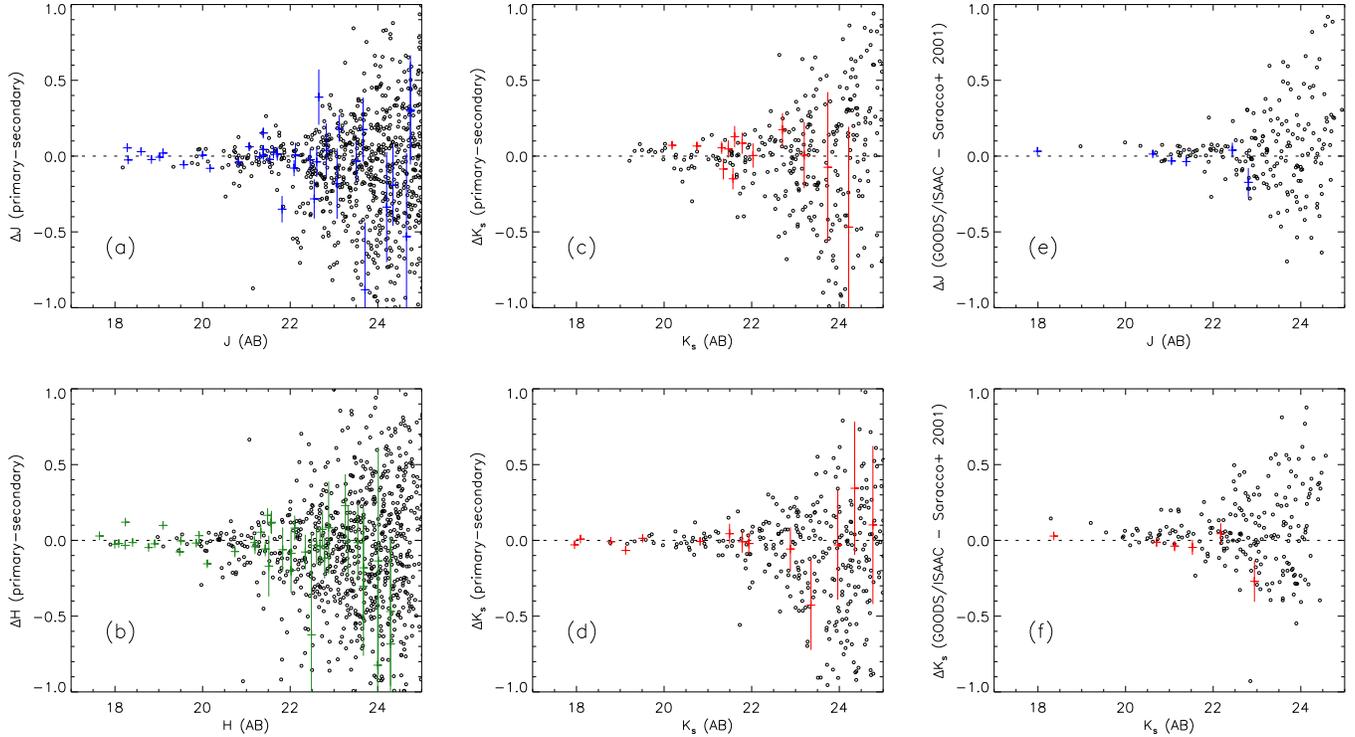}
  \caption{Photometric comparison of primary and secondary detections
  in overlapping regions in $J$ and $H$ \emph{(a,b)}, similarly, based
  on the ``supplementary'' survey fields F25 and F26 in $K_\mathrm{s}$ band
  \emph{(c,d)}, and, with respect to the photometric catalog by
  \cite{2001A&A...375....1S} in $J$ and $K_\mathrm{s}$ \emph{(e,f)}.
  Point-like sources are indicated by cross symbols with $1 \sigma$
  vertical error bars for the magnitude difference, whereas extended
  sources correspond to the circle symbols.\label{fig:photcmp1}}
\end{figure*}

\begin{figure*}
  \includegraphics[width=\textwidth]{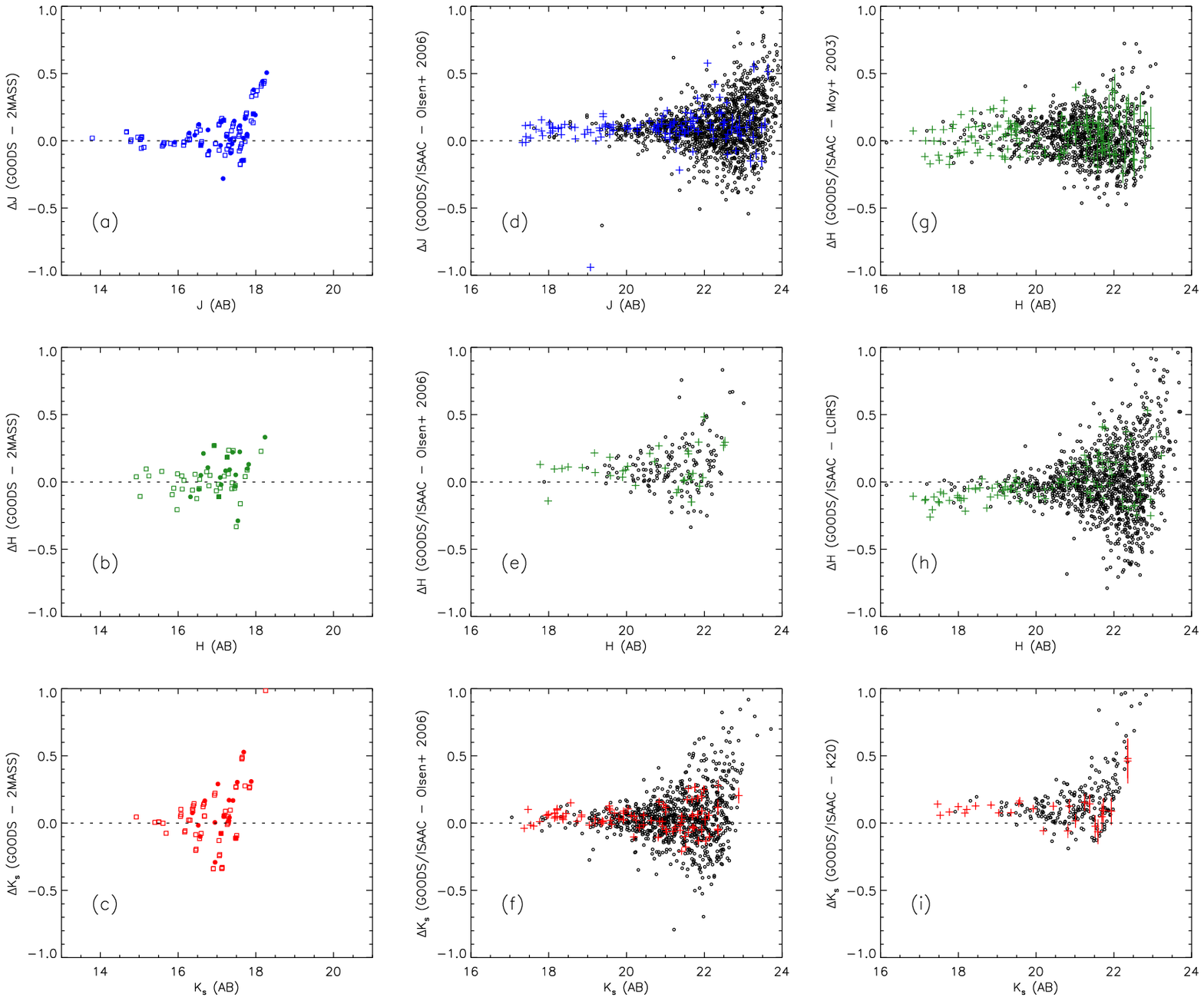}
  \caption{Photometric comparison with respect to public NIR
  catalogs of the CDF-S region.  Each panel corresponds to one
  reference data set in one filter as indicated in the label, namely,
  \emph{(a--c)}~Two Micron All Sky Survey \citep{2006AJ....131.1163S},
  \emph{(d--f)}~ESO imaging survey \citep{2006A&A...452..119O},
  \emph{(g)}~$H$-band observations of the CDF-S by
  \cite{2003A&A...403..493M}, \emph{(h)}~Las Campanas Infrared survey
  \citep{2002ApJ...570...54C}, and, \emph{(i)}~K20 survey
  \citep{2002A&A...392..395C}. The differences of magnitudes of
  sources as measured on the GOODS/ISAAC images and as given in the
  respective reference catalogs are displayed as a function of the
  measured GOODS/ISAAC magnitude in the AB system.  In case of 2MASS,
  dot symbols correspond to measurements performed directly on the
  calibrated ISAAC survey fields while square symbols denote
  measurements on the ``auxiliary'' SOFI images which were used to
  homogenize the photometric solution (see text).  For the comparison
  with the other --much deeper-- catalogs crosses and circles
  correspond to point-like and extended sources, respectively, with
  1-sigma error estimates on the magnitude difference being indicated
  for point-sources if exceeding the size of the plot
  symbol.\label{fig:fluxcmp2}}
\end{figure*}

\begin{figure*}
\centering
  \includegraphics[width=\textwidth]{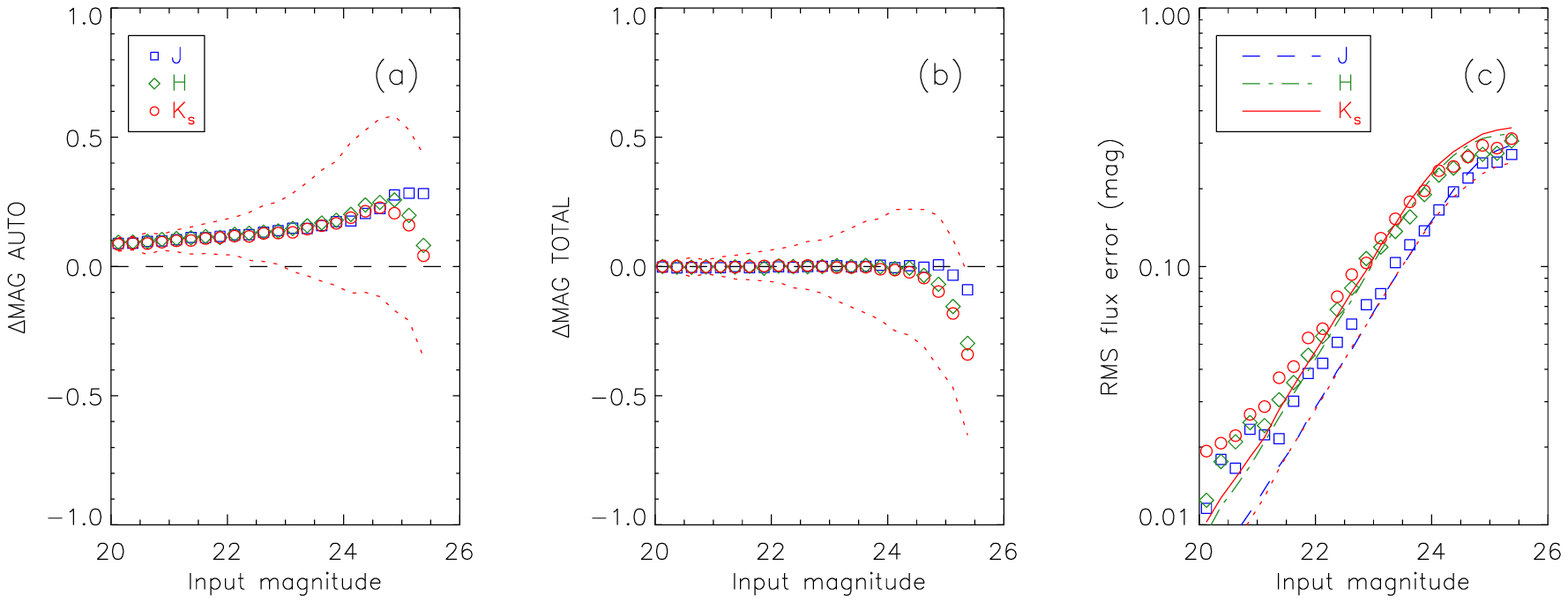}
  \caption{Performance of photometric estimators as inferred from
  image simulations.  Ratio of recovered flux and input flux
  (magnitude scale) for \emph{(a)}~Kron-like apertures (MAG AUTO), and
  \emph{(b)}~total, i.~e.{} aperture-corrected, flux (MAG TOTAL).  The
  binned average is plotted for $J$, $H$, and $K_\mathrm{s}$ with
  square, diamond, and circle symbols, respectively. The dotted line
  indicates the RMS scatter per bin for $K_\mathrm{s}$.  Panel
  \emph{(c)} contrasts the actual flux errors computed in the RMS
  sense per bin (square, diamond, and circle symbols) with the flux
  error estimate according to Eq.~\ref{fluxerr} (dashed, dash-dotted,
  and continuous line).  The standard SExtractor flux error estimate
  for $K_\mathrm{s}$ is shown as dotted line. \label{fig-JHKsimu}}
\end{figure*}

\subsection{Validation of the photometric calibration}
\label{sect:photval}

We have conducted several checks to verify the consistency of the
photometric calibration of the final survey images starting with an
internal check based on multiple detections, followed by the
comparison with an independently reduced and calibrated subset of the
same raw data, and, finally, by comparison with external NIR
photometric data having been published in the literature.

At first, we have analyzed sources having multiple detections to
verify the internal consistency of the photometric calibration of the
final survey images.  To this end, photometric differences of multiple
detections of non-blended sources were inspected as shown in
Fig.~\ref{fig:photcmp1}\emph{a--d}.  Total magnitudes have been used
in order to minimize any possible effect due to seeing variations
between different tiles.  We do not see any systematic discrepancies
in this test, rather, at large, we found that the scatter appears to
be compatible with the formal error bars.  For $J$ and $H$, the
sources are located in the overlapping region of adjacent survey
tiles, whereas for $K_\mathrm{s}$ band, we took advantage of the
``supplementary'' survey fields F25 and F26 which mark a part of the
survey area that is covered twice.  Consequently, in case of $J$ and
$H$, the amount of scatter about the zero line is enhanced because the
sources lie towards the border of the image tiles where the exposure
time of the jitter observations is effectively reduced with respect to
the central part of the image, and, in addition, where possible
flatfielding errors may be expected to increase.  Regarding the two
fields in $K_\mathrm{s}$, the larger scatter of F25 over F26 is due to
the fact that F25$K_\mathrm{s}$ is about 0.4~mag shallower than F26
$K_\mathrm{s}$ (cf.{} Sect.~\ref{sect:limdepth}).  Generally, the
overall scatter of measured fluxes with respect to true fluxes is
expected to be smaller than the scatter shown here.

Next, we have investigated whether the adopted strategy of data
reduction and calibration procedures affect the final photometric
results.  To this end, we have used the $J$ and $K_\mathrm{s}$
photometric data from \cite{2001A&A...375....1S} which is based on the
same set of raw data that went into the production of tile F16 of the
GOODS/ISAAC survey and allows to carry out a comparison that is
unaffected by varying observing conditions or different instrumental
characteristics.  Here, total magnitudes were used in order to be
compatible with the combination of corrected 2\arcsec{} apertures and
BEST apertures used by \cite{2001A&A...375....1S}. The results,
displayed in Fig.~\ref{fig:photcmp1}\emph{e--f}, show that both data
sets are photometrically compatible, except for very few outliers.

Furthermore, we have checked our survey's photometry against external
NIR catalogs of the CDF-S region which have been obtained and
calibrated independently from the GOODS/ISAAC data
(Fig.~\ref{fig:fluxcmp2}).  To this end, common sources were
identified using a maximum pairing distance of 1\arcsec{} from which
only isolated i.~e.{} non-blended sources were selected.  For the
comparison with the Two Micron All Sky Survey
\citep[2MASS,][]{2006AJ....131.1163S}, we also made use of the SOFI
data set which had been used before to establish the survey's global
photometric solution in order to extend the magnitude range for the
comparison towards the bright end.  In this case, just point sources
were considered. Fluxes were measured in apertures of varying size
(while maximizing the signal-to-noise ratio) and were corrected for
aperture losses to a common aperture diameter of 10\arcsec{} which is
virtually equivalent to a total magnitude given the typical PSF.  We
applied the transformation between the 2MASS and LCO photometric
system as determined by \cite{2001AJ....121.2851C}, and used
$J-J_\mathrm{2MASS}=0.022$, $H-H_\mathrm{2MASS}=0.012$, and
$K_\mathrm{s}-K_\mathrm{s,2MASS}=0.015$, whereas neglecting any color
terms.  The trend of measured source fluxes being apparently fainter
than 2MASS fluxes at magnitudes faintward of $~17.5$ AB is the result
of a Malmquist-like selection effect and is also present in the
comparison with other data sets that are significantly shallower than
GOODS/ISAAC.  Taking this effect into account, the data exhibits no
evidence for a bias in the photometric calibration with respect to
2MASS.  The scatter of the data is basically consistent with the
formal flux errors which are clearly dominated by 2MASS.  It is worth
noting that consistent aperture corrections turned out to be crucial
in this test.  In fact, a lack thereof results in a significant bias
at the $\ga$\,0.1~mag level which is visible despite the significant
scatter.

As we did not apply corrections for nonlinearity in the course of data
reduction, neither for ISAAC nor for SOFI data, differential
nonlinearity may in principle affect any photometric comparison.  To
address this issue we followed the same procedure as described in
Sect~\ref{sect:globphotcal} and discarded relatively bright,
potentially nonlinear objects from the photometric analysis
(independently in the ISAAC as well as in the SOFI data), thereby
reducing nonlinearity effects in our data to nominally $\la$\,0.01~mag
and, thus, making them basically negligible with respect to the other
sources of error.  In terms of 2MASS the reference sources used here
are relatively faint, which means that the photometric uncertainties
are dominated by background noise and nonlinearity is expected to be
insignificant.

For the following comparisons with catalogs which are comparatively
deeper than 2MASS, we used SExtractor's auto-scaling magnitudes for
formal consistency with the type of magnitudes published.  However,
photometric apertures were not matched for lack of detailed
information about the apertures and the associated flux corrections
for the reference data sets.  Thus, a systematic effect depending on
the data set is to be expected.  The comparison with the EIS-DEEP data
set in $JHK_\mathrm{s}$ is based on the most recent processing done by
\cite{2006A&A...452..119O}.  The differences of the AB corrections
adopted in their and in our work are corrected for.  It turns out that
magnitudes from GOODS/ISAAC and EIS-DEEP are consistent modulo a
constant offset of $\approx\! 0.1$~mag for $J$ and $H$, and certainly
$<$\,0.1~mag for $K_\mathrm{s}$.  The apparent scatter is larger than
what the individual flux error estimates suggest -- a finding that
applies to this and the following comparisons as well.  The magnitude
differences with respect to the $H$-band observations of the CDF-S by
\cite{2003A&A...403..493M} are noticeably more dispersed while not
exhibiting any systematic offset.  The comparison with the Las
Campanas Infrared survey
\citep[LCIRS,][]{2002ApJ...570...54C} features a slight trend with
magnitude in such a way that bright sources ($H\la 18$) appear
brighter in GOODS/ISAAC while faint sources appear to be unbiased or
even slightly fainter than in the LCIRS.
The magnitudes of the K20 galaxy survey \citep[]{2002A&A...392..395C}
appear to be brighter by $\approx\! 0.1$~mag than the
respective GOODS/ISAAC magnitude.

As a result, we do not find that photometric differences with respect
to various independent data sets exhibit any coherent trend being
attributable to the GOODS/ISAAC survey.  Rather, it seems to be
plausible to suppose that significant photometric differences are
mostly due to the photometric calibration, the exact definition of
photometric apertures, the effect of seeing variations and, possibly,
the lack of correction thereof of individual reference data sets.  A
mismatch of photometric apertures can make up a systematic difference
of $\sim\! 0.1$~mag. For example, correcting the GOODS/ISAAC
magnitudes for aperture losses by using total magnitudes completely
eliminates the systematic offset with respect to K20.  After all, the
excellent photometric agreement with 2MASS in $J$ and $K_\mathrm{s}$
provides strong evidence that the global photometric error of the
GOODS/ISAAC survey tiles is significantly below the 0.1~mag level.

\section{Source catalogs}
\label{catalogs}

\subsection{Source detection and photometry}
\label{detect-photom}

We use SExtractor \citep{1996A&AS..117..393B} for source detection
adopting a spatial filtering with a 2-dimensional Gaussian of 2.5
pixels FWHM to match with the PSF of the best seeing images.  A
detection threshold of 1.35 times the local background RMS has been
applied with a minimum detection area of one pixel.  This choice
corresponds to a $5\sigma$ significance once the spatial filtering is
taken into account.  The choice of detection threshold and minimum
detection area has been gauged based on the analysis of catalog
completeness and contamination for a grid of detection parameters
(cf.{} Sect.~\ref{complete-contam}).  For the given spatial filtering,
we have tested in particular if a minimum detection area of more than
one pixel is favorable. However, it turned out that this does not
increase the catalog completeness at a given contamination level,
neither for point-sources nor for extended sources.  The background
map has been estimated in square cells of 200 pixels on a side using a
$5 \times 5$ cells median filtering. For photometry a local background
annulus in square shape with 20 pixels width was used.

We have adopted the prescription of
\citet[Sect.~5.2]{2003AJ....125.1107L} to define the
aperture-corrected total flux and a flux from which colors are
computed based on the $K_\mathrm{s}$ band apertures.  For isolated
sources the total flux measurement is based on SExtractor's
auto-scaling elliptical apertures (MAG AUTO) inspired by
\cite{1980ApJS...43..305K} corrected for the flux loss due to the
finite aperture size.  For the parameters which control the size of
the auto-scaling apertures, we have used the default values, that is
$k=2.5$ for the scaling parameter and $R_\mathrm{min}=3.5$ for the
lower bound of the aperture size (in isophotal units).  To apply the
image-specific aperture corrections as given in Sect.~\ref{genprops},
we have converted the elliptical aperture area into the corresponding
circular aperture's diameter.  In order to minimize the mutual flux
contamination for blended sources, in this case the total flux
measurement is based on a reduced circular aperture whose area is half
the isophotal area.  To prevent too small and too large apertures that
are more error-prone, the aperture diameter is bound by 0.7\arcsec{}
and 2.0\arcsec, respectively.  The ``color flux'' is isophotal (as
defined in $K_\mathrm{s}$) if the source is isolated and if the
isophotal area ranges between the size of a circular aperture of
0.7\arcsec{} and 2.0\arcsec{} diameter. For sources whose isophotal
areas exceed these limits, the aperture flux at 0.7\arcsec{} or
2.0\arcsec{} is assigned instead.  For blended sources, the color flux
is equally defined as for the total flux, based on the flux measured
within a ``reduced'' circular aperture.

We computed flux errors taking into account the background
fluctuations, both for the source aperture and for the background
region, and the contribution due to the discrete nature of the
detector electrons. To this end, the noise model established in
Sect.~\ref{sect:limdepth} was evaluated at the spatial scales
corresponding to the aperture size, $A$, and the size of the
background region, $B$, respectively.  With the discreteness noise
which has been estimated based on the local gain, $G$, the flux error
is given by
\begin{equation}
\Delta f = 
\sqrt{ \sigma_\mathrm{sky}^2(A) +  \sigma_\mathrm{sky}^2(B) + \frac{f}{G} }.
\label{fluxerr}
\end{equation}

We investigated the performance of flux and flux error
measurements by means of simulated images, that is actual survey tiles
into which stellar sources of known flux had been embedded (cf.{}
Sect.~\ref{simus}).  Fig.~\ref{fig-JHKsimu}\emph{a} shows that
SExtractor's MAG AUTO measurement suffers a systematic flux loss of
$\approx\! 0.1$~mag at 20~mag that increases towards fainter fluxes,
amounting $\approx\! 0.2$~mag (at 24~mag) for isolated sources
throughout all bands.  The apparent reduction of the average flux loss
faintward of 24.5~mag is due to the selection effect which renders the
distribution to be significantly skewed towards larger measured flux
(Malmquist bias), and since this bias is determined by the
signal-to-noise ratio, the effect for the $J$ band is less pronounced
than for $H$ and $K_\mathrm{s}$.  Fig.~\ref{fig-JHKsimu}\emph{b}
demonstrates that the measured total flux, denoted MAG TOTAL, is
unbiased.

Fig.~\ref{fig-JHKsimu}\emph{c} illustrates that the flux error
computed according to Eq.~\ref{fluxerr} is able to represent the true
flux errors quite accurately over a wide range of fluxes.  In
contrast, SExtractor's standard flux error (MAGERR AUTO)
systematically underestimates the true flux errors by at least a
factor of $\sim\! 1.5$ (in mag) for all fluxes.  This discrepancy
underlines that it is essential to take into account the actual local
background fluctuations for realistic flux error computations instead
of simply scaling the pixel-to-pixel variance.  In detail, the
estimated flux error is in fairly good agreement faintward of
$\approx\! 22$~mag.  The apparent overestimation of the flux errors
for the faintest fluxes ($\ga$\,24~mag, especially for H and
$K_\mathrm{s}$) is probably an artefact due to the strong selection
effect in this regime.  Towards bright fluxes, errors are increasingly
underestimated, for instance, by about a factor of two at
$K_\mathrm{s}\approx 20$, because systematic effects that are beyond
the simple statistical error model, such as source blending, start to
become dominant.

\begin{figure}
  \resizebox{\hsize}{!}{\includegraphics{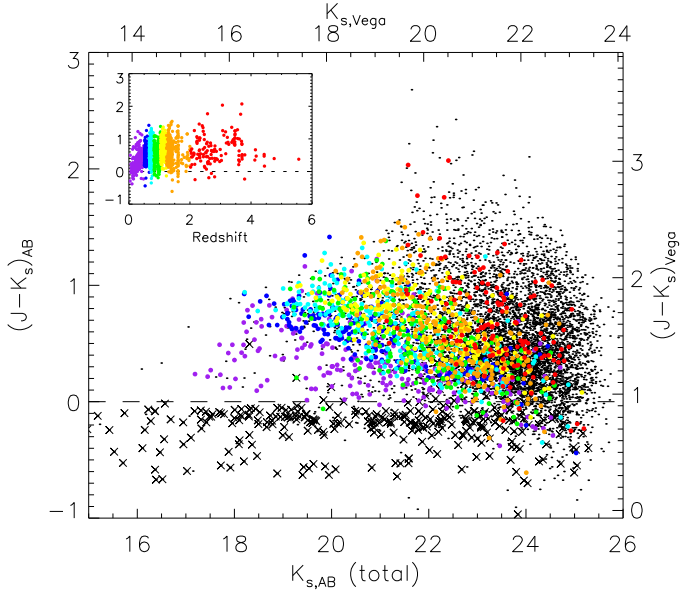}}
  \caption{NIR color-magnitude diagram (main plot) and $J-K_\mathrm{s}$
  color-redshift distribution (inset).  The color-coding is according to
  spectroscopic redshift using seven intervals between
  0, 0.5, 0.67, 0.82, 1.04, 1.23, 2, and 6, as shown in the
  inset. Sources without spectroscopic redshift are shown as black dots.
  Sources which appear point-like in the HST/ACS $z$-band image
  are marked by crosses.\label{color-mag-plot}}
\end{figure}

\begin{figure*}
\centering
  \includegraphics[width=\textwidth]{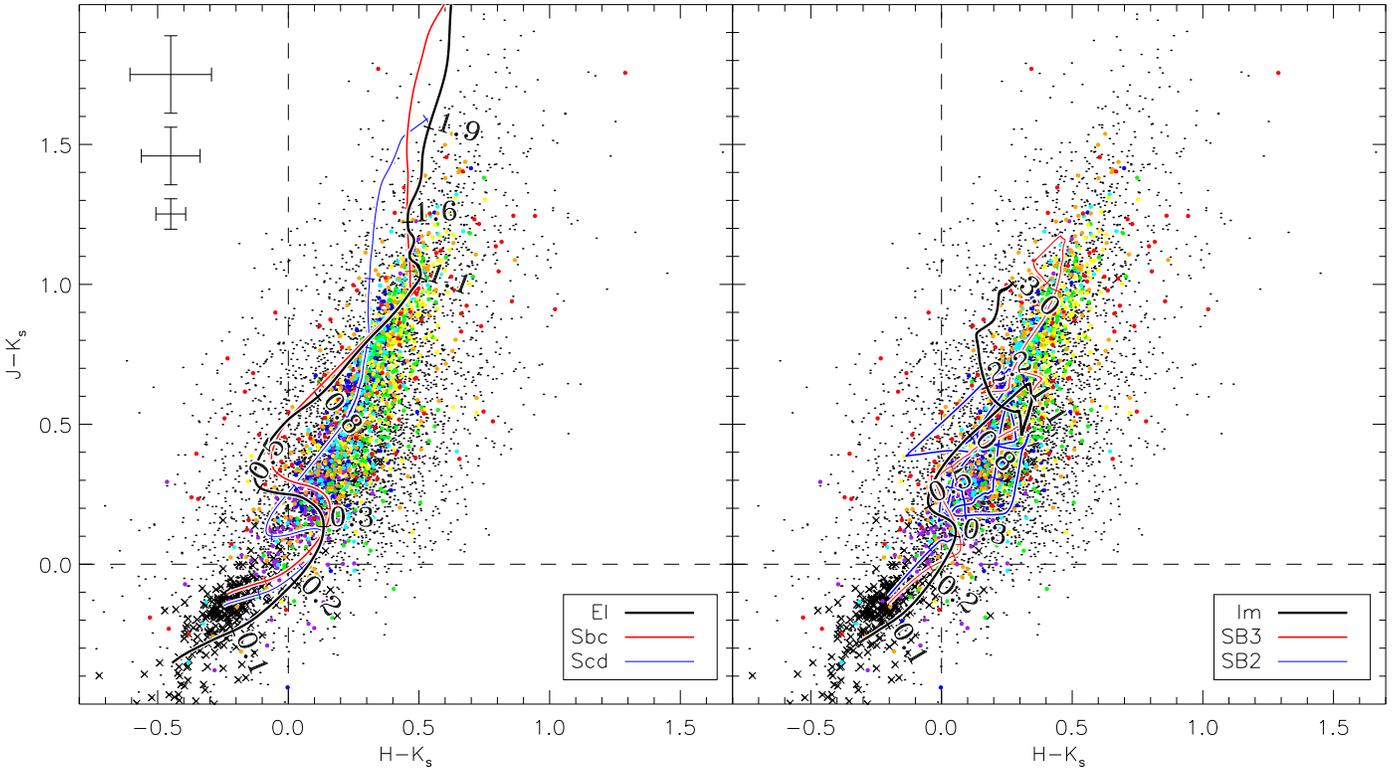}
  \caption{Observed NIR color distribution overlaid with
  tracks of El, Sbc, Scd (left panel), and Im, SB3, and SB2 (right panel) galaxy spectral templates
  for a redshift range between 0.01 and 3 (as indicated along the
  curves).  The color-coding according to spectroscopic redshift
  has been adopted from Fig.~\ref{color-mag-plot}.  Average color
  errors ($1\sigma$) being representative for sources with $K_\mathrm{s}=24, 23,
  22$ are displayed top left.  Sources which appear point-like in the
  HST/ACS $z$-band image are marked by crosses.\label{color-color-plot}}
\end{figure*}

\subsection{The Color Catalog}
\label{colorcat}

To create the $K_\mathrm{s}$ selected NIR color catalog, we ran
SExtractor for each of the 78 tiles in double image mode with the
$K_\mathrm{s}$ band mosaic for detection and with the respective tile
for measurement using the photometric scheme described in the previous
section.  This approach guarantees consistent and accurate flux
measurements by using unique apertures defined in $K_\mathrm{s}$ for
all filters and by allowing to take into account the variation of the
effective PSF and the sky background noise from image to image.  The
detections from all tiles were merged to obtain the final catalog of
unique sources.  In case of sources which were detected multiple times
in the same band -- in regions of tile overlap -- the best
signal-to-noise measurement evaluated based on the local limiting
magnitude was kept (cf.{} Sect.~\ref{sect:limdepth}).  Eventually,
differential aperture corrections were applied to the measured colors
using Table~\ref{tileapcors}.  The resulting catalog reports
$J-K_\mathrm{s}$ and $H-K_\mathrm{s}$ and the corresponding $1\sigma$
uncertainties for 7079 sources which have solid, i.~e.{} $\ge\!
5\sigma$, measurements in all three bands.  Note that corrections for
galactic extinction were not applied.  The catalog data can be
downloaded directly from the ESO archive by navigating to the URL
given in Sect.~\ref{url}.

The catalog columns are as follows. Columns no.~2 to 11 are directly adopted from
SExtractor run on the $K_\mathrm{s}$ images.
\begin{description}
\item (1) NUMBER -- sequential source number for reference.
\item (2, 3) ALPHA\_J2000, DELTA\_J2000 -- Source position R.A., Declination (J2000).
\item (4) ISOAREAF\_IMAGE -- isophotal area in pixel units.
\item (5) KRON\_RADIUS -- Scaling of the automatic elliptical aperture inspired by \cite{1980ApJS...43..305K}.
\item (6) FWHM\_IMAGE -- FWHM measured in pixels units.
\item (7) ELLIPTICITY -- Source ellipticity.
\item (8--10) A\_WORLD, B\_WORLD, THETA\_WORLD -- Elliptical shape parameters: semi-major axis, semi-minor axis and orientation (in degree).
\item (11) FLAGS -- SExtractor flags, possibly combined in a logical OR fashion. 1 - object has close neighbors, 2 - object was blended, 4 - object is saturated, 16 - aperture incomplete.
\item (12, 13) J\_MAG\_COLOR, J\_MAGERR\_COLOR -- $J$ magnitude and associated error from which colors are computed.
\item (14--17) As before but for $H$ and $K_\mathrm{s}$.
\item (18, 19) K\_MAG\_TOTAL, K\_MAGERR\_TOTAL -- Total $K_\mathrm{s}$ magnitude assuming a point source profile and associated error.
\item (20, 21) J\_K, J\_K\_ERR -- $J-K_\mathrm{s}$ color and associated error. Differentially corrected for aperture losses.
\item (22, 23) H\_K, H\_K\_ERR -- $H-K_\mathrm{s}$ color and associated error.
\item (24) J\_FRAME -- Original survey tile from which the $J$ measurement was extracted. This allows to trace back to the original data if needed, 
for instance for the creation of image cut-outs.
\item (25, 26) As before but for $H$ and $K_\mathrm{s}$.
\end{description}
Given that aperture corrections derived for point sources are applied
to mostly faint galaxies, in general, this is likely to yield an
underestimate of the total flux (column 18), since the light profile
for galaxies will be more extended than that of point sources.  The
amount of this effect will be quantified as a function of the extent
of galaxies in Sect.~\ref{sbsel} using image simulations.

\subsection{Color distributions}

To validate the observational data the color distribution is examined
and put in perspective with redshift data taken from the literature.
To this end spectroscopic redshifts were compiled from a number of
published data sets, namely
\cite{2000A&A...359..489C},
\cite{2001MNRAS.328..150C},
\cite{2004A&A...428.1043L},
\cite{2004ApJ...613..200S},
\cite{2004ApJS..155..271S},
\cite{2004ApJ...601L...5V},
\cite{2005A&A...437..883M},
\cite{2008A&A...478...83V}, and
\cite{2009A&A...494..443P}.

Fig.~\ref{color-mag-plot} shows the $J-K_\mathrm{s}$ color-magnitude
of $K_\mathrm{s}$-selected sources in the CDF-S.  Owing to their bluer
$J-K_\mathrm{s}$ color, the locus of stellar sources appears sharply
offset from the bulk of redshifted galaxies which, in contrast,
populate a large range of colors.  In sufficiently deep observations
red NIR color selection, $J-K_\mathrm{s} > 1.34$ (corresponding to
$J_\mathrm{s}-K_\mathrm{s}\ga 2.3$ in the Vega system), is an
efficient method to find high-redshift, i.~e.{} $z\ga 2$, galaxies
\citep{2003ApJ...587L..79F}.  In our data, very red objects having
$J-K_\mathrm{s} >1.34$ are found in the $K_\mathrm{s}$ range between
21 and 24.5 peaking at $K_\mathrm{s}\sim 22.7$.  However, we note that
to date there are very few spectroscopic redshifts for distant red
galaxies in this field \citep[see, e.~g.,][]{2006ApJ...649L..71K}.

The NIR color-color diagram is shown in Fig.~\ref{color-color-plot}
with color tracks of four Hubble-types (El, Sbc, Scd, Im) and two
starburst galaxies (SB2, SB3) overlaid for which the spectral
templates of \cite{1980ApJS...43..393C} and two templates of starburst
galaxies from \cite{1996ApJ...467...38K} were redshifted to
between 0.1 and 3 and folded with the actual instrumental throughput.
Part of the BPZ software package of \cite{2000ApJ...536..571B} has
been applied for this calculation.  The smooth tracks of the El, Sbc,
and Scd-types, originate from the sampling of their uniform spectral
energy distribution which is dominated by the absorption-modulated
continuum.  In contrast, the young stellar populations of the
starburst types (SB3, SB2), and to some extend also the Im-type, with
their strong spectral emission features result in more wrinkled
tracks.

The noticeable loop between redshift 0.3 and 0.5, at which the
$H-K_\mathrm{s}$ color gets bluer with increasing redshift is typical
for an old stellar population and therefore most prominent for the El,
Sbc and Scd types.  It occurs when the restframe $K_\mathrm{s}$ band
samples the strong absorption bands shortward of 1.5~$\mu$m mainly due
to TiO.  For an old stellar population seen at redshifts beyond one,
$J-K_\mathrm{s}$ is strongly increasing because the restframe $J$ flux
is continuously fading due to the strong stellar atmospheric
absorptions mainly induced by metals like \ion{Mg}{I}, \ion{Fe}{I},
\ion{Ca}{I} etc. Towards higher redshift above $\sim\! 1.7$ this trend
becomes even more exaggerated as the 4000-\AA/Balmer break starts to
enter the $J$ band.  The redshift dependence of the NIR colors for
redshifts above 1.5 illustrates the merit of NIR data for the
determination of accurate photometric redshifts when the
characteristic absorption features have moved out of the optical
bands.  A very young stellar population seen at high redshift can not
reproduce the very red NIR color as seen in the observations.
Therefore, the detected very red objects can be interpreted as distant
galaxies hosting an evolved stellar population. Otherwise, heavily
dust-absorbed star burst galaxies could give rise to such very red
colors.

\begin{figure*}
\centering
  \includegraphics[width=\textwidth]{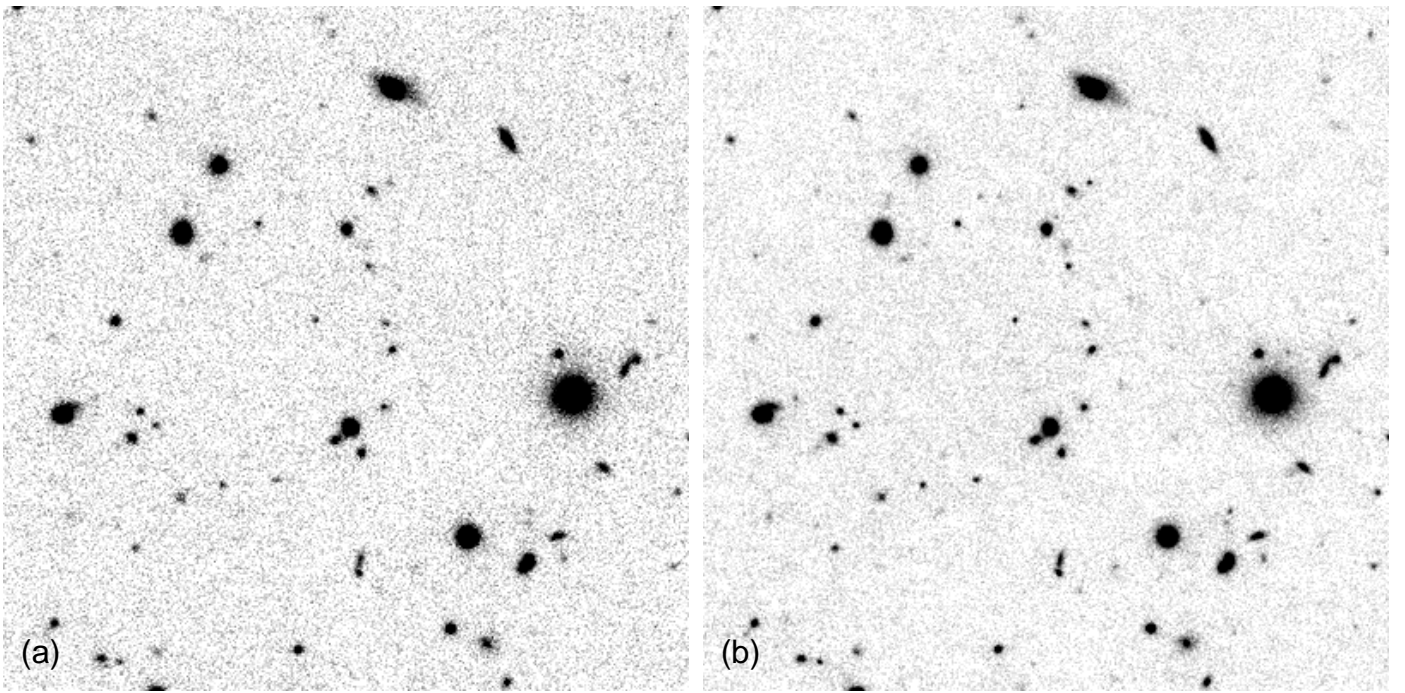}
  \caption{Central $1\arcmin\times 1\arcmin$ cut-out of \emph{(a)} the
  6-hours GOODS/ISAAC survey field F14 in $K_\mathrm{s}$, and
  \emph{(b)} the 25.3-hours co-added UDF/ISAAC $K_\mathrm{s}$ image,
  both displayed at linear scaling.  The effective PSF has
  0.46\arcsec{} and 0.36\arcsec{} FWHM, and the $5\sigma$ depth for
  point sources is 24.93 and 25.60~mag, respectively.
\label{fig-f14ks_udf_image}}
\end{figure*}

\begin{figure}
  \resizebox{\hsize}{!}{\includegraphics{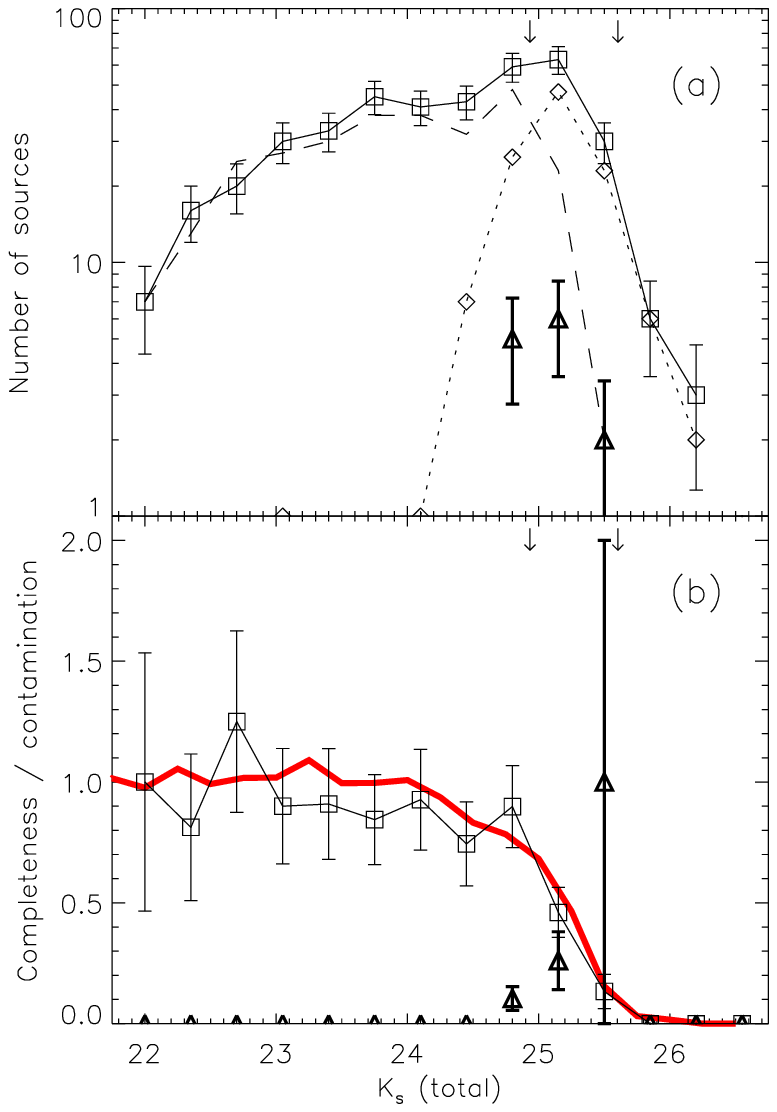}}
  \caption{\emph{(a)} $K_\mathrm{s}$ source counts in the GOODS field
  F14 with respect to the 0.67-mag deeper UDF/ISAAC data using exactly
  the same region. Plot symbols are as follows: all sources detected
  in the UDF/ISAAC image -- squares, true positives, that is, objects
  cross-identified between F14 and UDF/ISAAC -- dashed line, missed
  sources, that is, objects that are only detected in UDF/ISAAC --
  dotted line, and spurious detections, that is, objects that are
  detected in the GOODS F14 data but not in UDF/ISAAC -- triangles.
  Error bars are Poissonian. The small arrows on the top indicate the
  $5\sigma$ limiting magnitude of point sources for the two images.
  \emph{(b)} Resulting relative source completeness (squares) and
  relative contamination (triangles) in comparison to the completeness
  estimate based on image simulations (thick line,
  red).\label{fig-udf_f14_deteff}}
\end{figure}

\section{Analysis}
\label{analysis}

\subsection{Completeness and contamination}
\label{complete-contam}

\subsubsection{Empirical estimation}

Under the ESO programme 73.A-0764, lead by I.~Labb{\'e}, a very deep
21-hours ISAAC $K_\mathrm{s}$ integration of the Hubble Ultra Deep
Field (UDF) inside of the CDF-S region was obtained.  The pointing is
centered on F14 whereas the instrument was rotated by ca.\ 45 degrees
to optimally match the Hubble UDF observations
(Fig.~\ref{fig:acsz-layout}).  We have reduced the raw data, which are
publicly available from the ESO science archive, using similar
procedures as for the whole GOODS/ISAAC survey data reduction
(Sect.~\ref{datareduction-calibration}).  Furthermore, in order to
improve the background subtraction for faint sources, we have applied
a third pass in the OB processing using a source mask generated from
the co-added OBs.  The photometric zero points were bootstrapped
from the F14$K_\mathrm{s}$ image.  We have co-added the calibrated OBs
including the F14$K_\mathrm{s}$ GOODS/ISAAC data resulting in the very
deep image coving 5.40~arcmin${}^2$ of the UDF region with a total
integration time of 91182 sec, an effective PSF of 0.36\arcsec{} FWHM,
and a $5\sigma$ depth for point sources of 25.60~mag which is 0.67~mag
deeper than the respective survey image (see also
Fig.~\ref{fig:stackpar3}\emph{b}).

We have utilized this image as a reference to empirically study the
completeness and contamination, i.~e.{} the fractional number of
missed sources and spurious detections, in the GOODS tile F14 in the
$K_\mathrm{s}$ band.  As this is a typical tile in terms of total
exposure time, image quality etc.\ (see Sect.~\ref{genprops}), the
results can be considered to be representative for the whole survey in
$K_\mathrm{s}$ band, and -- to a minor degree -- for the $J$ and $H$
bands.  To identify spurious detections and missed sources, we have
cross-correlated the source catalogs of the two images using a $1''$
matching distance.  Fig.~\ref{fig-udf_f14_deteff} displays the source
counts aggregated within 0.35-mag bins according to the total flux and
the resulting completeness and contamination levels with respect to
the reference image.  The relative completeness was calculated from
the ratio of the number of cross-identified sources and the total
number of sources in the reference image both counted with respect to
the total flux measured in the reference image.  Hence, the resulting
numbers purely represent the detection efficiency without being
affected by either spurious detections or the migration effect due to
flux measurement errors (Eddington bias).  Counts faintward of
$K_\mathrm{s}=24.8$ start to become significantly incomplete, dropping
to a completeness of 46\% at $K_\mathrm{s}=25.15$.  At this flux
level, the relative contamination which is defined as the ratio of
counts of spurious sources over counts of true sources, both based on
the total flux measured on F14, amounts to 26\%.  Despite the
increasing statistical uncertainty towards fainter fluxes, it is
evident that the contamination becomes a dominant effect beyond the
50\% completeness magnitude.

\subsubsection{Image simulations}
\label{simus}

We have executed extensive image simulations in order to characterize
the source detection efficiency i.~e.{} the completeness level of the
resulting source catalogs as a function of source flux for each tile.

The simulations were set up as follows: for each survey tile image,
first, an isolated unsaturated star was selected and its $7.5\arcsec
\times 7.5\arcsec$ image was cut out to create a representative
template that is suitable to quantify the completeness for point
sources in that tile.  In order to estimate the completeness for faint
galaxies, we have smoothed the stellar images using a Gaussian kernel
of 0.35\arcsec{} FWHM, which has been adjusted so that the
distribution of central surface brightness of simulated faint objects
($\sim\! 24$~mag AB) matches the observational data. Although this is
certainly not sufficient to simulate the observed range of galaxy
types, for example low surface brightness galaxies, this approach
allows the completeness to be estimated at first order without
resorting to extra morphological parameters.  A more accurate
treatment of the statistical incompleteness taking into account a
proper mix of morphological types is beyond the scope of this work but
will be addressed in a forthcoming paper which will present a
comparison with physical models of galaxy evolution.  Then, the
template image was dimmed to a flux in the magnitude range between 20
and 27 which was drawn at random from a power-law distribution whose
slope of 0.26 (per magnitude) resembles the typical number counts in
the field.  That followed, the dimmed simulation template was co-added
to the image at a random position.  Per image, 500 realizations were
simulated, each of which with 100 simulated sources.  This corresponds
to a relatively small fraction of artificial sources to real sources
in the images of 1:8.5 on average so that the extra crowding and the
statistical incompleteness due to blending induced thereby is
negligible.  Source detection and photometry was carried out on the
simulated images in exactly the same way as for the observational data
using identical detection parameters.  For each tile, we have
delimited the analysis to the central 75\% of the image area within
which the exposure time is approximately uniform.
Fig.~\ref{fig-udf_f14_deteff}\emph{b} shows that the completeness
resulting from the simulations is consistent with the independent
empirical estimate.  This finding demonstrates the viability of the
adopted simulation approach in spite of its inherent simplicity of how
galaxies are modeled.

Based on the image simulations, we have characterized each tile in
terms of the completeness function for point sources and for galaxies.
On average the galaxy completeness is about 0.3~mag brighter than the
point source completeness.  The strong correlation of $m_{50}$,
defined as the total flux for which the point source completeness
drops to 50\%, with limiting magnitude can be parameterized by $m_{50}
\approx m_\mathrm{lim}(5\sigma) +0.25$.

\begin{figure}
  \resizebox{\hsize}{!}{\includegraphics{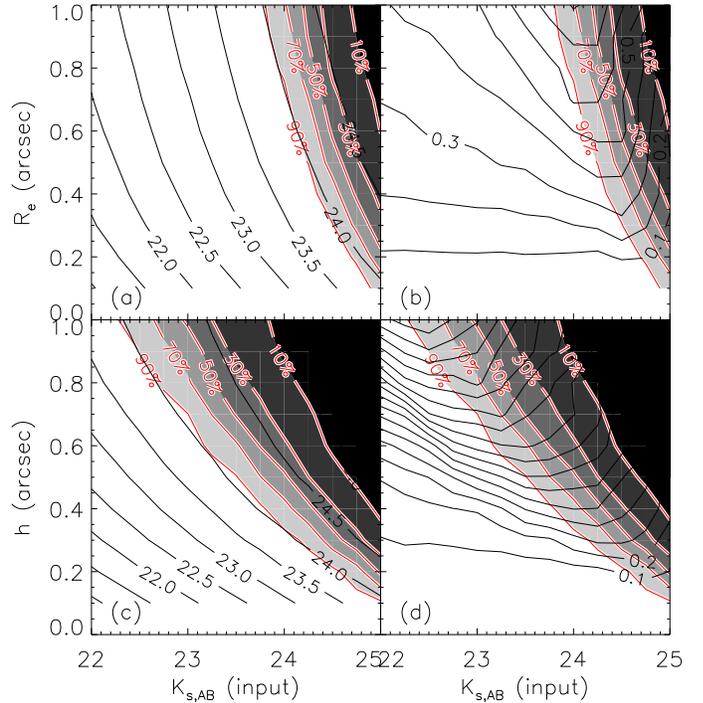}}
  \caption{Completeness, peak surface brightness and flux loss for
  spheroids (top panels) and disks (bottom panels) as inferred from
  image simulations for the GOODS field F14.  $R_\mathrm{e}$ denotes
  the effective radius for spheroids and $h$ the scale length of
  disks. Grey shading displays the completeness in levels between 10\%
  and 90\% in each plot.  Peak surface brightness (in mag
  arcsec${}^{-2}$) is overlaid on the left hand side \emph{(a,~c)},
  and flux loss (in steps of 0.1 mag starting from 0.1 mag) is shown
  on the right hand side \emph{(b,~d)}.\label{fig-f14ks-complt-fluxloss}}
\end{figure}

\begin{figure}
  \resizebox{\hsize}{!}{\includegraphics{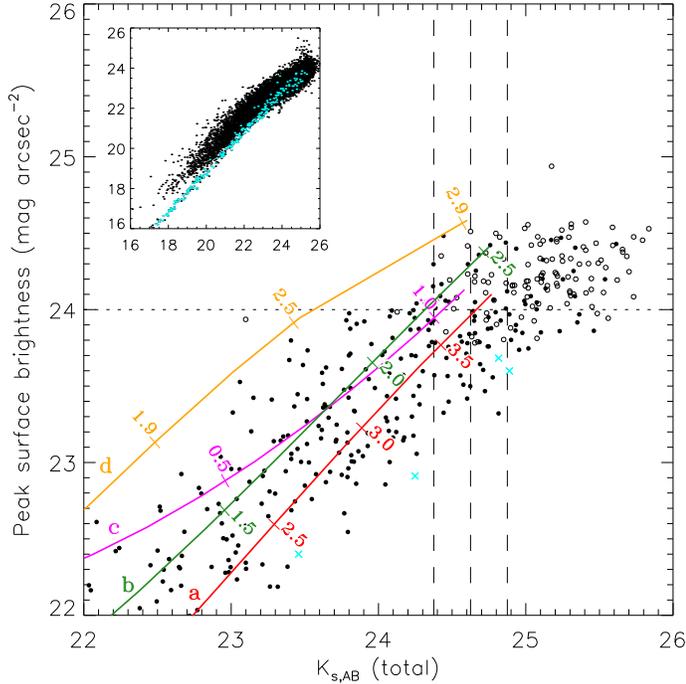}}
  \caption{Peak surface brightness as a function of
  $K_\mathrm{s}$ magnitude as defined in the text. Detections in GOODS
  field F14 (solid dots) opposed to missed sources (open circles) that
  were identified in the deeper UDF/ISAAC image.  The few stars in the
  field are marked by cross symbols (cyan color).  The insert shows
  the equivalent diagram for the GOODS survey as a whole with point
  sources being highlighted (cyan).  Overlaid tracks, labelled from
  \emph{a} to \emph{d}, correspond to individual galaxies as if they
  were observed at the redshifts denoted along the curves.
  \label{fig-csb-ksmag}}
\end{figure}

\subsubsection{Surface brightness selection}
\label{sbsel}

Surface brightness selection effects are inevitable in any galaxy
sample \citep[e.~g.][]{1995ApJ...455...50L} as well as the loss of
flux for extended sources.  To quantify the two effects, first,
spheroids and disks were modelled using deVaucouleurs and exponential
profiles, respectively, and were embedded into the $K_\mathrm{s}$
image of field F14.  As a surface brightness measure, we adopt the
surface brightness as inferred from the detected peak flux of
each source image.  Fig.~\ref{fig-f14ks-complt-fluxloss}\emph{a} and
\emph{c} show that the surface brightness serves as primary selection
parameter, almost independently of total flux, source extent, and type
of profile. In this case, the 90\% completeness limit corresponds to a
peak surface brightness of 24.0 mag $\mathrm{arcsec}^{-2}$, while
the completeness drops below 50\% for 24.5 mag $\mathrm{arcsec}^{-2}$.
For other images, however, the numeric values are expected to vary
according to image depth. Fig.~\ref{fig-f14ks-complt-fluxloss}\emph{b}
and \emph{d} illustrate the flux loss (for MAG TOTAL) featuring a
strong increase with increasing source extent.  Part of the
substantial flux loss of large disks is due to source blending which
inhibits the proper measurement of the total flux.

The surface brightness selection effect at 24.0 mag
$\mathrm{arcsec}^{-2}$ also becomes apparent in the observed
distribution of galaxies (Fig.~\ref{fig-csb-ksmag}).  Thanks to the
deep integration of the UDF region, the properties of the sources
being missed in the GOODS survey can be directly examined.  In the two
faintest bins in which galaxy counts are being accumulated, marked by
vertical dashed lines, centered at $K_\mathrm{s}=24.5$ and 24.75, the
bias against low surface brightness objects is evident.  Consequently,
the resulting total counts in this range may be biased,
but it is difficult to quantify the net effect.  For objects
brighter than $K_\mathrm{s}\sim\! 24.4$~mag, however, there is
apparently no significant selection effect.

To further examine the selection of faint galaxies depending on their
physical and morphological properties, we extracted
images of moderately distant objects ($z$ between 0.15 and 0.6),
scaled them according to redshift taking into account cosmological
dimming and the K-correction, and computed surface brightness and flux
as if they were actually observed. Fig.~\ref{fig-csb-ksmag} displays
the resulting tracks of galaxies with the following
characteristics: \emph{(a)} $2.7M^\ast$ spheroidal, \emph{(b)}
$0.7M^\ast$ disk, exponentially dominated profile, \emph{(c)}
$0.07M^\ast$, low-surface brightness, sub-dominant central bulge
otherwise irregular morphology, \emph{(d)} $2.8M^\ast$ nearly face-on
spiral, disk-dominated with central bulge.  The luminous spheroidal,
\emph{a}, which has the most centrally concentrated profile of the
sample, is least affected by surface brightness selection.  In
contrast, the spiral galaxy, seen nearly face-on, \emph{d}, appears
ca.{} 1.3 mag fainter in surface brightness which would be hardly
detectable beyond $z=2.5$ although the luminosities of the two
galaxies are similar.

\subsection{Number counts}
\label{numbercounts}

\begin{figure*}
\centering
\includegraphics[width=\textwidth]{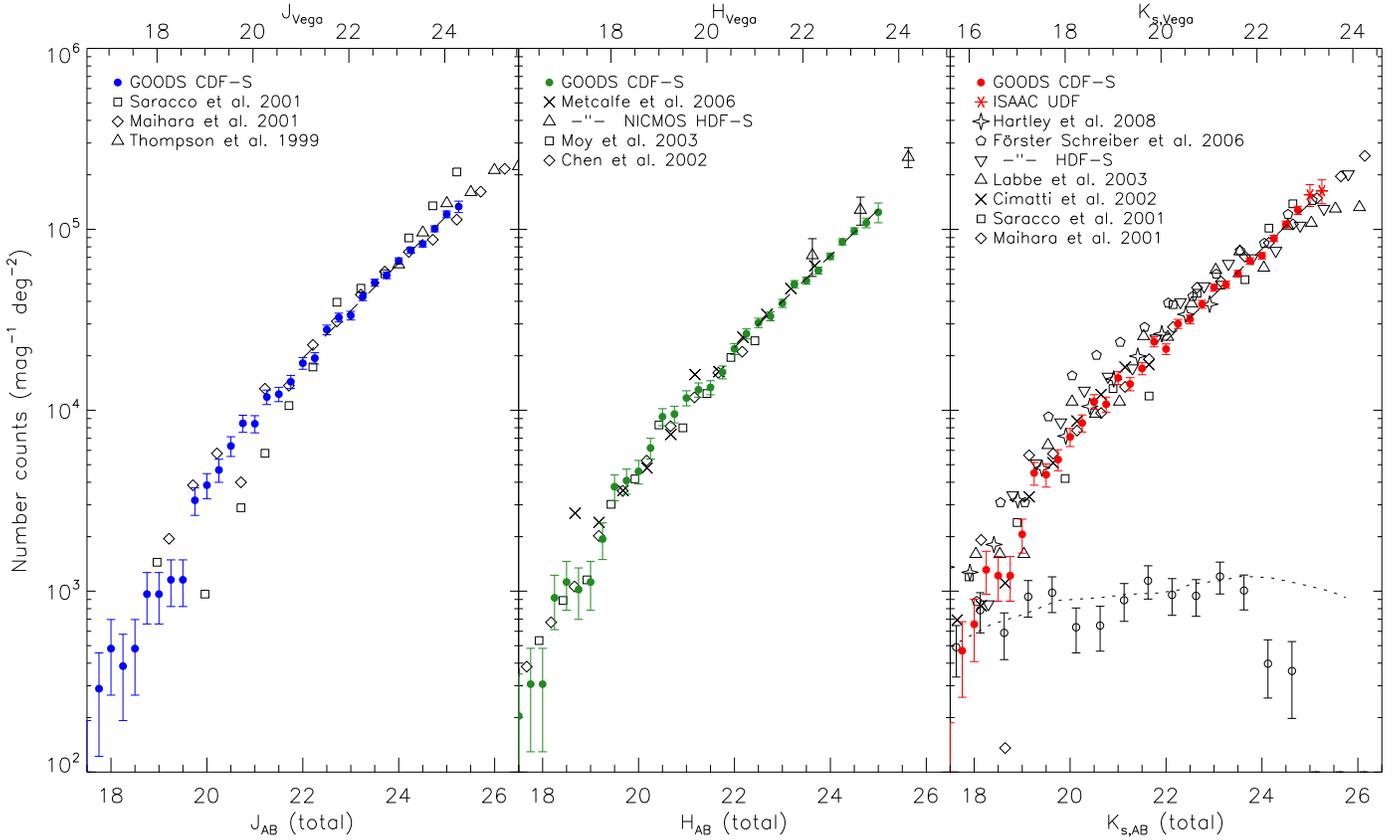}
\caption{\label{JHKcounts}Differential galaxy number counts in the
  GOODS/CDF-S region in the $J$, $H$, and $K_\mathrm{s}$ bands (left,
  middle, and right panel, respectively) corrected for incompleteness
  and displayed with Poissonian (1$\sigma$) error bars.  The two
  faintest data points in $K_\mathrm{s}$ (asterisk symbols) were
  inferred from the very deep UDF/ISAAC image. The dashed lines
  indicate the faint-end exponential fits with logarithmic slopes
  between 0.24 and 0.27~mag${}^{-1}$ (see text).  Results from other
  NIR surveys selected from the literature are displayed for
  comparison. Generally, incompleteness-corrected data is shown except
  for the HST/NICMOS results for which incompleteness at the depth of
  the GOODS/ISAAC data is negligible. Note that the data labeled
  \cite{2001A&A...375....1S} refers to the results obtained from the
  CDF-S, the data of \cite{1999AJ....117...17T} shown here is limited
  to $J\ge 24$ for which the relatively small field provides a
  sufficient absolute number of counts. The apparent bump around
  $K_\mathrm{s}\sim 20$ in the data of \cite{2006AJ....131.1891F} is
  owing to the contribution from the MS\,1054-03 cluster of galaxies
  on which the survey field is centered and should not be seen as
  discrepancy with the other data sets.  The $K_\mathrm{s}$ data from
  \cite{2006AJ....131.1891F} HDF-S refers to the results of
  \cite{2003AJ....125.1107L} but it is apparently not identical,
  however, as to the comparison with our results this incongruity
  seems to be irrelevant. In the right panel, the stellar
  $K_\mathrm{s}$-counts in the GOODS/CDF-S (circles) are shown in
  contrast to the simulation of \cite{2005A&A...436..895G} based on a
  model of the Galaxy (dotted line).}
\end{figure*}

Differential galaxy number counts were independently determined for
each of the three NIR bands as follows.  We have counted the number of
sources in 0.25-mag bins with respect to the total flux (as defined in
Sect.~\ref{detect-photom}) whereas stellar sources had been excluded
on the basis of their point-like appearance in the GOODS/ACS $z$-band
mosaic (see below). 
First, each tile was processed separately and the raw counts
were corrected for incompleteness according to the completeness
function for galaxies which had been established using simulations
(Sect.~\ref{complete-contam}).  At this point the central part of each
tile, 75\% in terms of area, was used over which the exposure is
approximately constant.  Then, for each magnitude bin the
contributions from individual tiles having at least 50\% point source
completeness were combined. In this way the variance of the effective
area as a function of depth was accounted for.  The correction factor
is effectively bound and, consequently, only the deepest tiles
contribute to the faintest bins (which are $J$ and
$K_\mathrm{s}=25.25$ and $H=25.0$).  The actually applied completeness
correction factors are about 1.6 for the faintest GOODS/CDF-S data
points in all three bands, and 1.7 for the data point at
$K_\mathrm{s}=25.25$ which was extracted from the UDF/ISAAC image.
The resulting differential counts are displayed in
Fig.~\ref{JHKcounts} and reported in Table~\ref{table:counts} where
cumulative counts, $N(<m)$, are quoted in addition.

As identification criterion for stars, a cut in $R_{50}$ was used
in conjunction with the flux limit $z_\mathrm{AB}\le 24.5$ (MAG\_AUTO) to
restrict the selection to the flux range where the stellarity
criterion is able to clearly identify the stellar locus.  Then, it was
verified that all sources have $J-K_\mathrm{s}<0$, which is a very
good criterion for the whole range of stellar types except for the
very cool but comparatively rare L and T dwarf types
\citep[e.~g.][]{2002AJ....123.3409H}, leading to the removal of 5
objects most of which have been independently spectroscopically
identified as QSO (cf.{} Fig.~\ref{color-mag-plot}).  
The number counts of the resulting sample of 295
stars are perfectly concordant with the Galaxy model as simulated by
\cite{2005A&A...436..895G} (Fig.~\ref{JHKcounts}, right panel).  Beyond
$K_\mathrm{s}\sim 24$, the sample becomes incomplete due to the
imposed flux limit in $z$ band, as expected.  Furthermore, using the
angular correlation function, we have checked that the
magnitude-limited stellar sample ($17< K_\mathrm{s}< 23$) is
uncorrelated over the full range of angular scales probed by the
survey ($\theta \sim\; 0.0005\ldots 0.1\degr$), which, besides,
confirms the photometric homogeneity of the data.  Altogether, the
tests provide no evidence for a significant contamination of the
stellar sample from e.~g.{} compact galaxies. Thus, the resulting
galaxy number counts are simply affected by the incomplete subtraction
of stellar sources beyond $K_\mathrm{s}\sim\; 24$.  The amplitude of
this effect, however, is just about 2\% which is less than the size of
the plot symbols in Fig.~\ref{JHKcounts}.

The faint-end galaxy number counts ($m \ge 22$) follow an exponential
with logarithmic slope of $\alpha=(0.262\pm 0.007)\,\mathrm{mag}^{-1}$
for $J$ and, marginally shallower, $\alpha=(0.254\pm
0.008)\,\mathrm{mag}^{-1}$, for $H$ and $K_\mathrm{s}$.  The
best-fitting amplitudes (in a least squares sense) are 3.31, 5.77, and
$6.40\times 10^{-2}\,\mathrm{mag}^{-1}\,\mathrm{deg}^{-2}$ for $J$,
$H$, and $K_\mathrm{s}$, respectively.  Down to the faintest fluxes we
do not find an indication for a significant change in the slope of the
number counts.  According to the size of the survey area which exceeds
previous surveys of similar depth by about an order of magnitude, the
statistical errors turn out to be almost negligible over a large range
of fluxes. At the faintest fluxes, the statistical errors of the data
points must slightly increase because of the decrease of the effective
area.  Generally, the results in the low-flux regime may be
systematically affected by uncertainties in the incompleteness
correction and possible contamination due to spurious detections. To
minimize the possible impact, we have limited the analysis to $\ga
5\sigma$ significant sources.  As a consequence of this choice,
correction factors are quite moderate and the contamination level
remains low ($\sim\! 25\%$).

Comparing our results with previously published data of deep surveys
selected from the literature, we find a good agreement in general
(Fig.~\ref{JHKcounts}).  However, closer examination reveals some
discrepancies much in the same way as discrepancies have been found
amongst previous data sets
\citep[e.~g.][]{2001A&A...375....1S,2006MNRAS.370.1257M}.  For
instance, \cite{2001A&A...375....1S} reported, based on the deep
imaging of the F16 field in the CDF-S region, comparatively steep
$J$-band number counts with systematically more faint ($J > 24$)
galaxies than we have found.  By contrast the number counts resulting
from the Subaru Deep Survey \citep{2001PASJ...53...25M} are consistent
with our results in the $J$-band (and similarly in the
$K_\mathrm{s}$-band), likewise the raw number counts in the Hubble
Deep Field-North obtained with HST/NICMOS using the F110W filter
\citep{1999AJ....117...17T}.  In the $H$-band the results of two
imaging surveys which have been conducted in the CDF-S region
previously, by \cite{2003A&A...403..493M} and
\cite{2002A&A...392..395C}, are fully compatible with our findings.
Also the number counts resulting from the completely independent study
by \cite{2006MNRAS.370.1257M} based on imaging of the $7\arcmin\times
7\arcmin$ William Herschel Deep Field using the Omega Prime camera on
the 3.5-m Calar Alto telescope, Spain, are perfectly in agreement with
our data.  The extremely deep $H$-band number counts reported by
\cite{2006MNRAS.370.1257M} based on observations of the Hubble Deep Field-South (HDF-S)
using NICMOS seem to be slightly in excess with respect to our
findings, however, taking into account the small size of the HDF-S,
$0.95\times 0.95~\mathrm{arcmin}^2$, the two results are consistent
within the $1\sigma$ Poissonian errors.  Our $K_\mathrm{s}$-band
results are compatible with the UKIDSS UDS, that is the ultra-deep
survey of the suite of infrared deep sky surveys using the United
Kingdom Infrared Telescope \citep{2008MNRAS.391.1301H}, the faint
infrared extragalactic survey (FIRES) in the MS\,1054-03 field and in
the HDF-S \citep{2006AJ....131.1891F,2003AJ....125.1107L}, the K20
survey \cite{2002A&A...392..395C}, the $K_\mathrm{s}$-band
observations of \cite{2001A&A...375....1S} in the CDF-S and the Subaru
Deep Survey \citep{2001PASJ...53...25M} once the formal uncertainties
are taken into account.  Concerning the faint end of the
$K_\mathrm{s}$ counts, though individual data points are formally
consistent within the error bars, the slope that we find is steeper
than what has been reported for the FIRES survey, namely
$\alpha\approx 0.15$ by \cite{2003AJ....125.1107L}, and $\alpha=0.20$
by \cite{2006AJ....131.1891F}.  However, the comparison of number
counts results must be exercised with caution.  At faint flux levels,
close to the limiting magnitude of the surveys, the statistical
corrections and their difficult to assess uncertainties usually
dominate the error budget.  In general, any photometric mismatch,
e.~g.{} due to discrepancies in the passband, the details of the flux
measurement scheme, or the correction to total flux, may result in
differences of the number counts.  Furthermore, the presence of cosmic
large-scale structure (LSS) causes a variation of number counts from
field to field in excess to the usually quoted fluctuation according
to Poissonian statistics.

To quantify the effect of cosmic variance on our results, we have
assumed an angular correlation of power-law form with the canonical
index, i.~e.{} $w(\theta)= A\theta^{-0.8}$, and a correlation
amplitude based on the measured angular clustering \citep[see,
e.~g.,][]{2003ApJ...588...50D}.  Taking into account the actual survey
area, we find the fractional cosmic variance of the cumulative counts
for $m <22$, 23, and 24~mag (AB) to be 3.4, 3.2, and 3.0\%
($1\sigma$), respectively, in addition to the respective Poissonian
fluctuations of 2.5, 1.8, and 1.7\%. Put differently, cosmic variance
increases the Poissonian errors on the (cumulative) number counts by
factors between $\sim\! 1.7$ and 2.1.  According to the power-law
model, Poissonian fluctuations decrease more rapidly with increasing
survey area ($\Omega$) than the LSS fluctuations, namely $\propto
\Omega^{-0.5}$ vs.{} $\propto \Omega^{-0.2}$.  Therefore, the
contribution of cosmic variance to the total error budget will be
significant also for next generation deep surveys.  For number counts
down to 24~mag in a $1\degr\times 1\degr$ field, for instance, one
expects about 0.34\% Poissonian and 1.6\% LSS fluctuations.

\section{Summary}
\label{summary}

We presented the deep VLT/ISAAC NIR imaging survey in the GOODS-South
field. The survey was conducted as part of the ESO/GOODS program and
the resulting data products were released to the public via the ESO
science archive. The survey covers an area of 173~arcmin${}^2$ in $J$
and $K_\mathrm{s}$, and 160~arcmin${}^2$ in the $H$ band.  The
``median'' survey depth is 25.1 in $J$ and 24.7 both in $H$ and
$K_\mathrm{s}$ in terms of the $5\sigma$ total limiting AB magnitude
for point sources.  The excellent image quality is characterized by a
PSF with FWHM between 0.34\arcsec{} and 0.65\arcsec.  The astrometric
calibration is accurate to $\sim\! 0.06\arcsec$ RMS over the whole
field. The absolute astrometric accuracy is limited by the accuracy of
the GSC\,2 on which the astrometry is ultimately based.  The
photometric calibration was verified in various ways including tests
against independent NIR photometric measurements such as 2MASS.  The
internal accuracy of the photometric calibration is $\le\! 0.017$~mag
($1\sigma$).  The overall photometric accuracy, including systematic
effects, is up to $\sim\! 0.05$~mag for all three bands.  We described
the survey layout, observations, data reduction and calibration, and
the quality control process. The data products were characterized in
terms of PSF width, curve of growth, and limiting depth.  A
$K_\mathrm{s}$-selected catalog of 7079 objects sources with
$JHK_\mathrm{s}$ photometry was defined for which color distributions
were presented with respect to available redshift data.

We characterized the properties of extracted source catalogs in terms
of their fractional completeness and contamination as a function of
magnitude. For this purpose additional 21 hours of $K_\mathrm{s}$
imaging data of the Hubble Ultra Deep field, which are available from
the ESO archive, were processed and the resulting image, which is
$\sim\! 0.7$~mag deeper than the respective survey image, was
analyzed. These results were combined with extensive image
simulations.  For the number counts of faint galaxies we found very
similar logarithmic slopes in all three bands, ranging at the faint
end between 0.24 and 0.27~mag${}^{-1}$ without an indication for a
change in the slope.  Our new results are largely compatible with the
ones from previous surveys.  However, the GOODS/ISAAC data establish
the surface density of faint galaxies in the universe with
unprecedented accuracy owing to its unique combination of survey area
and depth which is unrivaled by any other NIR survey.  In fact, this
survey represents the deepest NIR imaging data set to date covering a
contiguous region well above 100~$\mathrm{arcmin}^2$. In the $H$ band,
there is to our knowledge no ground based survey paralleling
GOODS/ISAAC in depth.

\begin{acknowledgements}
This long term large program would have not been successful without
the dedication and excellent support of the ESO Paranal staff. We are
also grateful to F.~Comeron and the ESO User Support group for their
continuous assistance in scheduling the observations and monitoring
their status.
\end{acknowledgements}

\bibliography{biblio}

\begin{thebibliography}{49}
\expandafter\ifx\csname natexlab\endcsname\relax\def\natexlab#1{#1}\fi

\bibitem[{Amico {et~al.}(2002)Amico, Cuby, Devillard, Jung, \&
  Lidman}]{2002IsaacDataReductionGuide}
Amico, P., Cuby, J., Devillard, N., Jung, Y., \& Lidman, C. 2002, {ISAAC Data
  Reduction Guide 1.5}, European Organisation for Astronomical Research in the
  Southern Hemisphere

\bibitem[{{Ben{\'{\i}}tez}(2000)}]{2000ApJ...536..571B}
{Ben{\'{\i}}tez}, N. 2000, \apj, 536, 571

\bibitem[{{Bertin} \& {Arnouts}(1996)}]{1996A&AS..117..393B}
{Bertin}, E. \& {Arnouts}, S. 1996, \aaps, 117, 393

\bibitem[{{Carpenter}(2001)}]{2001AJ....121.2851C}
{Carpenter}, J.~M. 2001, \aj, 121, 2851

\bibitem[{{Chen} {et~al.}(2002){Chen}, {McCarthy}, {Marzke}, {Wilson},
  {Carlberg}, {Firth}, {Persson}, {Sabbey}, {Lewis}, {McMahon}, {Lahav},
  {Ellis}, {Martini}, {Abraham}, {Oemler}, {Murphy}, {Somerville}, {Beckett},
  \& {Mackay}}]{2002ApJ...570...54C}
{Chen}, H.-W., {McCarthy}, P.~J., {Marzke}, R.~O., {et~al.} 2002, \apj, 570, 54

\bibitem[{{Cimatti} {et~al.}(2002){Cimatti}, {Mignoli}, {Daddi}, {Pozzetti},
  {Fontana}, {Saracco}, {Poli}, {Renzini}, {Zamorani}, {Broadhurst},
  {Cristiani}, {D'Odorico}, {Giallongo}, {Gilmozzi}, \&
  {Menci}}]{2002A&A...392..395C}
{Cimatti}, A., {Mignoli}, M., {Daddi}, E., {et~al.} 2002, \aap, 392, 395

\bibitem[{{Coleman} {et~al.}(1980){Coleman}, {Wu}, \&
  {Weedman}}]{1980ApJS...43..393C}
{Coleman}, G.~D., {Wu}, C.-C., \& {Weedman}, D.~W. 1980, \apjs, 43, 393

\bibitem[{{Cowie} {et~al.}(1994){Cowie}, {Gardner}, {Hu}, {Songaila}, {Hodapp},
  \& {Wainscoat}}]{1994ApJ...434..114C}
{Cowie}, L.~L., {Gardner}, J.~P., {Hu}, E.~M., {et~al.} 1994, \apj, 434, 114

\bibitem[{{Cristiani} {et~al.}(2000){Cristiani}, {Appenzeller}, {Arnouts},
  {Nonino}, {Arag{\'o}n-Salamanca}, {Benoist}, {da Costa}, {Dennefeld},
  {Rengelink}, {Renzini}, {Szeifert}, \& {White}}]{2000A&A...359..489C}
{Cristiani}, S., {Appenzeller}, I., {Arnouts}, S., {et~al.} 2000, \aap, 359,
  489

\bibitem[{{Croom} {et~al.}(2001){Croom}, {Warren}, \&
  {Glazebrook}}]{2001MNRAS.328..150C}
{Croom}, S.~M., {Warren}, S.~J., \& {Glazebrook}, K. 2001, \mnras, 328, 150

\bibitem[{{Daddi} {et~al.}(2004){Daddi}, {Cimatti}, {Renzini}, {Fontana},
  {Mignoli}, {Pozzetti}, {Tozzi}, \& {Zamorani}}]{2004ApJ...617..746D}
{Daddi}, E., {Cimatti}, A., {Renzini}, A., {et~al.} 2004, \apj, 617, 746

\bibitem[{{Daddi} {et~al.}(2003){Daddi}, {R{\"o}ttgering}, {Labb{\'e}},
  {Rudnick}, {Franx}, {Moorwood}, {Rix}, {van der Werf}, \& {van
  Dokkum}}]{2003ApJ...588...50D}
{Daddi}, E., {R{\"o}ttgering}, H.~J.~A., {Labb{\'e}}, I., {et~al.} 2003, \apj,
  588, 50

\bibitem[{{Dickinson} {et~al.}(2003){Dickinson}, {Giavalisco}, \& {The Goods
  Team}}]{2003mglh.conf..324D}
{Dickinson}, M., {Giavalisco}, M., \& {The Goods Team}. 2003, in The Mass of
  Galaxies at Low and High Redshift, ed. R.~{Bender} \& A.~{Renzini}, 324--+

\bibitem[{{F{\"o}rster Schreiber} {et~al.}(2006){F{\"o}rster Schreiber},
  {Franx}, {Labb{\'e}}, {Rudnick}, {van Dokkum}, {Illingworth}, {Kuijken},
  {Moorwood}, {Rix}, {R{\"o}ttgering}, \& {van der Werf}}]{2006AJ....131.1891F}
{F{\"o}rster Schreiber}, N.~M., {Franx}, M., {Labb{\'e}}, I., {et~al.} 2006,
  \aj, 131, 1891

\bibitem[{{Franx} {et~al.}(2003){Franx}, {Labb{\'e}}, {Rudnick}, {van Dokkum},
  {Daddi}, {F{\"o}rster Schreiber}, {Moorwood}, {Rix}, {R{\"o}ttgering}, {van
  de Wel}, {van der Werf}, \& {van Starkenburg}}]{2003ApJ...587L..79F}
{Franx}, M., {Labb{\'e}}, I., {Rudnick}, G., {et~al.} 2003, \apjl, 587, L79

\bibitem[{{Giavalisco}(2005)}]{2005NewAR..49..440G}
{Giavalisco}, M. 2005, New Astronomy Review, 49, 440

\bibitem[{{Giavalisco} {et~al.}(2004){Giavalisco}, {Ferguson}, {Koekemoer},
  {Dickinson}, {Alexander}, {Bauer}, {Bergeron}, {Biagetti}, {Brandt},
  {Casertano}, {Cesarsky}, {Chatzichristou}, {Conselice}, {Cristiani}, {Da
  Costa}, {Dahlen}, {de Mello}, {Eisenhardt}, {Erben}, {Fall}, {Fassnacht},
  {Fosbury}, {Fruchter}, {Gardner}, {Grogin}, {Hook}, {Hornschemeier}, {Idzi},
  {Jogee}, {Kretchmer}, {Laidler}, {Lee}, {Livio}, {Lucas}, {Madau},
  {Mobasher}, {Moustakas}, {Nonino}, {Padovani}, {Papovich}, {Park},
  {Ravindranath}, {Renzini}, {Richardson}, {Riess}, {Rosati}, {Schirmer},
  {Schreier}, {Somerville}, {Spinrad}, {Stern}, {Stiavelli}, {Strolger},
  {Urry}, {Vandame}, {Williams}, \& {Wolf}}]{2004ApJ...600L..93G}
{Giavalisco}, M., {Ferguson}, H.~C., {Koekemoer}, A.~M., {et~al.} 2004, \apjl,
  600, L93

\bibitem[{{Girardi} {et~al.}(2005){Girardi}, {Groenewegen}, {Hatziminaoglou},
  \& {da Costa}}]{2005A&A...436..895G}
{Girardi}, L., {Groenewegen}, M.~A.~T., {Hatziminaoglou}, E., \& {da Costa}, L.
  2005, \aap, 436, 895

\bibitem[{{Hartley} {et~al.}(2008){Hartley}, {Lane}, {Almaini}, {Cirasuolo},
  {Foucaud}, {Simpson}, {Maddox}, {Smail}, {Conselice}, {McLure}, \&
  {Dunlop}}]{2008MNRAS.391.1301H}
{Hartley}, W.~G., {Lane}, K.~P., {Almaini}, O., {et~al.} 2008, \mnras, 391,
  1301

\bibitem[{{Hawarden} {et~al.}(2001){Hawarden}, {Leggett}, {Letawsky},
  {Ballantyne}, \& {Casali}}]{2001MNRAS.325..563H}
{Hawarden}, T.~G., {Leggett}, S.~K., {Letawsky}, M.~B., {Ballantyne}, D.~R., \&
  {Casali}, M.~M. 2001, \mnras, 325, 563

\bibitem[{{Hawley} {et~al.}(2002){Hawley}, {Covey}, {Knapp}, {Golimowski},
  {Fan}, {Anderson}, {Gunn}, {Harris}, {Ivezi{\'c}}, {Long}, {Lupton},
  {McGehee}, {Narayanan}, {Peng}, {Schlegel}, {Schneider}, {Spahn}, {Strauss},
  {Szkody}, {Tsvetanov}, {Walkowicz}, {Brinkmann}, {Harvanek}, {Hennessy},
  {Kleinman}, {Krzesinski}, {Long}, {Neilsen}, {Newman}, {Nitta}, {Snedden}, \&
  {York}}]{2002AJ....123.3409H}
{Hawley}, S.~L., {Covey}, K.~R., {Knapp}, G.~R., {et~al.} 2002, \aj, 123, 3409

\bibitem[{{Hunt} {et~al.}(1998){Hunt}, {Mannucci}, {Testi}, {Migliorini},
  {Stanga}, {Baffa}, {Lisi}, \& {Vanzi}}]{1998AJ....115.2594H}
{Hunt}, L.~K., {Mannucci}, F., {Testi}, L., {et~al.} 1998, \aj, 115, 2594

\bibitem[{{Kinney} {et~al.}(1996){Kinney}, {Calzetti}, {Bohlin}, {McQuade},
  {Storchi-Bergmann}, \& {Schmitt}}]{1996ApJ...467...38K}
{Kinney}, A.~L., {Calzetti}, D., {Bohlin}, R.~C., {et~al.} 1996, \apj, 467, 38

\bibitem[{{Koranyi} {et~al.}(1998){Koranyi}, {Kleyna}, \&
  {Grogin}}]{1998PASP..110.1464K}
{Koranyi}, D.~M., {Kleyna}, J., \& {Grogin}, N.~A. 1998, \pasp, 110, 1464

\bibitem[{{Kriek} {et~al.}(2006){Kriek}, {van Dokkum}, {Franx}, {Quadri},
  {Gawiser}, {Herrera}, {Illingworth}, {Labb{\'e}}, {Lira}, {Marchesini},
  {Rix}, {Rudnick}, {Taylor}, {Toft}, {Urry}, \& {Wuyts}}]{2006ApJ...649L..71K}
{Kriek}, M., {van Dokkum}, P.~G., {Franx}, M., {et~al.} 2006, \apjl, 649, L71

\bibitem[{{Kron}(1980)}]{1980ApJS...43..305K}
{Kron}, R.~G. 1980, \apjs, 43, 305

\bibitem[{{Labb{\'e}} {et~al.}(2003){Labb{\'e}}, {Franx}, {Rudnick},
  {Schreiber}, {Rix}, {Moorwood}, {van Dokkum}, {van der Werf},
  {R{\"o}ttgering}, {van Starkenburg}, {van de Wel}, {Kuijken}, \&
  {Daddi}}]{2003AJ....125.1107L}
{Labb{\'e}}, I., {Franx}, M., {Rudnick}, G., {et~al.} 2003, \aj, 125, 1107

\bibitem[{{Le F{\`e}vre} {et~al.}(2004){Le F{\`e}vre}, {Vettolani}, {Paltani},
  {Tresse}, {Zamorani}, {Le Brun}, {Moreau}, {Bottini}, {Maccagni}, {Picat},
  {Scaramella}, {Scodeggio}, {Zanichelli}, {Adami}, {Arnouts}, {Bardelli},
  {Bolzonella}, {Cappi}, {Charlot}, {Contini}, {Foucaud}, {Franzetti},
  {Garilli}, {Gavignaud}, {Guzzo}, {Ilbert}, {Iovino}, {McCracken}, {Mancini},
  {Marano}, {Marinoni}, {Mathez}, {Mazure}, {Meneux}, {Merighi}, {Pell{\`o}},
  {Pollo}, {Pozzetti}, {Radovich}, {Zucca}, {Arnaboldi}, {Bondi}, {Bongiorno},
  {Busarello}, {Ciliegi}, {Gregorini}, {Mellier}, {Merluzzi}, {Ripepi}, \&
  {Rizzo}}]{2004A&A...428.1043L}
{Le F{\`e}vre}, O., {Vettolani}, G., {Paltani}, S., {et~al.} 2004, \aap, 428,
  1043

\bibitem[{{Lilly} {et~al.}(1995){Lilly}, {Le Fevre}, {Crampton}, {Hammer}, \&
  {Tresse}}]{1995ApJ...455...50L}
{Lilly}, S.~J., {Le Fevre}, O., {Crampton}, D., {Hammer}, F., \& {Tresse}, L.
  1995, \apj, 455, 50

\bibitem[{{Maihara} {et~al.}(2001){Maihara}, {Iwamuro}, {Tanabe}, {Taguchi},
  {Hata}, {Oya}, {Kashikawa}, {Iye}, {Miyazaki}, {Karoji}, {Yoshida}, {Totani},
  {Yoshii}, {Okamura}, {Shimasaku}, {Saito}, {Ando}, {Goto}, {Hayashi},
  {Kaifu}, {Kobayashi}, {Kosugi}, {Motohara}, {Nishimura}, {Noumaru},
  {Ogasawara}, {Sasaki}, {Sekiguchi}, {Takata}, {Terada}, {Yamashita}, {Usuda},
  \& {Tokunaga}}]{2001PASJ...53...25M}
{Maihara}, T., {Iwamuro}, F., {Tanabe}, H., {et~al.} 2001, \pasj, 53, 25

\bibitem[{{Mason} {et~al.}(2008){Mason}, {Lombardi}, {Lidman}, \&
  {Jaunsen}}]{Mason2008}
{Mason}, E., {Lombardi}, G., {Lidman}, C., \& {Jaunsen}, A.~O. 2008, in The
  2007 ESO Instrument Calibration Workshop, ed. A.~{Kaufer} \& F.~{Kerber}
  (Springer-Verlag Berlin Heidelberg), 439--442

\bibitem[{{Metcalfe} {et~al.}(2006){Metcalfe}, {Shanks}, {Weilbacher},
  {McCracken}, {Fong}, \& {Thompson}}]{2006MNRAS.370.1257M}
{Metcalfe}, N., {Shanks}, T., {Weilbacher}, P.~M., {et~al.} 2006, \mnras, 370,
  1257

\bibitem[{{Mignoli} {et~al.}(2005){Mignoli}, {Cimatti}, {Zamorani}, {Pozzetti},
  {Daddi}, {Renzini}, {Broadhurst}, {Cristiani}, {D'Odorico}, {Fontana},
  {Giallongo}, {Gilmozzi}, {Menci}, \& {Saracco}}]{2005A&A...437..883M}
{Mignoli}, M., {Cimatti}, A., {Zamorani}, G., {et~al.} 2005, \aap, 437, 883

\bibitem[{{Moorwood} {et~al.}(1999){Moorwood}, {Cuby}, {Ballester},
  {Biereichel}, {Brynnel}, {Conzelmann}, {Delabre}, {Devillard}, {van
  Dijsseldonk}, {Finger}, {Gemperlein}, {Lidman}, {Herlin}, {Huster},
  {Knudstrup}, {Lizon}, {Mehrgan}, {Meyer}, {Nicolini}, {Silber}, {Spyromilio},
  \& {Stegmeier}}]{1999Msngr..95....1M}
{Moorwood}, A., {Cuby}, J.-G., {Ballester}, P., {et~al.} 1999, The Messenger,
  95, 1

\bibitem[{{Moorwood} {et~al.}(1998){Moorwood}, {Cuby}, \&
  {Lidman}}]{1998Msngr..91....9M}
{Moorwood}, A., {Cuby}, J.-G., \& {Lidman}, C. 1998, The Messenger, 91, 9

\bibitem[{{Moy} {et~al.}(2003){Moy}, {Barmby}, {Rigopoulou}, {Huang},
  {Willner}, \& {Fazio}}]{2003A&A...403..493M}
{Moy}, E., {Barmby}, P., {Rigopoulou}, D., {et~al.} 2003, \aap, 403, 493

\bibitem[{{Oke} \& {Gunn}(1983)}]{1983ApJ...266..713O}
{Oke}, J.~B. \& {Gunn}, J.~E. 1983, \apj, 266, 713

\bibitem[{{Olsen} {et~al.}(2006){Olsen}, {Miralles}, {da Costa}, {Benoist},
  {Vandame}, {Rengelink}, {Rit{\'e}}, {Scodeggio}, {Slijkhuis}, {Wicenec}, \&
  {Zaggia}}]{2006A&A...452..119O}
{Olsen}, L.~F., {Miralles}, J.-M., {da Costa}, L., {et~al.} 2006, \aap, 452,
  119

\bibitem[{{Persson} {et~al.}(1998){Persson}, {Murphy}, {Krzeminski}, {Roth}, \&
  {Rieke}}]{1998AJ....116.2475P}
{Persson}, S.~E., {Murphy}, D.~C., {Krzeminski}, W., {Roth}, M., \& {Rieke},
  M.~J. 1998, \aj, 116, 2475

\bibitem[{{Popesso} {et~al.}(2009){Popesso}, {Dickinson}, {Nonino}, {Vanzella},
  {Daddi}, {Fosbury}, {Kuntschner}, {Mainieri}, {Cristiani}, {Cesarsky},
  {Giavalisco}, {Renzini}, \& {The Goods Team}}]{2009A&A...494..443P}
{Popesso}, P., {Dickinson}, M., {Nonino}, M., {et~al.} 2009, \aap, 494, 443

\bibitem[{{Renzini} {et~al.}(2003){Renzini}, {Cesarsky}, {Cristiani}, {da
  Costa}, {Fosbury}, {Hook}, {Leibundgut}, {Rosati}, \&
  {Vandame}}]{2003mglh.conf..332R}
{Renzini}, A., {Cesarsky}, C., {Cristiani}, S., {et~al.} 2003, in The Mass of
  Galaxies at Low and High Redshift, ed. R.~{Bender} \& A.~{Renzini}, 332--+

\bibitem[{{Saracco} {et~al.}(2001){Saracco}, {Giallongo}, {Cristiani},
  {D'Odorico}, {Fontana}, {Iovino}, {Poli}, \&
  {Vanzella}}]{2001A&A...375....1S}
{Saracco}, P., {Giallongo}, E., {Cristiani}, S., {et~al.} 2001, \aap, 375, 1

\bibitem[{{Skrutskie} {et~al.}(2006){Skrutskie}, {Cutri}, {Stiening},
  {Weinberg}, {Schneider}, {Carpenter}, {Beichman}, {Capps}, {Chester},
  {Elias}, {Huchra}, {Liebert}, {Lonsdale}, {Monet}, {Price}, {Seitzer},
  {Jarrett}, {Kirkpatrick}, {Gizis}, {Howard}, {Evans}, {Fowler}, {Fullmer},
  {Hurt}, {Light}, {Kopan}, {Marsh}, {McCallon}, {Tam}, {Van Dyk}, \&
  {Wheelock}}]{2006AJ....131.1163S}
{Skrutskie}, M.~F., {Cutri}, R.~M., {Stiening}, R., {et~al.} 2006, \aj, 131,
  1163

\bibitem[{{Strolger} {et~al.}(2004){Strolger}, {Riess}, {Dahlen}, {Livio},
  {Panagia}, {Challis}, {Tonry}, {Filippenko}, {Chornock}, {Ferguson},
  {Koekemoer}, {Mobasher}, {Dickinson}, {Giavalisco}, {Casertano}, {Hook},
  {Blondin}, {Leibundgut}, {Nonino}, {Rosati}, {Spinrad}, {Steidel}, {Stern},
  {Garnavich}, {Matheson}, {Grogin}, {Hornschemeier}, {Kretchmer}, {Laidler},
  {Lee}, {Lucas}, {de Mello}, {Moustakas}, {Ravindranath}, {Richardson}, \&
  {Taylor}}]{2004ApJ...613..200S}
{Strolger}, L.-G., {Riess}, A.~G., {Dahlen}, T., {et~al.} 2004, \apj, 613, 200

\bibitem[{{Szokoly} {et~al.}(2004){Szokoly}, {Bergeron}, {Hasinger}, {Lehmann},
  {Kewley}, {Mainieri}, {Nonino}, {Rosati}, {Giacconi}, {Gilli}, {Gilmozzi},
  {Norman}, {Romaniello}, {Schreier}, {Tozzi}, {Wang}, {Zheng}, \&
  {Zirm}}]{2004ApJS..155..271S}
{Szokoly}, G.~P., {Bergeron}, J., {Hasinger}, G., {et~al.} 2004, \apjs, 155,
  271

\bibitem[{{Thompson} {et~al.}(1999){Thompson}, {Storrie-Lombardi}, {Weymann},
  {Rieke}, {Schneider}, {Stobie}, \& {Lytle}}]{1999AJ....117...17T}
{Thompson}, R.~I., {Storrie-Lombardi}, L.~J., {Weymann}, R.~J., {et~al.} 1999,
  \aj, 117, 17

\bibitem[{{van der Wel} {et~al.}(2004){van der Wel}, {Franx}, {van Dokkum}, \&
  {Rix}}]{2004ApJ...601L...5V}
{van der Wel}, A., {Franx}, M., {van Dokkum}, P.~G., \& {Rix}, H.-W. 2004,
  \apjl, 601, L5

\bibitem[{{Vandame}(2004)}]{2004ThesisVandame}
{Vandame}, B. 2004, {Processing Multi-wavelength and Wide-Field Imaging data:
  Application to the ESO Imaging Survey} (Nice University)

\bibitem[{{Vanzella} {et~al.}(2008){Vanzella}, {Cristiani}, {Dickinson},
  {Giavalisco}, {Kuntschner}, {Haase}, {Nonino}, {Rosati}, {Cesarsky},
  {Ferguson}, {Fosbury}, {Grazian}, {Moustakas}, {Rettura}, {Popesso},
  {Renzini}, {Stern}, \& {The Goods Team}}]{2008A&A...478...83V}
{Vanzella}, E., {Cristiani}, S., {Dickinson}, M., {et~al.} 2008, \aap, 478, 83

\end{thebibliography}
\bibliographystyle{aa}

%
%
\clearpage \onecolumn
\begin{landscape}
\begin{longtable}{lcccccccccccc}
\caption{\label{tileparams}Properties of the final survey tiles.}\\
\hline\hline
Tile & Band & Integration Time & No.{} Images & Start of Observations & End of Observations & FWHM     & Lim.{} Depth  & Ap.{} Diam. & $\sigma_0$  & $a$ & $b$         & Gain \\
     &      & (sec)         &             &                       &                     &(\arcsec) & ($5\sigma$) & (\arcsec) & ($10^{-2}$) &     & ($10^{-2}$) &($\mathrm{e}^{-}/\mathrm{DU}$) \\
\hline
\endfirsthead
\caption{continued.}\\
\hline\hline
Tile & Band & Integration Time & No.{} Images & Start of Observations & End of Observations & FWHM     & Lim.{} Depth  & Ap.{} Diam. & $\sigma_0$  & $a$ & $b$         & Gain \\
     &      & (sec)         &             &                       &                     &(\arcsec) & ($5\sigma$) & (\arcsec) & ($10^{-2}$) &     & ($10^{-2}$) &($\mathrm{e}^{-}/\mathrm{DU}$) \\
\hline
\endhead
\hline
\endfoot
   F01 &     $J$ & 10560 &  88 & 2005-12-25T01:54:21 & 2006-01-12T04:11:57 &  0.601 & 24.94 &  0.741 &  4.02 &  1.07 &  5.43 &  43747 \\
   F01 &    $K_\mathrm{s}$ & 17910 & 199 & 2005-12-25T04:18:42 & 2006-01-14T04:12:11 &  0.495 & 24.52 &  0.671 &  7.01 &  1.14 &  4.43 &  73435 \\
   F02 &     $J$ & 10920 &  91 & 2006-02-13T00:20:53 & 2006-02-16T02:11:28 &  0.588 & 25.02 &  0.681 &  4.29 &  0.92 &  8.48 &  43245 \\
   F02 &    $K_\mathrm{s}$ & 14040 & 156 & 2005-11-19T01:49:15 & 2005-11-21T08:13:49 &  0.556 & 24.40 &  0.701 &  7.81 &  1.08 &  6.22 &  56026 \\
   F03 &     $J$ & 12600 & 105 & 2002-11-07T05:25:56 & 2003-10-21T08:36:51 &  0.430 & 25.29 &  0.581 &  4.03 &  1.04 &  6.95 &  46777 \\
   F03 &     $H$ & 17880 & 149 & 2005-01-19T00:37:54 & 2005-01-27T03:24:42 &  0.536 & 24.29 &  0.681 &  7.57 &  1.11 &  7.16 &  60683 \\
   F03 &    $K_\mathrm{s}$ & 20340 & 226 & 2003-10-19T04:58:13 & 2004-02-10T02:25:10 &  0.490 & 24.28 &  0.611 &  7.53 &  1.07 & 13.10 &  76404 \\
   F04 &     $J$ & 12480 & 104 & 2003-10-06T05:45:02 & 2003-10-20T07:07:15 &  0.446 & 25.22 &  0.591 &  4.13 &  1.11 &  5.16 &  43220 \\
   F04 &     $H$ & 25200 & 210 & 2003-10-17T03:53:32 & 2004-02-03T01:59:21 &  0.473 & 24.71 &  0.611 &  6.07 &  1.07 &  7.18 &  87769 \\
   F04 &    $K_\mathrm{s}$ &  9810 & 109 & 2003-10-17T06:58:20 & 2003-10-18T07:58:33 &  0.509 & 24.10 &  0.671 & 10.16 &  1.23 &  5.63 &  39734 \\
   F05 &     $J$ & 12360 & 103 & 2002-12-26T02:00:41 & 2003-01-17T03:45:47 &  0.393 & 25.33 &  0.561 &  4.19 &  1.13 &  4.00 &  42934 \\
   F05 &     $H$ & 17520 & 146 & 2004-12-23T04:44:03 & 2005-09-19T09:51:56 &  0.400 & 25.07 &  0.551 &  5.48 &  1.06 &  6.43 &  59704 \\
   F05 &    $K_\mathrm{s}$ & 16200 & 180 & 2002-12-23T05:05:37 & 2003-12-05T02:44:47 &  0.564 & 24.37 &  0.711 &  7.51 &  1.08 &  6.29 &  62863 \\
   F06 &     $J$ & 14400 & 120 & 2004-11-25T05:52:43 & 2005-01-18T05:00:13 &  0.418 & 25.18 &  0.601 &  4.59 &  1.09 &  5.13 &  38905 \\
   F06 &     $H$ & 18000 & 300 & 2006-11-08T02:16:59 & 2007-01-03T03:36:50 &  0.368 & 24.90 &  0.551 &  6.60 &  1.17 &  4.02 &  55599 \\
   F06 &    $K_\mathrm{s}$ & 21330 & 237 & 2004-10-28T02:18:45 & 2004-11-25T05:51:34 &  0.396 & 24.84 &  0.581 &  6.47 &  1.13 &  6.39 &  90159 \\
   F08 &     $J$ & 11880 &  99 & 2003-10-04T04:00:43 & 2003-10-21T04:15:30 &  0.451 & 25.58 &  0.651 &  3.01 &  1.14 &  3.57 &  43846 \\
   F08 &     $H$ & 16800 & 140 & 2003-02-18T00:59:19 & 2004-01-29T03:09:53 &  0.366 & 24.97 &  0.521 &  6.82 &  1.09 &  5.71 &  55228 \\
   F08 &    $K_\mathrm{s}$ & 21510 & 239 & 2002-12-17T04:07:16 & 2003-02-15T01:44:40 &  0.383 & 24.84 &  0.561 &  6.94 &  1.17 &  3.04 &  72870 \\
   F09 &     $J$ & 12600 & 105 & 2001-11-06T04:38:03 & 2001-11-21T04:53:24 &  0.401 & 25.51 &  0.471 &  3.96 &  1.07 &  6.55 &  47591 \\
   F09 &     $H$ & 18000 & 150 & 2001-11-07T05:50:01 & 2001-12-20T04:21:47 &  0.398 & 24.95 &  0.531 &  5.81 &  1.03 &  8.41 &  74606 \\
   F09 &    $K_\mathrm{s}$ & 21510 & 239 & 2001-10-01T05:31:32 & 2001-11-16T03:17:39 &  0.419 & 24.93 &  0.571 &  6.01 &  1.03 &  7.62 &  89172 \\
   F10 &     $J$ & 11880 &  66 & 2000-10-14T05:44:47 & 2000-11-12T05:29:28 &  0.524 & 25.09 &  0.711 &  3.77 &  1.14 &  5.66 &  49834 \\
   F10 &     $H$ & 18120 & 151 & 2001-12-20T04:44:28 & 2001-12-26T01:46:02 &  0.493 & 24.69 &  0.671 &  6.48 &  1.03 &  6.82 &  80606 \\
   F10 &    $K_\mathrm{s}$ & 23800 & 238 & 2000-10-11T04:46:36 & 2001-01-04T02:21:23 &  0.458 & 24.90 &  0.621 &  5.32 &  1.02 &  9.50 & 102343 \\
   F11 &     $J$ & 11520 &  64 & 2000-11-13T02:20:20 & 2000-11-13T06:09:33 &  0.488 & 25.27 &  0.651 &  3.80 &  1.13 &  4.93 &  47437 \\
   F11 &     $H$ & 17760 & 148 & 2001-12-21T05:38:41 & 2001-12-26T03:11:06 &  0.482 & 24.90 &  0.631 &  5.49 &  1.08 &  5.77 &  71877 \\
   F11 &    $K_\mathrm{s}$ & 30900 & 309 & 2000-11-13T06:10:33 & 2002-01-18T03:23:51 &  0.470 & 24.94 &  0.611 &  5.12 &  1.05 &  8.19 & 133790 \\
   F13 &     $J$ & 10800 &  90 & 2002-10-26T07:51:45 & 2002-12-26T04:33:03 &  0.414 & 25.29 &  0.601 &  4.24 &  1.10 &  4.37 &  41405 \\
   F13 &     $H$ & 14400 & 120 & 2002-11-29T02:42:50 & 2002-11-29T07:42:10 &  0.497 & 24.76 &  0.591 &  6.31 &  1.10 &  7.16 &  67692 \\
   F13 &    $K_\mathrm{s}$ & 21150 & 235 & 2002-10-26T03:30:49 & 2002-12-26T05:58:32 &  0.445 & 24.79 &  0.551 &  7.11 &  1.12 &  7.00 &  85965 \\
   F14 &     $J$ & 12600 & 105 & 2001-11-06T03:20:35 & 2001-11-19T04:23:35 &  0.431 & 25.44 &  0.641 &  3.46 &  1.15 &  3.17 &  58402 \\
   F14 &     $H$ & 14280 & 119 & 2001-09-30T07:36:38 & 2001-12-19T06:55:45 &  0.415 & 24.82 &  0.611 &  6.23 &  1.11 &  5.46 &  71657 \\
   F14 &    $K_\mathrm{s}$ & 21690 & 241 & 2001-09-30T08:55:26 & 2001-11-14T02:24:35 &  0.457 & 24.93 &  0.661 &  5.55 &  1.09 &  4.76 & 102667 \\
   F15 &     $J$ & 11340 &  63 & 2002-01-19T01:35:29 & 2002-02-13T02:18:33 &  0.485 & 25.02 &  0.671 &  4.95 &  1.18 &  2.71 &  40869 \\
   F15 &     $H$ & 17880 & 149 & 2001-12-25T02:28:31 & 2002-01-15T04:28:48 &  0.655 & 24.45 &  0.771 &  6.13 &  1.04 &  6.09 &  85444 \\
   F15 &    $K_\mathrm{s}$ & 27100 & 271 & 2002-01-21T01:01:25 & 2002-02-13T01:20:40 &  0.517 & 24.85 &  0.681 &  5.51 &  1.14 &  4.13 & 125058 \\
   F16 &     $J$ & 10620 &  59 & 1999-10-28T04:48:32 & 1999-11-17T05:43:59 &  0.443 & 25.04 &  0.611 &  5.30 &  1.11 &  4.28 &  28490 \\
   F16 &     $H$ & 17760 & 148 & 2001-12-26T04:35:19 & 2002-01-15T03:07:45 &  0.545 & 24.73 &  0.701 &  5.69 &  1.07 &  5.51 &  81466 \\
   F16 &    $K_\mathrm{s}$ & 30100 & 301 & 1999-10-21T03:44:53 & 1999-10-25T08:34:53 &  0.533 & 24.14 &  0.671 &  8.56 &  1.01 &  8.45 &  74460 \\
   F18 &     $J$ & 18000 & 150 & 2004-11-26T03:39:18 & 2005-02-25T02:23:17 &  0.499 & 25.09 &  0.681 &  4.09 &  1.09 &  5.48 &  51109 \\
   F18 &     $H$ & 14520 & 242 & 2006-12-07T05:28:19 & 2007-01-04T03:26:14 &  0.565 & 24.52 &  0.701 &  6.76 &  1.09 &  4.91 &  59944 \\
   F18 &    $K_\mathrm{s}$ & 21600 & 240 & 2004-10-31T06:04:11 & 2004-11-25T08:20:55 &  0.417 & 24.73 &  0.591 &  6.43 &  1.11 &  8.18 &  88837 \\
   F19 &     $J$ & 16200 & 135 & 2002-09-15T07:06:00 & 2002-09-23T06:14:29 &  0.623 & 25.17 &  0.801 &  3.00 &  1.18 &  3.24 &  54756 \\
   F19 &     $H$ & 17280 & 144 & 2005-01-20T00:39:02 & 2005-08-22T09:30:29 &  0.608 & 24.48 &  0.801 &  5.94 &  1.14 &  3.14 &  61070 \\
   F19 &    $K_\mathrm{s}$ & 20160 & 224 & 2002-01-25T00:46:57 & 2002-02-16T01:34:16 &  0.508 & 24.57 &  0.731 &  6.49 &  1.21 &  2.78 &  88384 \\
   F20 &     $J$ & 12600 & 105 & 2001-11-13T03:15:44 & 2001-11-22T08:05:37 &  0.460 & 25.47 &  0.641 &  3.44 &  1.08 &  4.34 &  59277 \\
   F20 &     $H$ & 17640 & 147 & 2001-12-27T01:05:57 & 2002-01-16T04:06:24 &  0.512 & 24.76 &  0.681 &  5.84 &  1.06 &  5.50 &  78636 \\
   F20 &    $K_\mathrm{s}$ & 21510 & 239 & 2001-11-13T01:50:20 & 2001-11-21T02:45:54 &  0.386 & 25.01 &  0.561 &  6.07 &  1.11 &  4.41 &  93412 \\
   F21 &     $J$ & 15120 & 126 & 2001-11-06T05:54:47 & 2001-11-24T04:35:16 &  0.396 & 25.58 &  0.551 &  3.52 &  1.09 &  4.24 &  63229 \\
   F21 &     $H$ & 13800 & 115 & 2001-12-27T02:46:08 & 2001-12-31T02:59:03 &  0.467 & 24.68 &  0.641 &  6.66 &  1.06 &  5.72 &  59856 \\
   F21 &    $K_\mathrm{s}$ & 21420 & 238 & 2001-11-07T07:36:11 & 2001-11-19T03:42:05 &  0.381 & 25.02 &  0.561 &  5.99 &  1.10 &  4.84 &  90812 \\
   F22 &     $J$ & 15000 & 125 & 2003-01-21T03:25:07 & 2003-02-12T01:42:25 &  0.552 & 25.06 &  0.701 &  3.57 &  1.06 &  6.65 &  57417 \\
   F22 &     $H$ & 17520 & 146 & 2003-02-05T02:02:37 & 2003-02-11T02:50:22 &  0.520 & 24.50 &  0.711 &  6.78 &  1.11 &  4.97 &  81391 \\
   F22 &    $K_\mathrm{s}$ & 21150 & 235 & 2003-01-13T03:09:16 & 2003-01-26T03:54:18 &  0.526 & 24.52 &  0.701 &  6.42 &  1.15 &  5.09 &  88797 \\
   F23 &     $J$ &  9000 &  75 & 2004-01-04T04:33:04 & 2004-01-14T03:38:33 &  0.462 & 25.10 &  0.691 &  4.48 &  1.13 &  3.01 &  35427 \\
   F23 &     $H$ & 14040 & 117 & 2004-12-24T04:51:43 & 2004-12-26T03:33:11 &  0.339 & 25.03 &  0.511 &  6.48 &  1.14 &  3.03 &  50175 \\
   F23 &    $K_\mathrm{s}$ & 21420 & 238 & 2004-01-14T03:41:59 & 2004-02-16T02:24:07 &  0.451 & 24.70 &  0.671 &  6.98 &  1.16 &  2.85 &  86321 \\
   F24 &     $J$ &  9840 &  82 & 2002-10-17T07:27:23 & 2003-10-21T07:19:34 &  0.543 & 25.08 &  0.711 &  4.27 &  1.14 &  3.70 &  38142 \\
   F24 &     $H$ & 17760 & 148 & 2004-11-17T00:36:07 & 2004-11-27T04:12:23 &  0.424 & 24.73 &  0.581 &  7.09 &  1.13 &  6.01 &  74197 \\
   F24 &    $K_\mathrm{s}$ & 20880 & 232 & 2002-10-11T04:44:27 & 2003-10-21T05:56:21 &  0.505 & 24.57 &  0.651 &  6.92 &  1.11 &  6.69 &  81231 \\
   F25 &    $K_\mathrm{s}$ &  7110 &  79 & 2002-01-23T00:42:30 & 2002-02-17T01:24:14 &  0.579 & 23.84 &  0.801 & 11.04 &  1.19 &  2.52 &  29234 \\
  F25n &     $J$ & 12600 & 105 & 2002-09-24T05:27:09 & 2002-10-08T07:58:48 &  0.516 & 25.37 &  0.701 &  3.07 &  1.05 &  4.90 &  52255 \\
  F25n &     $H$ & 14280 & 119 & 2004-11-28T05:41:58 & 2005-01-27T01:56:11 &  0.456 & 24.64 &  0.661 &  6.96 &  1.11 &  3.89 &  62361 \\
  F25n &    $K_\mathrm{s}$ & 21600 & 240 & 2002-09-15T03:54:15 & 2002-10-06T06:56:45 &  0.419 & 24.70 &  0.621 &  6.48 &  1.17 &  5.41 &  83748 \\
   F26 &    $K_\mathrm{s}$ & 10440 & 116 & 2002-01-13T00:49:07 & 2002-01-25T03:40:48 &  0.475 & 24.25 &  0.681 &  9.12 &  1.15 &  3.87 &  43470 \\
  F26n &     $J$ & 12240 & 102 & 2002-09-24T06:41:56 & 2002-10-15T08:36:08 &  0.557 & 25.12 &  0.741 &  3.75 &  1.12 &  3.53 &  45074 \\
  F26n &     $H$ & 18000 & 150 & 2004-11-27T04:16:43 & 2004-11-28T05:37:10 &  0.512 & 24.53 &  0.701 &  6.42 &  1.16 &  4.52 &  80019 \\
  F26n &    $K_\mathrm{s}$ & 21240 & 236 & 2002-09-24T04:06:29 & 2002-10-10T09:09:30 &  0.568 & 24.55 &  0.721 &  6.77 &  1.15 &  3.87 &  85442 \\
   F29 &     $J$ &  9000 &  75 & 2004-02-02T03:11:25 & 2004-09-01T09:38:25 &  0.603 & 24.95 &  0.771 &  4.19 &  1.09 &  4.23 &  32443 \\
   F29 &     $H$ & 18060 & 301 & 2006-09-13T08:00:56 & 2006-11-03T03:58:45 &  0.408 & 25.28 &  0.601 &  4.62 &  1.12 &  4.00 &  72630 \\
   F29 &    $K_\mathrm{s}$ & 16650 & 185 & 2004-02-12T01:03:50 & 2004-08-01T10:44:49 &  0.492 & 24.61 &  0.711 &  6.87 &  1.16 &  3.28 &  71920 \\
   F30 &     $J$ & 12360 & 103 & 2003-11-17T08:05:29 & 2004-01-01T03:18:39 &  0.587 & 25.09 &  0.711 &  3.93 &  1.05 &  4.86 &  47691 \\
   F30 &     $H$ & 15180 & 253 & 2006-09-15T08:08:05 & 2006-11-06T06:21:48 &  0.460 & 24.81 &  0.631 &  6.10 &  1.17 &  3.87 &  51625 \\
   F30 &    $K_\mathrm{s}$ & 24210 & 269 & 2002-12-22T02:35:02 & 2004-02-02T03:09:13 &  0.413 & 24.75 &  0.601 &  6.71 &  1.18 &  5.24 &  83247 \\
   F31 &     $J$ & 10800 &  90 & 2004-02-21T00:33:23 & 2004-03-03T01:23:46 &  0.571 & 24.87 &  0.721 &  5.04 &  1.06 &  4.57 &  39051 \\
   F31 &     $H$ & 16920 & 282 & 2006-09-15T07:19:30 & 2006-09-24T06:05:01 &  0.612 & 24.65 &  0.711 &  4.99 &  0.95 &  8.59 &  70757 \\
   F31 &    $K_\mathrm{s}$ & 21060 & 234 & 2003-12-30T02:56:37 & 2004-03-09T01:15:45 &  0.408 & 24.61 &  0.591 &  7.82 &  1.15 &  3.83 &  57063 \\
\end{longtable}
%
%
\clearpage
\begin{longtable}{lcccccccccccccccc}
\caption{\label{tileapcors}Aperture corrections of the final survey tiles for aperture diameters between 
0.7\arcsec and 8.38\arcsec assuming point source profiles.}\\
\hline\hline
Tile & Band & $\delta m_\mathrm{ap}$ & $\delta m_\mathrm{ap}$ & $\delta m_\mathrm{ap}$ & $\delta m_\mathrm{ap}$ & $\delta m_\mathrm{ap}$ & $\delta m_\mathrm{ap}$ & $\delta m_\mathrm{ap}$ & $\delta m_\mathrm{ap}$ & $\delta m_\mathrm{ap}$ & $\delta m_\mathrm{ap}$ & $\delta m_\mathrm{ap}$ & $\delta m_\mathrm{ap}$ & $\delta m_\mathrm{ap}$ & $\delta m_\mathrm{ap}$ & $\delta m_\mathrm{ap}$ \\
 & & (0.70\arcsec) & (0.84\arcsec) & (1.00\arcsec) & (1.19\arcsec) & (1.43\arcsec) & (1.70\arcsec) & (2.03\arcsec) & (2.42\arcsec) & (2.89\arcsec) & (3.46\arcsec) & (4.13\arcsec) & (4.92\arcsec) & (5.88\arcsec) & (7.02\arcsec) & (8.38\arcsec) \\

\hline
\endfirsthead
\caption{continued.}\\
\hline\hline
Tile & Band & $\delta m_\mathrm{ap}$ & $\delta m_\mathrm{ap}$ & $\delta m_\mathrm{ap}$ & $\delta m_\mathrm{ap}$ & $\delta m_\mathrm{ap}$ & $\delta m_\mathrm{ap}$ & $\delta m_\mathrm{ap}$ & $\delta m_\mathrm{ap}$ & $\delta m_\mathrm{ap}$ & $\delta m_\mathrm{ap}$ & $\delta m_\mathrm{ap}$ & $\delta m_\mathrm{ap}$ & $\delta m_\mathrm{ap}$ & $\delta m_\mathrm{ap}$ & $\delta m_\mathrm{ap}$ \\
 & & (0.70\arcsec) & (0.84\arcsec) & (1.00\arcsec) & (1.19\arcsec) & (1.43\arcsec) & (1.70\arcsec) & (2.03\arcsec) & (2.42\arcsec) & (2.89\arcsec) & (3.46\arcsec) & (4.13\arcsec) & (4.92\arcsec) & (5.88\arcsec) & (7.02\arcsec) & (8.38\arcsec) \\

\hline
\endhead
\hline
\endfoot
F\,01 & $J$ & 0.979 & 0.760 & 0.578 & 0.433 & 0.322 & 0.237 & 0.174 & 0.127 & 0.092 & 0.067 & 0.047 & 0.031 & 0.019 & 0.010 & 0.005 \\
F\,01 & $K_\mathrm{s}$ & 0.773 & 0.594 & 0.450 & 0.337 & 0.247 & 0.182 & 0.134 & 0.098 & 0.071 & 0.051 & 0.036 & 0.024 & 0.015 & 0.009 & 0.004 \\
F\,02 & $J$ & 0.848 & 0.642 & 0.476 & 0.350 & 0.256 & 0.188 & 0.139 & 0.103 & 0.077 & 0.055 & 0.038 & 0.027 & 0.016 & 0.009 & 0.004 \\
F\,02 & $K_\mathrm{s}$ & 0.761 & 0.562 & 0.411 & 0.295 & 0.210 & 0.150 & 0.111 & 0.082 & 0.060 & 0.044 & 0.031 & 0.018 & 0.011 & 0.006 & 0.003 \\
F\,03 & $J$ & 0.636 & 0.484 & 0.365 & 0.276 & 0.210 & 0.161 & 0.122 & 0.092 & 0.069 & 0.051 & 0.037 & 0.025 & 0.015 & 0.008 & 0.004 \\
F\,03 & $H$ & 0.853 & 0.657 & 0.495 & 0.370 & 0.274 & 0.205 & 0.152 & 0.113 & 0.085 & 0.063 & 0.046 & 0.031 & 0.017 & 0.010 & 0.004 \\
F\,03 & $K_\mathrm{s}$ & 0.734 & 0.556 & 0.412 & 0.303 & 0.222 & 0.162 & 0.118 & 0.086 & 0.062 & 0.043 & 0.031 & 0.020 & 0.013 & 0.007 & 0.003 \\
F\,04 & $J$ & 0.662 & 0.510 & 0.389 & 0.295 & 0.223 & 0.169 & 0.128 & 0.096 & 0.072 & 0.054 & 0.039 & 0.027 & 0.016 & 0.009 & 0.004 \\
F\,04 & $H$ & 0.714 & 0.549 & 0.415 & 0.312 & 0.235 & 0.177 & 0.134 & 0.102 & 0.077 & 0.058 & 0.042 & 0.029 & 0.018 & 0.010 & 0.005 \\
F\,04 & $K_\mathrm{s}$ & 0.683 & 0.504 & 0.367 & 0.263 & 0.187 & 0.135 & 0.098 & 0.072 & 0.050 & 0.035 & 0.024 & 0.016 & 0.010 & 0.005 & 0.002 \\
F\,05 & $J$ & 0.592 & 0.458 & 0.351 & 0.270 & 0.207 & 0.160 & 0.121 & 0.091 & 0.069 & 0.050 & 0.037 & 0.024 & 0.016 & 0.009 & 0.004 \\
F\,05 & $H$ & 0.544 & 0.412 & 0.308 & 0.228 & 0.169 & 0.126 & 0.096 & 0.071 & 0.052 & 0.038 & 0.025 & 0.017 & 0.010 & 0.006 & 0.003 \\
F\,05 & $K_\mathrm{s}$ & 0.830 & 0.625 & 0.457 & 0.328 & 0.235 & 0.165 & 0.116 & 0.086 & 0.065 & 0.045 & 0.031 & 0.020 & 0.013 & 0.007 & 0.003 \\
F\,06 & $J$ & 0.611 & 0.462 & 0.351 & 0.264 & 0.197 & 0.149 & 0.113 & 0.086 & 0.064 & 0.046 & 0.032 & 0.021 & 0.014 & 0.008 & 0.003 \\
F\,06 & $H$ & 0.494 & 0.370 & 0.277 & 0.207 & 0.155 & 0.116 & 0.088 & 0.067 & 0.049 & 0.034 & 0.022 & 0.015 & 0.009 & 0.005 & 0.002 \\
F\,06 & $K_\mathrm{s}$ & 0.520 & 0.382 & 0.270 & 0.193 & 0.139 & 0.101 & 0.074 & 0.051 & 0.036 & 0.025 & 0.017 & 0.011 & 0.007 & 0.004 & 0.002 \\
F\,08 & $J$ & 0.661 & 0.497 & 0.372 & 0.279 & 0.209 & 0.159 & 0.121 & 0.093 & 0.072 & 0.054 & 0.036 & 0.024 & 0.015 & 0.009 & 0.004 \\
F\,08 & $H$ & 0.439 & 0.326 & 0.245 & 0.185 & 0.139 & 0.103 & 0.078 & 0.057 & 0.041 & 0.030 & 0.021 & 0.014 & 0.009 & 0.005 & 0.002 \\
F\,08 & $K_\mathrm{s}$ & 0.532 & 0.404 & 0.306 & 0.228 & 0.173 & 0.131 & 0.100 & 0.075 & 0.051 & 0.036 & 0.026 & 0.017 & 0.011 & 0.006 & 0.003 \\
F\,09 & $J$ & 0.505 & 0.393 & 0.304 & 0.237 & 0.186 & 0.147 & 0.117 & 0.092 & 0.070 & 0.053 & 0.038 & 0.026 & 0.017 & 0.010 & 0.005 \\
F\,09 & $H$ & 0.574 & 0.439 & 0.336 & 0.258 & 0.200 & 0.154 & 0.121 & 0.094 & 0.070 & 0.052 & 0.037 & 0.026 & 0.017 & 0.010 & 0.004 \\
F\,09 & $K_\mathrm{s}$ & 0.560 & 0.419 & 0.316 & 0.241 & 0.185 & 0.147 & 0.115 & 0.087 & 0.064 & 0.044 & 0.032 & 0.022 & 0.014 & 0.008 & 0.004 \\
F\,10 & $J$ & 0.832 & 0.632 & 0.475 & 0.351 & 0.258 & 0.189 & 0.138 & 0.102 & 0.072 & 0.052 & 0.034 & 0.023 & 0.014 & 0.008 & 0.003 \\
F\,10 & $H$ & 0.699 & 0.515 & 0.378 & 0.272 & 0.194 & 0.140 & 0.103 & 0.075 & 0.053 & 0.036 & 0.025 & 0.016 & 0.010 & 0.006 & 0.002 \\
F\,10 & $K_\mathrm{s}$ & 0.638 & 0.469 & 0.348 & 0.252 & 0.181 & 0.130 & 0.096 & 0.071 & 0.052 & 0.037 & 0.026 & 0.017 & 0.011 & 0.006 & 0.003 \\
F\,11 & $J$ & 0.685 & 0.516 & 0.383 & 0.283 & 0.210 & 0.156 & 0.116 & 0.087 & 0.065 & 0.047 & 0.034 & 0.023 & 0.014 & 0.008 & 0.003 \\
F\,11 & $H$ & 0.665 & 0.502 & 0.370 & 0.271 & 0.200 & 0.147 & 0.109 & 0.081 & 0.059 & 0.043 & 0.030 & 0.019 & 0.012 & 0.007 & 0.003 \\
F\,11 & $K_\mathrm{s}$ & 0.661 & 0.496 & 0.367 & 0.269 & 0.196 & 0.143 & 0.105 & 0.078 & 0.057 & 0.042 & 0.029 & 0.019 & 0.012 & 0.007 & 0.003 \\
F\,13 & $J$ & 0.598 & 0.451 & 0.345 & 0.262 & 0.199 & 0.152 & 0.116 & 0.088 & 0.066 & 0.049 & 0.035 & 0.024 & 0.016 & 0.009 & 0.004 \\
F\,13 & $H$ & 0.616 & 0.466 & 0.349 & 0.259 & 0.193 & 0.143 & 0.107 & 0.080 & 0.059 & 0.040 & 0.028 & 0.019 & 0.012 & 0.007 & 0.003 \\
F\,13 & $K_\mathrm{s}$ & 0.483 & 0.356 & 0.265 & 0.199 & 0.151 & 0.116 & 0.090 & 0.067 & 0.049 & 0.033 & 0.024 & 0.016 & 0.010 & 0.006 & 0.003 \\
F\,14 & $J$ & 0.666 & 0.506 & 0.381 & 0.288 & 0.217 & 0.164 & 0.124 & 0.093 & 0.069 & 0.052 & 0.037 & 0.024 & 0.016 & 0.009 & 0.004 \\
F\,14 & $H$ & 0.608 & 0.455 & 0.332 & 0.242 & 0.176 & 0.128 & 0.096 & 0.072 & 0.053 & 0.038 & 0.026 & 0.017 & 0.011 & 0.006 & 0.003 \\
F\,14 & $K_\mathrm{s}$ & 0.641 & 0.470 & 0.335 & 0.237 & 0.167 & 0.119 & 0.086 & 0.060 & 0.042 & 0.029 & 0.019 & 0.013 & 0.008 & 0.004 & 0.002 \\
F\,15 & $J$ & 0.681 & 0.512 & 0.385 & 0.288 & 0.215 & 0.162 & 0.123 & 0.093 & 0.069 & 0.051 & 0.037 & 0.024 & 0.015 & 0.009 & 0.004 \\
F\,15 & $H$ & 1.016 & 0.783 & 0.588 & 0.434 & 0.316 & 0.227 & 0.164 & 0.121 & 0.089 & 0.064 & 0.045 & 0.030 & 0.019 & 0.010 & 0.004 \\
F\,15 & $K_\mathrm{s}$ & 0.714 & 0.534 & 0.395 & 0.289 & 0.210 & 0.155 & 0.114 & 0.086 & 0.065 & 0.049 & 0.034 & 0.023 & 0.014 & 0.008 & 0.003 \\
F\,16 & $J$ & 0.595 & 0.447 & 0.329 & 0.245 & 0.183 & 0.135 & 0.104 & 0.079 & 0.056 & 0.043 & 0.031 & 0.023 & 0.014 & 0.007 & 0.003 \\
F\,16 & $H$ & 0.808 & 0.610 & 0.451 & 0.328 & 0.238 & 0.173 & 0.131 & 0.097 & 0.065 & 0.046 & 0.032 & 0.021 & 0.013 & 0.007 & 0.003 \\
F\,16 & $K_\mathrm{s}$ & 0.904 & 0.705 & 0.534 & 0.401 & 0.300 & 0.223 & 0.162 & 0.117 & 0.083 & 0.062 & 0.047 & 0.032 & 0.020 & 0.012 & 0.005 \\
F\,18 & $J$ & 0.795 & 0.606 & 0.458 & 0.342 & 0.254 & 0.189 & 0.141 & 0.104 & 0.076 & 0.055 & 0.038 & 0.026 & 0.017 & 0.010 & 0.004 \\
F\,18 & $H$ & 0.836 & 0.641 & 0.487 & 0.365 & 0.274 & 0.206 & 0.155 & 0.117 & 0.085 & 0.062 & 0.043 & 0.030 & 0.019 & 0.011 & 0.005 \\
F\,18 & $K_\mathrm{s}$ & 0.584 & 0.434 & 0.320 & 0.237 & 0.178 & 0.138 & 0.104 & 0.079 & 0.059 & 0.043 & 0.025 & 0.016 & 0.010 & 0.006 & 0.003 \\
F\,19 & $J$ & 1.074 & 0.845 & 0.645 & 0.484 & 0.359 & 0.264 & 0.193 & 0.143 & 0.108 & 0.079 & 0.057 & 0.040 & 0.026 & 0.015 & 0.006 \\
F\,19 & $H$ & 1.050 & 0.823 & 0.627 & 0.469 & 0.346 & 0.253 & 0.188 & 0.140 & 0.106 & 0.079 & 0.060 & 0.043 & 0.024 & 0.013 & 0.006 \\
F\,19 & $K_\mathrm{s}$ & 0.800 & 0.605 & 0.448 & 0.328 & 0.237 & 0.172 & 0.124 & 0.092 & 0.066 & 0.046 & 0.032 & 0.021 & 0.013 & 0.007 & 0.003 \\
F\,20 & $J$ & 0.655 & 0.492 & 0.366 & 0.272 & 0.202 & 0.152 & 0.114 & 0.086 & 0.063 & 0.047 & 0.034 & 0.024 & 0.014 & 0.008 & 0.003 \\
F\,20 & $H$ & 0.758 & 0.571 & 0.426 & 0.315 & 0.233 & 0.174 & 0.133 & 0.102 & 0.080 & 0.062 & 0.048 & 0.035 & 0.022 & 0.009 & 0.004 \\
F\,20 & $K_\mathrm{s}$ & 0.506 & 0.377 & 0.280 & 0.207 & 0.153 & 0.114 & 0.086 & 0.064 & 0.048 & 0.033 & 0.022 & 0.015 & 0.009 & 0.005 & 0.002 \\
F\,21 & $J$ & 0.551 & 0.420 & 0.321 & 0.243 & 0.185 & 0.143 & 0.111 & 0.085 & 0.063 & 0.048 & 0.036 & 0.024 & 0.013 & 0.008 & 0.003 \\
F\,21 & $H$ & 0.704 & 0.534 & 0.402 & 0.299 & 0.223 & 0.168 & 0.126 & 0.095 & 0.071 & 0.053 & 0.038 & 0.022 & 0.014 & 0.008 & 0.003 \\
F\,21 & $K_\mathrm{s}$ & 0.514 & 0.384 & 0.284 & 0.209 & 0.152 & 0.110 & 0.080 & 0.057 & 0.042 & 0.029 & 0.020 & 0.014 & 0.009 & 0.005 & 0.002 \\
F\,22 & $J$ & 0.957 & 0.752 & 0.583 & 0.445 & 0.336 & 0.252 & 0.188 & 0.141 & 0.106 & 0.077 & 0.054 & 0.036 & 0.022 & 0.013 & 0.006 \\
F\,22 & $H$ & 0.833 & 0.633 & 0.472 & 0.345 & 0.247 & 0.176 & 0.127 & 0.094 & 0.066 & 0.046 & 0.033 & 0.022 & 0.014 & 0.008 & 0.003 \\
F\,22 & $K_\mathrm{s}$ & 0.837 & 0.642 & 0.484 & 0.358 & 0.261 & 0.188 & 0.136 & 0.103 & 0.073 & 0.056 & 0.039 & 0.026 & 0.017 & 0.010 & 0.004 \\
F\,23 & $J$ & 0.733 & 0.552 & 0.414 & 0.307 & 0.227 & 0.168 & 0.126 & 0.094 & 0.069 & 0.050 & 0.036 & 0.024 & 0.015 & 0.008 & 0.004 \\
F\,23 & $H$ & 0.476 & 0.364 & 0.279 & 0.215 & 0.169 & 0.136 & 0.110 & 0.087 & 0.067 & 0.048 & 0.034 & 0.024 & 0.016 & 0.009 & 0.004 \\
F\,23 & $K_\mathrm{s}$ & 0.636 & 0.468 & 0.338 & 0.240 & 0.170 & 0.122 & 0.089 & 0.066 & 0.050 & 0.036 & 0.023 & 0.013 & 0.008 & 0.005 & 0.002 \\
F\,24 & $J$ & 0.773 & 0.583 & 0.434 & 0.322 & 0.238 & 0.176 & 0.132 & 0.100 & 0.073 & 0.053 & 0.037 & 0.025 & 0.016 & 0.009 & 0.004 \\
F\,24 & $H$ & 0.533 & 0.394 & 0.290 & 0.215 & 0.160 & 0.119 & 0.088 & 0.067 & 0.047 & 0.030 & 0.021 & 0.014 & 0.009 & 0.005 & 0.002 \\
F\,24 & $K_\mathrm{s}$ & 0.690 & 0.515 & 0.380 & 0.277 & 0.201 & 0.147 & 0.107 & 0.078 & 0.058 & 0.038 & 0.026 & 0.018 & 0.011 & 0.006 & 0.003 \\
F\,25 & $K_\mathrm{s}$ & 0.997 & 0.774 & 0.584 & 0.432 & 0.314 & 0.224 & 0.162 & 0.117 & 0.083 & 0.058 & 0.040 & 0.027 & 0.017 & 0.010 & 0.004 \\
F\,25n & $J$ & 0.869 & 0.670 & 0.510 & 0.383 & 0.286 & 0.215 & 0.162 & 0.121 & 0.091 & 0.068 & 0.048 & 0.032 & 0.019 & 0.011 & 0.005 \\
F\,25n & $H$ & 0.704 & 0.532 & 0.399 & 0.297 & 0.223 & 0.169 & 0.128 & 0.099 & 0.075 & 0.054 & 0.039 & 0.025 & 0.015 & 0.009 & 0.004 \\
F\,25n & $K_\mathrm{s}$ & 0.634 & 0.478 & 0.355 & 0.262 & 0.193 & 0.143 & 0.108 & 0.082 & 0.061 & 0.044 & 0.031 & 0.020 & 0.013 & 0.007 & 0.003 \\
F\,26 & $K_\mathrm{s}$ & 0.769 & 0.588 & 0.448 & 0.335 & 0.246 & 0.179 & 0.131 & 0.095 & 0.069 & 0.049 & 0.032 & 0.023 & 0.015 & 0.008 & 0.004 \\
F\,26n & $J$ & 0.901 & 0.697 & 0.527 & 0.395 & 0.296 & 0.221 & 0.166 & 0.124 & 0.093 & 0.069 & 0.052 & 0.037 & 0.024 & 0.014 & 0.006 \\
F\,26n & $H$ & 0.835 & 0.642 & 0.484 & 0.361 & 0.268 & 0.200 & 0.150 & 0.112 & 0.084 & 0.061 & 0.043 & 0.029 & 0.019 & 0.011 & 0.005 \\
F\,26n & $K_\mathrm{s}$ & 0.793 & 0.597 & 0.437 & 0.316 & 0.227 & 0.164 & 0.118 & 0.085 & 0.063 & 0.044 & 0.031 & 0.020 & 0.012 & 0.007 & 0.003 \\
F\,29 & $J$ & 0.958 & 0.736 & 0.558 & 0.415 & 0.306 & 0.226 & 0.168 & 0.124 & 0.091 & 0.066 & 0.047 & 0.033 & 0.022 & 0.012 & 0.005 \\
F\,29 & $H$ & 0.521 & 0.381 & 0.281 & 0.207 & 0.154 & 0.117 & 0.090 & 0.069 & 0.050 & 0.039 & 0.028 & 0.020 & 0.012 & 0.007 & 0.003 \\
F\,29 & $K_\mathrm{s}$ & 0.733 & 0.546 & 0.402 & 0.291 & 0.209 & 0.154 & 0.113 & 0.081 & 0.059 & 0.044 & 0.031 & 0.020 & 0.012 & 0.007 & 0.003 \\
F\,30 & $J$ & 0.887 & 0.686 & 0.518 & 0.389 & 0.291 & 0.216 & 0.160 & 0.118 & 0.086 & 0.062 & 0.043 & 0.030 & 0.019 & 0.011 & 0.005 \\
F\,30 & $H$ & 0.638 & 0.480 & 0.358 & 0.266 & 0.199 & 0.150 & 0.113 & 0.085 & 0.064 & 0.046 & 0.033 & 0.021 & 0.014 & 0.008 & 0.003 \\
F\,30 & $K_\mathrm{s}$ & 0.555 & 0.411 & 0.298 & 0.216 & 0.157 & 0.116 & 0.087 & 0.065 & 0.048 & 0.035 & 0.024 & 0.014 & 0.009 & 0.005 & 0.002 \\
F\,31 & $J$ & 0.840 & 0.636 & 0.469 & 0.342 & 0.247 & 0.179 & 0.130 & 0.095 & 0.070 & 0.051 & 0.035 & 0.023 & 0.014 & 0.008 & 0.003 \\
F\,31 & $H$ & 1.020 & 0.799 & 0.615 & 0.466 & 0.348 & 0.258 & 0.189 & 0.138 & 0.100 & 0.072 & 0.052 & 0.035 & 0.022 & 0.013 & 0.006 \\
F\,31 & $K_\mathrm{s}$ & 0.599 & 0.456 & 0.341 & 0.252 & 0.185 & 0.136 & 0.101 & 0.075 & 0.055 & 0.040 & 0.026 & 0.017 & 0.011 & 0.006 & 0.003 \\
\end{longtable}
\end{landscape}

\begin{table}
\caption{Galaxy number counts in $J$, $H$, and $K_\mathrm{s}$ as function of
total magnitude, $m$, expressed in the AB photometric system.
Differential counts and $1\sigma$-errors, $n(m) \pm \Delta n$, are quoted per
unit mag per square degree. Cumulative counts, $N(>m)$, are per square
degree.\label{table:counts}}
\centering
\begin{tabular}{lrrr|rrr|rrr} \hline \hline
 & \multicolumn{3}{c|}{$J$} & \multicolumn{3}{c|}{$H$} & \multicolumn{3}{c}{$K_\mathrm{s}$} \\ \cline{2-4} \cline{5-7} \cline{8-10}
Mag.\ & $n(m)$ & $\Delta n$ & $N(<m)$ & $n(m)$ & $\Delta n$ & $N(<m)$ & $n(m)$ & $\Delta n$ & $N(<m)$ \\ \hline
17.00 &     --   &     --   &     --   &     --   &     --   &     --   &    187.4 &    132.5 &     46.8 \\
17.25 &    192.6 &    136.2 &     48.1 &    204.1 &    144.3 &     51.0 &     93.7 &     93.7 &     70.3 \\
17.50 &     96.3 &     96.3 &     72.2 &    204.1 &    144.3 &    102.1 &     93.7 &     93.7 &     93.7 \\
17.75 &    288.9 &    166.8 &    144.4 &    306.2 &    176.8 &    178.6 &    468.4 &    209.5 &    210.8 \\
18.00 &    481.5 &    215.3 &    264.8 &    306.2 &    176.8 &    255.1 &    655.8 &    247.9 &    374.7 \\
18.25 &    385.2 &    192.6 &    361.1 &    918.5 &    306.2 &    484.7 &   1311.5 &    350.5 &    702.6 \\
18.50 &    481.5 &    215.3 &    481.5 &   1122.6 &    338.5 &    765.4 &   1217.9 &    337.8 &   1007.1 \\
18.75 &    962.9 &    304.5 &    722.2 &   1020.5 &    322.7 &   1020.5 &   1217.9 &    337.8 &   1311.5 \\
19.00 &    962.9 &    304.5 &    962.9 &   1122.6 &    338.5 &   1301.2 &   2061.0 &    439.4 &   1826.8 \\
19.25 &   1155.5 &    333.6 &   1251.8 &   1939.0 &    444.8 &   1785.9 &   4496.7 &    649.0 &   2951.0 \\
19.50 &   1155.5 &    333.6 &   1540.7 &   3775.9 &    620.8 &   2729.9 &   4403.0 &    642.2 &   4051.7 \\
19.75 &   3177.7 &    553.2 &   2335.1 &   4082.1 &    645.4 &   3750.4 &   5339.8 &    707.3 &   5386.7 \\
20.00 &   3851.7 &    609.0 &   3298.0 &   4592.3 &    684.6 &   4898.5 &   7119.8 &    816.7 &   7166.6 \\
20.25 &   4674.1 &    680.9 &   4466.6 &   6181.6 &    807.0 &   6443.9 &   8497.0 &    913.4 &   9290.9 \\
20.50 &   6343.3 &    793.3 &   6052.4 &   9204.2 &    992.4 &   8744.9 &  11152.5 &   1045.0 &  12079.0 \\
20.75 &   8464.5 &    921.1 &   8168.5 &   9517.3 &   1002.5 &  11124.2 &  10786.6 &   1030.8 &  14775.7 \\
21.00 &   8417.2 &    914.9 &  10272.8 &  11691.1 &   1116.5 &  14047.0 &  15092.2 &   1224.9 &  18548.7 \\
21.25 &  11834.9 &   1081.4 &  13231.6 &  12978.5 &   1178.5 &  17291.7 &  13970.6 &   1169.0 &  22041.4 \\
21.50 &  12273.7 &   1111.6 &  16300.0 &  13363.6 &   1191.6 &  20632.6 &  17009.5 &   1297.3 &  26293.7 \\
21.75 &  14373.0 &   1199.2 &  19893.2 &  16229.6 &   1319.8 &  24690.0 &  23914.1 &   1535.0 &  32272.3 \\
22.00 &  18201.4 &   1357.0 &  24443.6 &  21834.4 &   1536.5 &  30148.6 &  21793.7 &   1465.3 &  37720.7 \\
22.25 &  19396.3 &   1396.6 &  29292.7 &  26466.7 &   1684.3 &  36765.2 &  30044.2 &   1729.9 &  45231.7 \\
22.50 &  27849.6 &   1685.1 &  36255.1 &  30430.8 &   1799.7 &  44372.9 &  31809.6 &   1772.8 &  53184.1 \\
22.75 &  32572.9 &   1824.4 &  44398.3 &  33121.6 &   1887.9 &  52653.4 &  38564.6 &   1939.3 &  62825.3 \\
23.00 &  33383.1 &   1839.5 &  52744.1 &  39008.2 &   2042.3 &  62405.4 &  47694.9 &   2167.0 &  74749.0 \\
23.25 &  42430.3 &   2076.1 &  63351.6 &  49751.0 &   2312.7 &  74843.1 &  49461.9 &   2207.0 &  87114.5 \\
23.50 &  50740.0 &   2278.4 &  76036.6 &  52058.2 &   2372.2 &  87857.7 &  56856.0 &   2378.3 & 101328.5 \\
23.75 &  55735.1 &   2396.4 &  89970.4 &  59219.8 &   2549.0 & 102662.7 &  66811.4 &   2649.3 & 118031.4 \\
24.00 &  66881.7 &   2623.6 & 106690.8 &  70936.5 &   2903.4 & 120396.8 &  71444.1 &   2870.9 & 135892.4 \\
24.25 &  76601.7 &   2880.8 & 125841.2 &  85300.8 &   3424.7 & 141722.0 &  89161.3 &   3459.9 & 158182.7 \\
24.50 &  82958.4 &   3124.9 & 146580.9 &  97756.2 &   4236.4 & 166161.0 & 106466.9 &   4435.9 & 184799.5 \\
24.75 & 100728.4 &   3775.7 & 171763.0 & 108646.9 &   6676.6 & 193322.8 & 127600.9 &   6680.3 & 216699.7 \\
25.00 & 121070.0 &   5210.8 & 202030.5 & 124232.9 &  15534.6 & 224381.0 & 155102.4 &  21395.0 & 255475.3 \\
25.25 & 133487.5 &   9676.3 & 235402.3 &     --   &     --   &     --   & 163124.3 &  24634.1 & 296256.3 \\
\hline
\end{tabular}
\end{table}

\end{document}